\documentclass[pdflatex,sn-mathphys-num]{sn-jnl}
\usepackage{longtable}
\usepackage{graphicx}%
\usepackage{multirow}%
\usepackage{amsmath,amssymb,amsfonts}%
\usepackage{amsthm}%
\usepackage{mathrsfs}%
\usepackage{float}
\usepackage[title]{appendix}%
\usepackage{xcolor}%
\usepackage{textcomp}%
\usepackage{manyfoot}%
\usepackage{booktabs}%
\usepackage{algorithm}%
\usepackage{algorithmicx}%
\usepackage{algpseudocode}%
\usepackage{listings}%
\usepackage{subcaption}
\usepackage{tcolorbox}
\theoremstyle{thmstyleone}%
\theoremstyle{thmstyletwo}%

\theoremstyle{thmstylethree}%
\usepackage{lineno}  % Load the lineno package

\begin{document}
% \linenumbers  % Enable line numbering for the entire document

\title[Article Title]{Scaling Truth: The Confidence Paradox in AI Fact-Checking}

% Tracking number (Nature): NATCOMPUTSCI-25-1167

\author*[1]{\fnm{Ihsan A.} \sur{Qazi}}\email{ihsan.qazi@lums.edu.pk}

\author[1]{\fnm{Zohaib} \sur{Khan}}
\author[1]{\fnm{Abdullah} \sur{Ghani}}
\author[1]{\fnm{Agha A.} \sur{Raza}}
\author[1]{\fnm{Zafar A.} \sur{Qazi}}
\author[1]{\fnm{Wassay} \sur{Sajjad}}
\author[2]{\fnm{Ayesha} \sur{Ali}}
\author[1]{\fnm{Asher} \sur{Javaid}}
\author[1]{\fnm{Muhammad Abdullah} \sur{Sohail}}
\author[1]{\fnm{Abdul H.} \sur{Azeemi}}

\affil*[1]{\orgdiv{Department of Computer Science}, \orgname{Lahore University of Management Sciences}, \orgaddress{\city{Lahore}, \postcode{54792}, \country{Pakistan}}}

\affil[2]{\orgdiv{Department of Economics}, \orgname{Lahore University of Management Sciences}, \orgaddress{\city{Lahore}, \postcode{54792}, \country{Pakistan}}}

\abstract{The rise of misinformation underscores the need for scalable and reliable fact-checking solutions. Large language models (LLMs) hold promise in automating fact verification, yet their effectiveness across global contexts remains uncertain. We systematically evaluate nine established LLMs across multiple categories (open/closed-source, multiple sizes, diverse architectures, reasoning-based) using 5,000 claims previously assessed by 174 professional fact-checking organizations across 47 languages. Our methodology tests model generalizability on claims postdating training cutoffs and four prompting strategies mirroring both citizen and professional fact-checker interactions, with over 240,000 human annotations as ground truth. Findings reveal a concerning pattern resembling the Dunning-Kruger effect: smaller, accessible models show high confidence despite lower accuracy, while larger models demonstrate higher accuracy but lower confidence. This risks systemic bias in information verification, as resource-constrained organizations typically use smaller models. Performance gaps are most pronounced for non-English languages and claims originating from the Global South, threatening to widen existing information inequalities. These results establish a multilingual benchmark for future research and provide an evidence base for policy aimed at ensuring equitable access to trustworthy, AI-assisted fact-checking.}
%Our study establishes benchmarks for developing solutions and policy frameworks that can inform development of equitable access to reliable fact-checking systems.




\maketitle

\clearpage
\section{Introduction}\label{sec1}

As large language models (LLMs) become increasingly integrated into everyday life, their influence on the global information ecosystem is both undeniable and expanding ~\cite{Burton2024, Jones2025,DeVerna2024,Leippold2024,Gilardi2023,Vetter2025,NEURIPS2020_1457c0d6,touvron2023llama2openfoundation}. Popular search engines, such as Google Search, have begun experimenting with AI-generated summaries to accelerate information retrieval for users \cite{Reid2024}.
Independent fact-checking organizations are likewise incorporating generative AI into their verification processes, to streamline the rapid evaluation of online claims \cite{Poynter2024,ReutersInstitute2024}.
However, these developments raise critical questions about the effectiveness of LLMs as fact-checking tools and their impacts on information integrity, public trust, and democratic processes.

First, while LLMs excel at processing and summarizing information, they are prone to \emph{hallucinations}—generating plausible, human-like yet entirely false content—which heightens the risk that misinformation could be amplified ~\cite{Nicola2025, Hao2024,Ji_2023}. Compounding this issue is the widespread perception-reality gap, where many users hold inflated expectations about AI accuracy or remain unaware of known weaknesses ~\cite{Steyvers_2025,brauner2024}. A striking example involved a professor from a reputable institution who unknowingly included an LLM-generated reference list in expert testimony, later discovered to contain non-existent citations \cite{Daniel2025}.
These incidents underscore how AI-generated misinformation can swiftly enter authoritative discourse, particularly when users trust AI outputs without verification.

Second, equitable access to advanced LLMs remains a significant challenge. Because more capable models are costly and less accessible, many users and smaller organizations rely on smaller, less reliable alternatives, heightening inaccuracy risks ~\cite{sathish2024,Whiting2025}.
Compounding this challenge, LLMs frequently exhibit performance disparities in multilingual contexts, particularly for underrepresented languages ~\cite{unequal-opportunities,Sheng_2019}. These concerns resonate with recent platform-level interventions---such as Meta's ``Community Notes''---which rely on user-generated annotations to correct or contextualize misleading posts \cite{Meta2025}. 
However, if contributors to these community-driven systems rely on LLM-generated content, automation bias—the tendency to over-rely on automated systems—can create feedback loops that amplify rather than curtail the spread of misinformation. This risk is heightened by the fact that AI-generated text is often indistinguishable from human-written content, while people consistently overestimate their ability to detect it \cite{Porter_2024,Jakesch2023,Spitale2023}.
Consequently, AI-generated falsehoods can spread widely before identification and correction.

Third, deploying AI-driven fact-checking systems presents further challenges, especially under regulatory frameworks such as the EU AI Act \cite{AIAct2024}.
This initiative, among other international efforts, identifies certain AI applications as high-risk when they can significantly influence public trust and democratic processes. The classification as high-risk arises from a risk-based assessment defined in the Act, which considers the potential for societal harm, implications for fundamental rights, and the need for accountability. Such AI systems must therefore meet stringent standards for reliability and fairness to mitigate the risks associated with misinformation. However, the absence of benchmark datasets that feature human-validated LLM responses to fact-checking claims remains a significant barrier to achieving consistent calibration and standardization of these systems.

Current evaluations of LLMs for fact-checking use researcher-crafted prompts and structured response formats rather than studying natural user interactions \cite{hoes_altay_bermeo_2023, Quelle_2024}.
These methods rely on predetermined response structures (such as mandated JSON responses with explanations) \cite{he2024doespromptformattingimpact, tam2024letspeakfreelystudy}, and utilize datasets with limited geographic scope \cite{ma-etal-2024-ex}, inadequately representing global misinformation complexities or realistic user interactions. Even when assessing multilingual capabilities, evaluations often overlook the critical aspect of native language prompting, often using English instructions even for non-English claims \cite{pelrine-etal-2023-towards, Yifan2023}.
% Current evaluations of LLMs for fact-checking prioritize controlled experiments with artificial prompting over real-world user scenarios. These approaches use structured outputs (such as JSON-formatted responses with explanations) \cite{he2024doespromptformattingimpact, tam2024letspeakfreelystudy}, and datasets limited in geographic scope \cite{ma-etal-2024-ex}, which may not reflect the global misinformation complexities or typical user interactions. Even when assessing multilingual capabilities, evaluations often overlook the critical aspect of native language prompting, often using English instructions even for non-English claims \cite{pelrine-etal-2023-towards, Yifan2023}.

Our work addresses these gaps by introducing a methodology that \textcolor{black}{mirrors} real-world user experiences with global misinformation. Leveraging a dataset of real-world claims sourced globally, across diverse linguistic and regional contexts, we evaluate LLMs using simple, unstructured prompts (e.g., ``Is this true?,'' ``Is this false?,'' ``Is this true or false?'') that mirror lay person queries \cite{dawn2025,TrippasSIGIR2024PromptLogAnalysis, Kelly2025ChatGPT, tam2024letspeakfreelystudy}. Moreover, we translate these prompts into the claim's language for multilingual cases, capturing the nuances of native-language interaction. We also evaluate using a system prompt (instructions that define the model's role and behavior) that adopts the persona of a professional fact-checker—inspired by International Fact-Checking Network (IFCN) guidelines \cite{IFCN2024}—and compare these results with typical user promptings. This approach assesses LLMs' practical utility as accessible fact-checking tools across diverse linguistic and regional contexts, offering insights into their real-world impact on misinformation—a perspective largely unexplored in previous research.

To this end, we evaluate nine established models spanning multiple categories (open/closed-source, multiple sizes, diverse architectures, reasoning-based) to assess their effectiveness as fact-checking tools. We employ four distinct prompting strategies and test the models on a high-quality, diverse dataset of 5,000 factual claims spanning multiple languages, domains, and countries across both the Global North and the Global South. To ensure accuracy and reliability, each model’s responses undergo review by multiple annotators. Beyond standard classification metrics, we employ additional measures—namely, \textit{Selective Accuracy}, \textit{Abstention-Friendly Accuracy} and \textit{Certainty Rate}—to offer a more holistic perspective on how these models balance confidence and correctness.

Our findings reveal critical trade-offs in model behavior and provide actionable insights into improving the robustness and fairness of LLMs for scalable misinformation mitigation. In particular, we uncover trade-offs in fact-checking between small and large LLMs reminiscent of the Dunning-Kruger effect ~\cite{Kruger_1999}—where smaller models display high confidence despite lower accuracy, while larger models show low confidence but high accuracy. This pattern has far-reaching implications for the information ecosystem. Specifically, when combined with the economic reality that resource-constrained organizations often rely on smaller models, it risks creating a systemic bias where the most accessible LLM based fact-checking tools are also the most miscalibrated. As automated fact-checking systems proliferate, this dynamic could amplify existing information inequalities between well-resourced and resource-constrained contexts, underscoring the need for both technical solutions in confidence calibration and policy frameworks that ensure equitable access to reliable verification tools. Such insights, in turn, help shape best practices for the ethical and responsible integration of generative AI into the global fact-checking landscape.

\section{Results}

\subsection{Study Design, Dataset, and Models}
\label{sec:study-design}
The dataset of true and false claims used in this study was compiled using Google Fact Check (GFC) Explorer \cite{GoogleFactCheckTools}; a tool that aggregates fact checks from various independent fact-checking organizations worldwide.
Using the GFC Claim Search API, we fetched claims verified by 250 fact-checking organizations globally, yielding 302,288 claims. Data cleaning removed 60,741 duplicate claims, identified within or across publishers.
While all fact-checking organizations use true/false categories, they vary significantly in their intermediate labels like ``lacks context,'' ``partially true,'' or ``misleading.'' Forcing these nuanced labels into binary categories risks misrepresenting ground truth.
We addressed this by selecting only claims with consistent true/false ratings across all fact-checking organizations, excluding cases with intermediate labels.

Using a structured codebook based on professional fact-checker criteria, trained annotators ensured only claims with unambiguous true/false ratings were included.
% \textcolor{blue}{Veracity labels across fact-checking organizations vary significantly in scale and meaning, and attempts to standardize nuanced labels such as ``partially true'', ``half-true'', ``misleading'', or ``sarcasm'' into a simplified label set risk misrepresenting their ground-truth veracity. To minimize this source of bias, we developed a structured codebook that adopted the definitional criteria used by professional fact-checking organizations to classify claims.  Trained human annotators used this codebook to apply consistent standards, selecting only those claims that were clearly rated as strictly true or false, excluding cases with ambiguous or intermediate assessments.} 
This resulted in 137,393 claims containing 125,215 false claims and 12,178 true claims.
From this, we constructed an evaluation dataset by randomly sampling 5,000 claims, weighted by language distribution.  We selected equal numbers of true and false claims per language when possible. Where true claims were insufficient, we included all available true claims and added randomly selected false claims to maintain the intended language distribution. The resulting evaluation set comprised 2,967 false and 2,033 true claims across 47 languages, fact-checked by 174 unique publishers, \textcolor{black}{yielding a dataset that mirrors the complexity and diversity of the real-world information ecosystem}. These claims spanned from March 2009 to June 2023. We evaluated nine LLMs on this dataset: Llama-2 (7B, 13B, 70B), Mistral (\texttt{Mistral-7B}, \texttt{Mixtral-8x7B}) ~\cite{jiang2023mistral7b,jiang2024mixtralexperts}, and OpenAI's GPT series (\texttt{GPT-3.5}, \texttt{GPT-4}, \texttt{GPT-4o}, \texttt{o1-preview}) \cite{OpenAIModels2024}. To address potential data contamination from claims possibly seen during training, we also test model generalizability on claims published after their respective knowledge cutoffs (see Supplementary Section A.5). 

\begin{figure}
    \centering
    \includegraphics[width=0.85\linewidth]{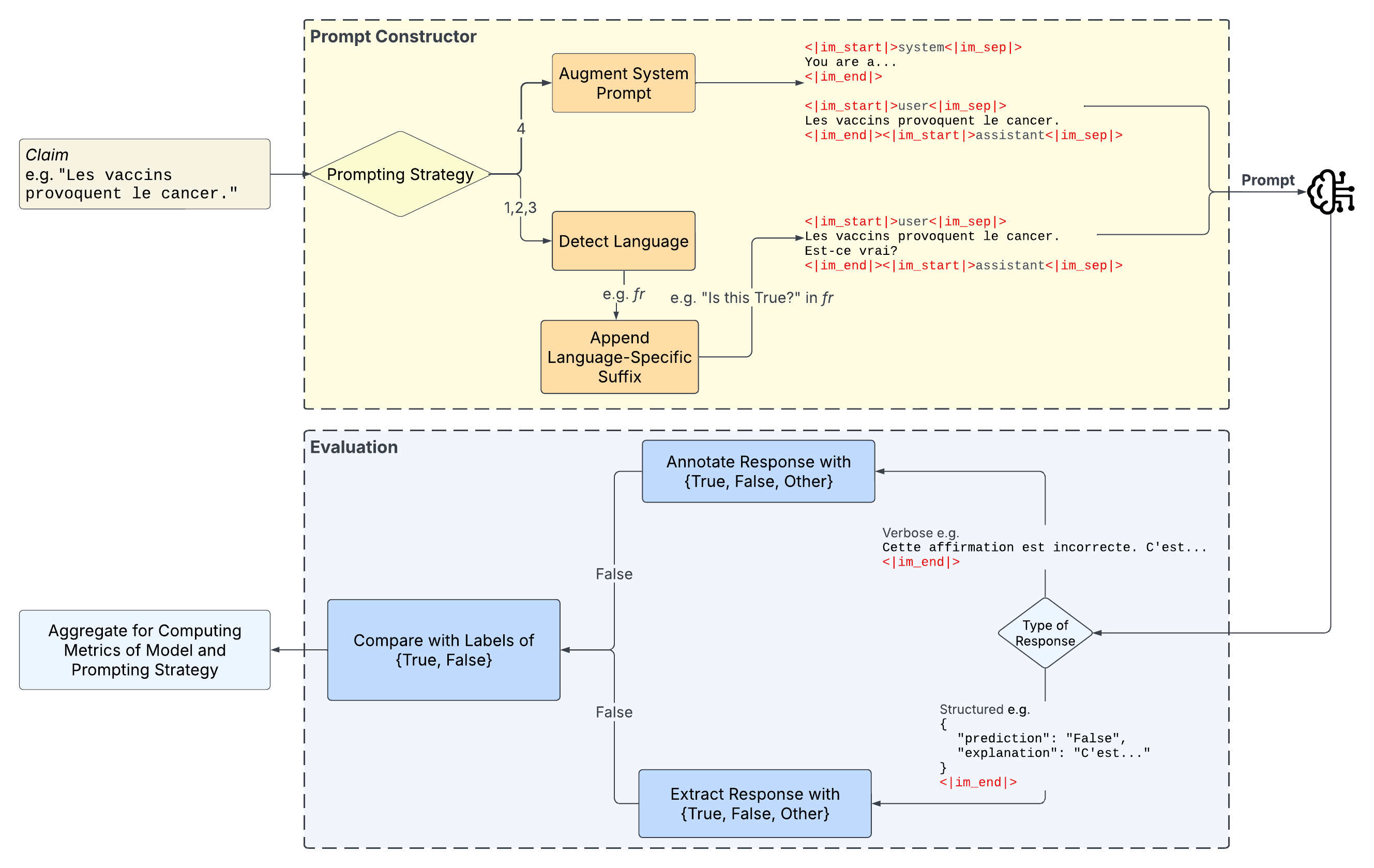}

    \caption{Overview of Prompting Strategies and Response Evaluation for LLM Fact-Checking. (Orange) The Prompt Constructor dynamically constructs a tailored prompt to elicit factual assessments from the LLM. The system first receives a factual claim, such as ``Les vaccins provoquent le cancer.'' Based on the selected prompting strategy (1, 2, or 3), the system appends a language-specific user instruction suffix---for example, ``Est-ce vrai?'' for French---ensuring the claim is framed appropriately for the model. Alternatively, if Prompting Strategy 4 is selected, the system prompt itself is modified to include the instruction (e.g., ``You are a\ldots''), while the user message contains only the raw claim without a language-specific suffix. In both cases, the system utilizes structured prompt tags like \texttt{<|im\_start|>} and \texttt{<|im\_end|>} to clearly define system, user, and assistant segments. The constructed prompt is then submitted to the LLM for inference. (Blue) The Evaluation module processes the LLM's response. Responses are annotated as `True', `False,' or `Other,' depending on the model's output. The model may return either verbose explanations (e.g., ``Cette affirmation est incorrecte\ldots'') or structured outputs in JSON format (e.g., \texttt{\{``prediction'': ``false'', ``explanation'': ``C'est\ldots''\}}). Responses are then compared against gold-standard labels (True or False) for correctness. Aggregation metrics are computed to assess model accuracy across different prompting strategies. This pipeline enables comprehensive evaluation by combining prompt variation, multilingual support, structured output processing, and flexible annotation strategies.}
    \label{fig:inference_pipeline}
\end{figure}

We evaluated LLMs using four prompting strategies. Research shows typical users employ simple, direct prompts without prompt engineering techniques and rarely use structured output formats, despite their impact on model performance \cite{dawn2025,TrippasSIGIR2024PromptLogAnalysis, Kelly2025ChatGPT, tam2024letspeakfreelystudy}.
For example, ``\emph{Hey @Grok, is this true?}'' has become a common query on the X platform \cite{dawn2025}.
Thus, to mirror real user behavior \textcolor{black}{and assess robustness against minor variations in phrasing,} we used three basic prompts with unstructured output direct truth inquiry (\texttt{``{claim}. Is this True?''}) (Prompt 1), direct falsity inquiry (\texttt{``{claim}. Is this False?''}) (Prompt 2), and binary choice (\texttt{``{claim}. Is this True or False?''}) (Prompt 3). The fourth strategy used a specialized system prompt (Prompt 4) adopting a professional fact-checker persona with IFCN-aligned guidelines \cite{IFCN2024} and structured JSON output (see Supplementary Fig. B1). Due to cost constraints, we evaluated \texttt{o1-preview} only with the specialized prompt.

% We evaluated these nine LLMs using four prompting strategies across structured and unstructured output formats. \textcolor{blue}{Recent work analyzing interactions of common users with LLMs suggests that they employ prompts that are straightforward and concise, and often do not make use of prompt-engineering techniques \cite{TrippasSIGIR2024PromptLogAnalysis, Kelly2025ChatGPT}.
% For example, ``Hey @Grok, is this true?'' has become a common query on the X platform \cite{dawn2025}.
% Moreover, basic users often do not restrict an LLM to output a verdict in a structured manner (e.g., JSON format) as used by experts or developers, even though it can impact a model’s downstream performance \cite{tam2024letspeakfreelystudy}}. We use three basic, direct queries without specifying any structured output to mirror such common interactions from non-AI experts: a direct truth inquiry (\texttt{\{claim\}. Is this True?''}) (Prompt 1), a direct falsity inquiry (\texttt{\{claim\}. Is this False?''})  (Prompt 2), and a binary choice prompt (\texttt{``\{claim\}. Is this True or False?''})  (Prompt 3). The fourth strategy implemented a specialized system prompt  (Prompt 4) that adopts the persona of a professional fact-checker using guidelines aligned with the IFCN standards \cite{IFCN2024} and specified a structured output format for responses; the prompt is provided in Section~\ref{fig:sys-prompt}. We evaluated the \texttt{o1-preview} model using only the specialized system prompt due to cost constraints. 

To mirror typical user interactions, the first three user-level prompts were presented in the claim's original language. A French claim, for example, was prompted with \texttt{``\{claim in French\}. Est-ce vrai?''} instead of \texttt{``\{claim in French\}. Is this True?''}. All non-English model responses were translated into English using the Google Translate API. Given the unstructured nature of LLM responses to these user-level prompts, two independent annotators categorized them as `True,' `False,' or `Other' (Fig.~\ref{fig:inference_pipeline}). A response was classified as `True' or `False' only when the LLM explicitly and unequivocally labeled the claim. Responses showing ambiguity or uncertainty, including terms like `probably', `possibly', or `partly', or failing to provide a definitive true/false assessment, were categorized as `Other.' We assessed inter-annotator agreement using Cohen's Kappa. Across 24 annotation sets (representing three prompts for each of the eight LLMs, each containing 5,000 claims), the mean Cohen's Kappa score was 0.84, with a minimum of 0.65, indicating substantial agreement. Annotation discrepancies were resolved through discussion, ensuring a consistent and reliable final dataset for analysis.

%\subsection{Performance gaps widen between conversational and specialized system prompting approaches across LLMs in multilingual claim verification}
%\subsection{Impact of conversational and specialized system prompting approaches on multilingual claim verification in LLMs}
\subsection{Conversational versus specialized prompting effects on multilingual LLM fact verification}

To evaluate the fact-checking performance of each model across the various prompts, we report three primary metrics: selective accuracy, abstention-friendly accuracy, and certainty rate. 
These metrics are rooted in the principles of selective classification, following the seminal work of Chow \cite{Chow1970IEEEPress, Chow1957IRETEC}, which allow models to abstain from making predictions when uncertain. Such an approach provides a more nuanced evaluation of model reliability and calibration, as established in prior work \cite{traub2024overcomingcommonflawsevaluation, madhusudhan2024llmsknowanswerinvestigating}. For additional robustness checks, we also report selective accuracy over claims common across all models (see Supplementary Fig. B3) as well as standard metrics of precision, recall, and F1-score, calculated over both selective (excluding `Other') and complete sets of claims.

Selective accuracy is the proportion of those definitive `True' or `False' responses that correctly matched the ground-truth veracity label.
Abstention-friendly accuracy measures the proportion of all claims where the LLM either provides a correct `True' or `False' verdict or abstains with an `Other' response, This metric is particularly valuable in sensitive domains like law, medicine, and security, where incorrect predictions can have serious consequences \cite{Weidinger2022Risks, lin2022truthfulqameasuringmodelsmimic, xu2025largelanguagemodelscyber, Yinheng2023LLMFinance}.
The certainty rate (\emph{coverage} in selective classification literature) represents the proportion of all claims for which the LLM provided a definitive `True' or `False' response.
% This methodological choice reflects recent findings that current LLMs lack an internally coherent sense of confidence, which could render explicit confidence queries misleading. \cite{Pawitan2025Confidence}.
\textcolor{black}{Our use of certainty rate as a proxy to assess the confidence of black-box LLMs serves as a methodological approach that addresses the limitations of existing confidence estimation techniques for real-world fact-checking tasks involving free-text evaluations.
%within a black-box, free-text evaluation context.
First, directly querying a model for its confidence score is unreliable due to their demonstrably poor calibration properties that compromise the validity of self-reported uncertainty estimates~\cite{yang2024trustllmsmitigateoverconfidence,Pawitan2025Confidence}.
%as existing confidence estimation techniques are ill-suited for our real-world evaluation framework.
%that addresses the limitations of existing confidence estimation techniques.
%as a number of existing alternatives are ill-suited for our real-world evaluation framework.
%, as current LLMs lack a calibrated sense of self-certainty~\cite{Pawitan2025Confidence}.
% Second, confidence metrics derived from token-level log probabilities are poorly suited to the unstructured, free-text outputs elicited by our user-level prompts. While effective for tasks with fixed-class outputs (e.g., a single `True' or `False' token), this method is intractable when a model can express its verdict in countless ways.
Second, while metrics derived from token-level log-probabilities are useful for fixed-label tasks (e.g., a single `True' or `False' token), they perform poorly on free-form generation. This degradation occurs because such metrics become undefined when multiple semantically equivalent phrasings are acceptable; a common characteristic of the unstructured, free-text outputs produced by our user-level prompts~\cite{Steyvers_2025}.
Moreover, the P(True) self-evaluation method~\cite{Steyvers_2025, kadavath2022languagemodelsmostlyknow}—a sophisticated workaround for free-text outputs—is poorly calibrated, especially for lengthy responses, as models are often overconfident when assessing their own outputs~\cite{kadavath2022languagemodelsmostlyknow}.}

\begin{figure}
    \centering
    \includegraphics[width=0.75\linewidth]{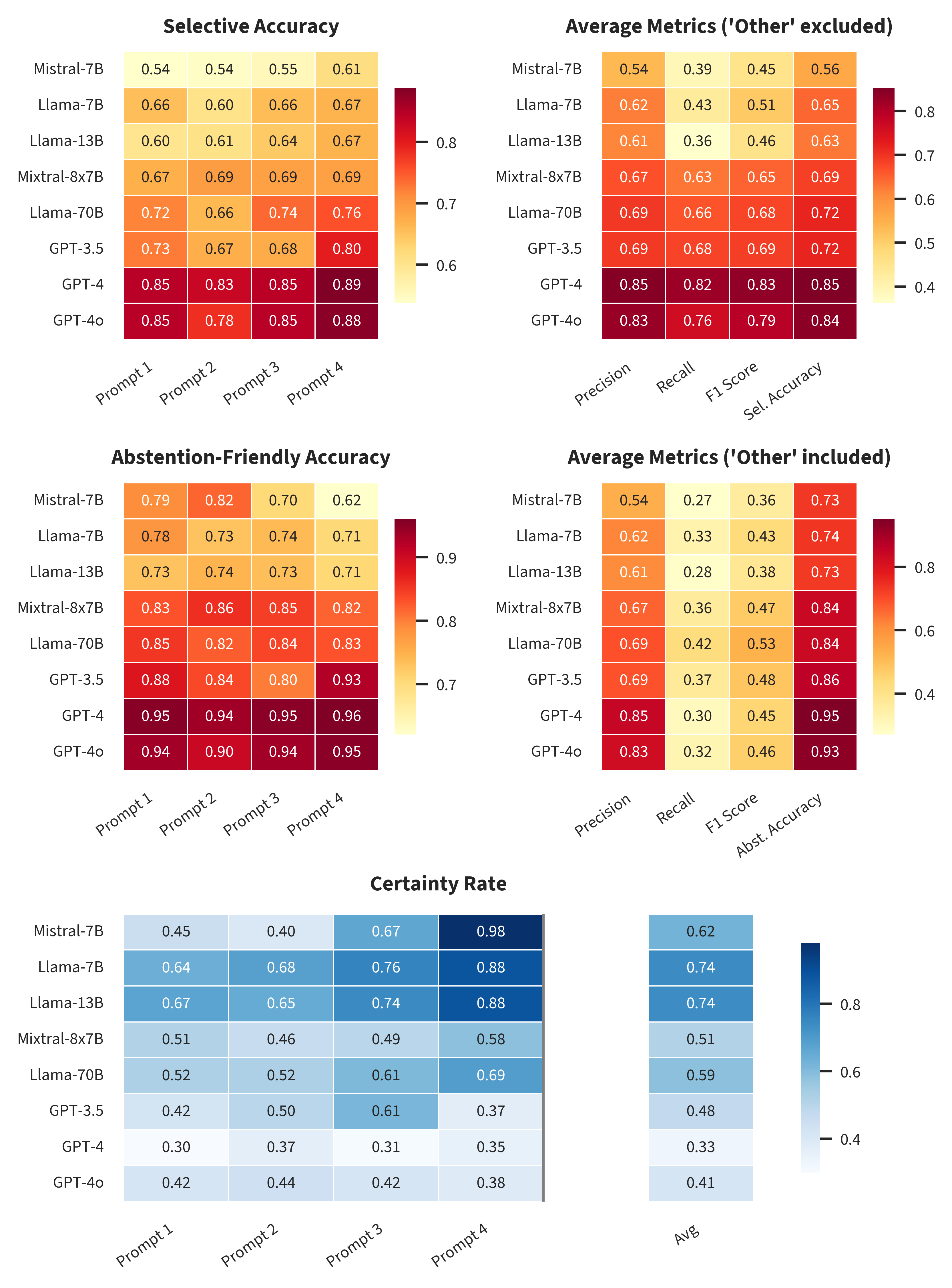}
    \caption{\textcolor{black}{Performance of LLMs across four prompting strategies. The figure provides a detailed breakdown of model performance. (Top-left) Heatmap of selective accuracy, showing the proportion of correct verdicts among all definitive `True'/`False' responses. (Top-right) Standard classification metrics (Precision, Recall, and F1 Score) and calculated exclusively on this same subset of definitive responses (`Other' excluded) and Selective Accuracy aggregated across prompts. (Middle-left) Heatmap of abstention-friendly accuracy, which measures the proportion of all claims where the model was either correct or appropriately abstained ('Other'). (Middle-right) Standard classification metrics calculated across all claims ('Other' included) and Abstention-Friendly Accuracy aggregated across prompts. Note that this penalizes the Recall and F1 Score of more cautious models that abstain frequently. (Bottom-left) Heatmap of certainty rate, representing the proportion of all claims for which models provided a definitive 'True'/'False' response. Refer to Supplementary Fig. B4, Supplementary Fig. B5, and Supplementary Fig. B6 for pair-wise statistical comparisons.}}
    \label{fig:model-prompting-influences}
\end{figure}

Prompting strategy significantly influences model performance, affecting both accuracy and confidence in responses (Fig.~\ref{fig:model-prompting-influences}).
The system-level prompt consistently achieves the highest selective accuracy across all models by providing structured reasoning guidelines that reduce ambiguity. 
%The system-level prompt consistently achieves the tends to show the highest selective accuracy across the four prompts for all models. This indicates that system-level prompts, by providing structured reasoning guidelines, reduce ambiguity and encourage models to better assess the claim’s veracity, leading to more reliable outputs. 
%
However, system prompts create a trade-off for smaller models: while improving selective accuracy, they reduce abstention-friendly responses. This suggests smaller models over-interpret structured directives, committing to verdicts even when abstention would be safer, unlike larger models that maintain more discerning abstention behavior.
%\textcolor{blue}{Interestingly, system prompts, while improving selective accuracy overall, reduce abstention-friendly responses for smaller models. This suggests that smaller models, when provided with structured guidelines, may over-interpret the directive to provide a definitive verdict, leading them to commit more often even in situations where a safer response would have been to abstain, compared to larger models, which exhibit a more discerning approach to abstention.}
%
Regarding certainty rates, smaller models (Llama-2, Mistral) show substantial increases in certainty rate with system prompts, while larger models (GPT-family) show minimal differences. For larger, more sophisticated models, the system prompt enhances judgment quality (accuracy) rather than simply increasing prediction quantity (certainty), likely due to a combination of their inherent capabilities and the prompt's structured guidance.
On average, larger models also demonstrate higher Precision than smaller models. While Recall shows a similar improving trend with model size when 'Other' responses are excluded, its value is lower when 'Other' responses are included, as larger models that abstain are penalized in this calculation.
%As for the certainty rate, relatively smaller models (Llama-2 and Mistral families) tend to show a substantial increase in certainty rate for system prompts compared to user-level prompts, while relatively larger models (GPT-family) do not exhibit a notable difference. This indicates that for such larger, more sophisticated models, the primary advantage of system prompts lies in enhancing the quality of their judgments (accuracy) rather than simply increasing their quantity (certainty), likely due to a combination of their inherent capabilities and the prompt's structured guidance. 
%

In summary, user-level prompts yield lower selective accuracy than a structured system prompt but maintain better abstention behavior in smaller models. Among user-level approaches, binary choice prompts  (``\{claim\}. Is this True or False?'') moderately improve both selective accuracy and confidence across most models, likely through framing effects that orient outputs toward explicit options.
%Overall, the findings show that user-level prompts tend to yield lower selective accuracy compared to curated, structured prompts (which are generally utilized for assessing large language model performance); \textcolor{blue}{however, for smaller models, these structured prompts can paradoxically decrease abstention-friendly accuracy despite increases in certainty.}
%Overall, the findings show that user-level prompts tend to yield lower selective accuracy compared to curated, structured prompts (which are generally utilized for assessing large language model performance); \textcolor{blue}{however, for smaller models, these structured prompts can paradoxically decrease abstention-friendly accuracy despite increases in certainty.}
%
% \textcolor{blue}{Among the user-level prompts, the binary choice prompt (``\{claim\}. Is this True or False?'') tends to increase the selective accuracy as well as the confidence of most models, though the increase in selective accuracy over the other user-level prompts is modest. This suggests that the binary choice prompt creates a framing effect, tending to orient the model's output towards one of the explicitly provided options.}

%\subsection{LLMs exhibit Dunning-Kruger effect: Larger models show higher accuracy but lower confidence in claim verification}
\subsection{Dunning-Kruger effect in LLM fact-checking: Capability inversely correlates with confidence}

\begin{figure}
    \centering
    \includegraphics[width=0.9\linewidth]{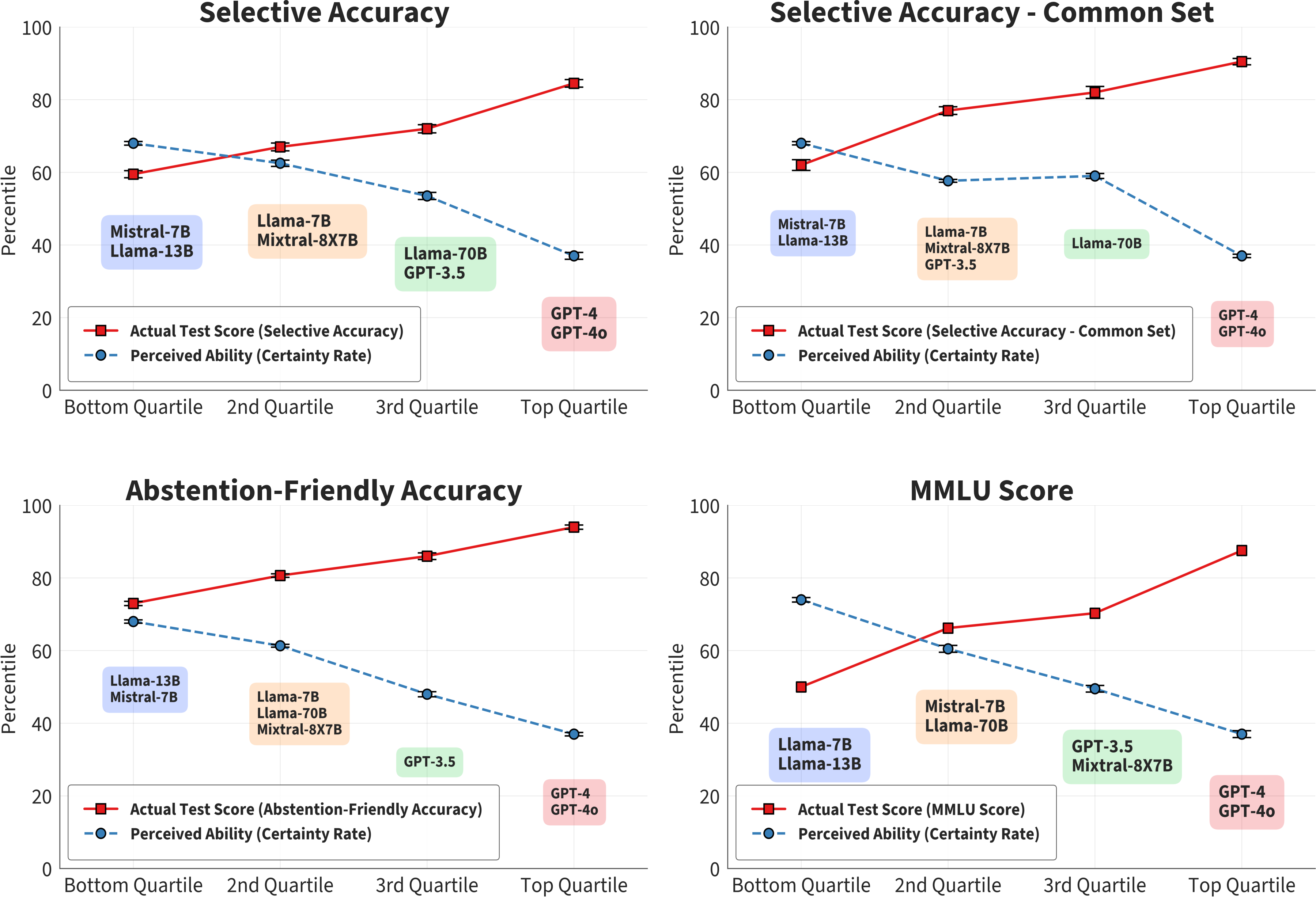}
    \vspace{1em}
    \caption{\textcolor{black}{The relationship between actual ability and perceived ability reveals a statistical pattern across model architectures similar to the Dunning-Kruger effect. (Top-left) Performance quartile analysis comparing actual ability (selective accuracy) to perceived ability (certainty rate) averaged across all prompts. Less capable models (\texttt{Llama-7B}, \texttt{Llama-13B}, \texttt{Mistral-7B}) systematically overestimate their abilities, while more advanced models (e.g., \texttt{GPT-4}, \texttt{GPT-4o}) tend to underestimate theirs.  This inverse relationship persists across multiple evaluation frameworks: when using a common claim set of claims for selective accuracy (top-right), when measuring abstention-friendly accuracy that penalizes confident misinformation propagation (bottom-left), and when benchmarking against MMLU scores (bottom-right).  The statistical pattern indicates that model scale correlates with calibrated self-assessment, with larger architectures producing confidence estimates that better align with their actual capabilities. Certainty rates and accuracy measures represent averaged values across prompts. Error bars show 95\% confidence intervals (two-sided tests without adjustments for multiple comparisons). Similar trends hold for each prompt (see Supplementary Fig. B7)}} 
    \label{fig:dunning-kruger}
\end{figure}

Model size creates a trade-off between selective accuracy and certainty rate (representing an LLM’s self-perceived confidence), which is illustrated in Fig.~\ref{fig:dunning-kruger}. Larger models, such as \texttt{GPT-4o}, achieve high selective accuracy, reaching up to 89\%, excelling in definitive judgments when they commit. However, these models also display a more pronounced tendency toward epistemic caution, evidenced by certainty rates below 40\% when confronting complex or ambiguous claims.
While this cautious response pattern may reflect better statistical calibration to knowledge limitations, these higher abstention rates can substantially impair operational efficiency in real-world fact-checking applications where timely decisions are crucial.
%\textcolor{black}{While this cautious response pattern} may reflect their understanding of knowledge limitations, their hesitancy to provide definitive judgments can substantially impair operational efficiency in real-world fact-checking applications where timely decisions are crucial.
Smaller models, like \texttt{Llama-7B}, demonstrate much higher certainty rates, reaching up to 88\%, providing definitive responses across a wider range of claims. However, their confidence is often misplaced due to limited training data and reasoning capabilities, resulting in lower selective accuracy, which falls around 60\%. While prompting strategies help mitigate some of these trade-offs, core differences between small and large models remain.

This tradeoff resembles the Dunning-Kruger effect, where people with lower competence overestimate their abilities, while those with higher competence underestimate theirs \cite{Kruger_1999}.
LLMs display similar statistical patterns when selective accuracy and certainty rates are used as measures of actual and perceived ability, respectively, with smaller models showing high confidence despite lower accuracy, while larger models hedge responses even when correct (Fig.~\ref{fig:dunning-kruger}).
Smaller models like \texttt{Llama-7B} frequently commit to definitive but incorrect answers, likely due to limited training data and reasoning abilities. In contrast, larger models like \texttt{GPT-4} exhibit greater hesitation with lower certainty rates. 
\textcolor{black}{Robustness checks confirm this pattern persists across multiple benchmarks: when evaluating only claims where all models provide definitive responses, when using MMLU scores as ability measures, and when assessing the proportion of correct or appropriately uncertain responses--a critical metric for preventing confident misinformation propagation. This cautious approach enables superior selective accuracy in larger models.}
However, this tendency may hinder their effectiveness in applications requiring rapid fact verification. This also reveals a core challenge in LLM design: smaller models, though decisive, risk propagating errors due to overconfidence, while larger models, though precise, exhibit excessive statistical caution. The Dunning-Kruger framework illuminates these patterns, reinforcing the need to optimize both certainty rate and selective accuracy for robust fact-checking systems.
%\textcolor{blue}{As a robustness check, we also find that the behavior persists when selective accuracy is computed across the set of claims all models are confident on (i.e., claims all model responses labeled as ``True''/``False''), when MMLU (Massive Multitask Language Understanding) scores are used as the benchmark for actual ability, and when actual ability is evaluated as the proportion of either correct or uncertain instances out of all claims, a critical benchmark for fact-checking frameworks marking instances where misinformation is not confidently propagated. This cautious approach contributes to their superior selective accuracy, as they are typically precise when confident.} However, this deliberative tendency may hinder their effectiveness in applications requiring rapid fact verification. This relationship between model size, confidence, and accuracy highlights a core challenge in LLM design: smaller models, though decisive, risk propagating errors due to overconfidence, while larger models, though precise, struggle with decisiveness from their cautious nature. The Dunning-Kruger framework illuminates these patterns, reinforcing the need to balance certainty rate and selective accuracy for robust fact-checking tools.

%\subsection{Reasoning models achieve both high accuracy and better confidence but are costly}

OpenAI's \texttt{o1-preview} model, leveraging enhanced reasoning capabilities, achieves both high selective accuracy (84\%), abstention-friendly accuracy (88\%) and certainty rate (72\%) with statistically significant differences ($p < 0.001$) compared to almost all other models (See Fig. ~\ref{fig:cr-ca-heatmap-sub}; for statistical comparisons, see Supplementary Fig. B15, Supplementary Fig. B16, and Supplementary Fig. B17). By narrowing the gap between accuracy and confidence, \texttt{o1-preview} establishes a new benchmark for robust claim verification and demonstrates how improvements in LLM reasoning capabilities can help overcome limitations of non-CoT based LLMs. These improvements suggest that architectural enhancements can help bridge this divide, potentially enabling AI-driven fact-checking systems that are both reliable and appropriately confident. This balance is crucial for developing AI systems capable of effectively supporting misinformation detection and verification workflows. While o1-preview delivers significant performance improvements, this comes at a substantial computational cost of \$88.75 per 1,000 claims (Figure~\ref{fig:inference-costs-sub}), limiting its viability for cost-sensitive applications. In contrast, smaller models like \texttt{Llama-2-7B} and \texttt{Mistral-7B} cost less than \$0.10 per 1,000 claims but offer limited accuracy. \texttt{GPT-4o}, at \$2.22 per 1,000 claims, strikes the best balance between cost and performance, appearing to be the most practical option for scalable, reliable fact-checking among the rest.

\begin{figure}[t] % Use placement specifiers as needed
    \centering % Center the contents of the figure
    % --- First Subfigure (Top Row) ---
    \begin{subfigure}[b]{0.75\linewidth} % Subfigure takes full width, aligns at bottom
        \centering
        \includegraphics[width=\linewidth]{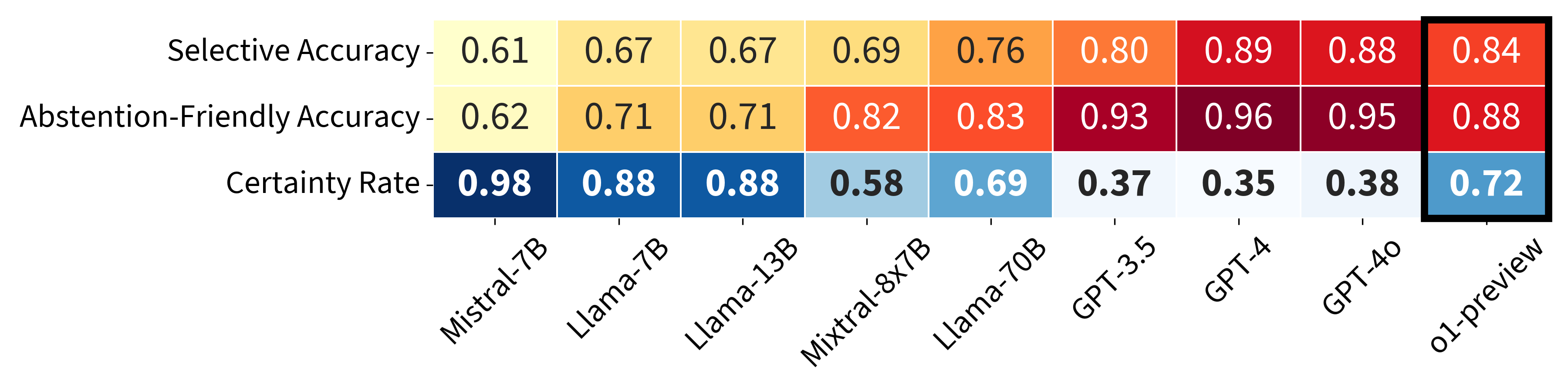}
        \caption{}
        \label{fig:cr-ca-heatmap-sub} % Renamed label slightly to avoid conflict if needed
    \end{subfigure}
    \vspace{-1em}
    % --- Second Subfigure (Bottom Row) ---
    \begin{subfigure}[b]{0.75\linewidth} % Subfigure takes 90% width, aligns at bottom
        \centering
        \includegraphics[width=\linewidth]{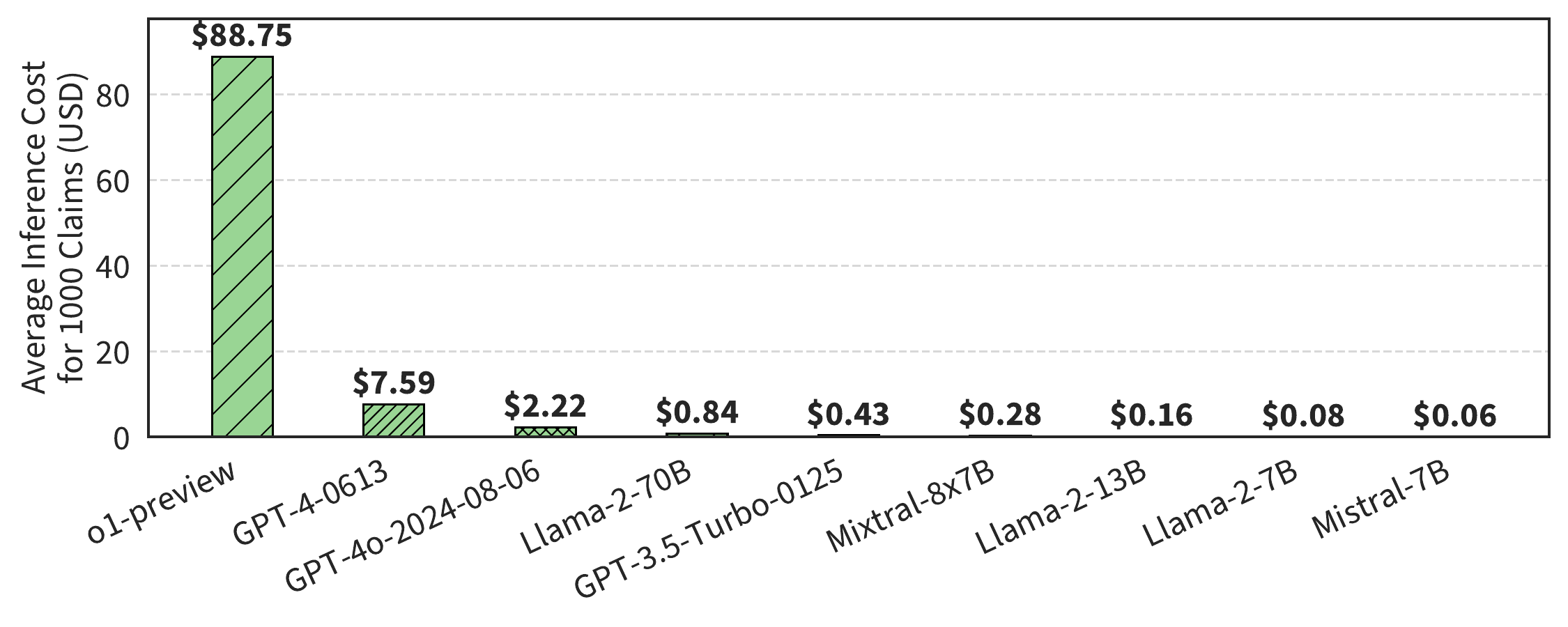} % Image width relative to subfigure width
        \caption{}
        \label{fig:inference-costs-sub} % Renamed label slightly
    \end{subfigure}
    % --- Optional: Add an overall caption for the combined figure ---
    \caption{Comparison of model performance and cost analysis. (a) Performance comparison across models using Prompt 4 versus \texttt{o1-preview}. \texttt{o1-preview} achieves both high accuracy and high certainty rate while narrowing the gap between them, indicating that reasoning-enhanced models may be effective for automatic fact-checking. (Top) Selective accuracy comparisons using McNemar's test (with adjustments for multiple-comparisons using the Holm-Bonferroni method) show statistically significant differences ($p < 0.001$) between \texttt{o1-preview} and all other models except \texttt{GPT-4o} and \texttt{GPT-4}. Analysis included correct labels (1) and incorrect labels (0), with sample sizes ranging from 1{,}156 to 4,394 claims (average $n = 2,267$). Only definitive responses (i.e., `True' or `False') common across pairs were included. (Bottom) Abstention-friendly and certainty rate comparisons ($n = 5000$ claims) using McNemar's test (with adjustments for multiple-comparisons using the Holm-Bonferroni method) also show statistically significant differences ($p < 0.001$) between \texttt{o1-preview} and all other models. (b) Estimated API-based inference costs (USD) for verifying 1,000 claims. Costs were calculated based on the actual token usage in our experiments and API pricing as of November 29, 2024. Open-source models were run using commercial inference APIs (i.e. Together AI and DeepInfra), and their respective pricing was used for cost estimation. Together, the panels illustrate the significant performance-cost trade-off.}
    \label{fig:combined-performance-cost} % New label for the combined figure
\end{figure}

% \begin{figure}
%     \centering
%     \includegraphics[width=\linewidth]{figures/prompt-4-performance.png}
%     \caption{Performance comparison across models using Prompt 4 versus \texttt{o1-preview}. \texttt{o1-preview} achieves both high selective accuracy and high certainty rate while narrowing the gap between them, indicating that reasoning-enhanced models may be effective for automatic fact-checking. (Top) Certainty rate comparisons ($n = 5000$ claims) using Welch's \textit{t}-test show statistically significant differences ($p < 0.001$) between \texttt{o1-preview} and all other models. (Bottom) selective accuracy comparisons using Welch's \textit{t}-test also show statistically significant differences ($p < 0.001$) between \texttt{o1-preview} and all other models. Analysis included correct labels (1) and incorrect labels (0), with sample sizes ranging from 1,763 to 4,902 claims (average $n = 3,188$). Only definitive responses (i.e., ``True'' or ``False'') were included in the latter case.}
%     \label{fig:cr-ca-heatmap}
% \end{figure}
% \begin{figure}
%     \centering
%     \includegraphics[width=0.9\linewidth]{figures/average_inference_cost.pdf}
%     \caption{Estimated inference costs (USD) for verifying 1,000 claims across evaluated LLMs, based on pricing as of November 29, 2024.}
%     \label{fig:inference-costs}
% \end{figure}

\subsection{Geographic and linguistic disparities in LLM fact-checking performance}
%\subsection{Model accuracy and confidence declines across less prevalent languages}

\begin{figure}
    \centering
    \includegraphics[width=0.9\linewidth]{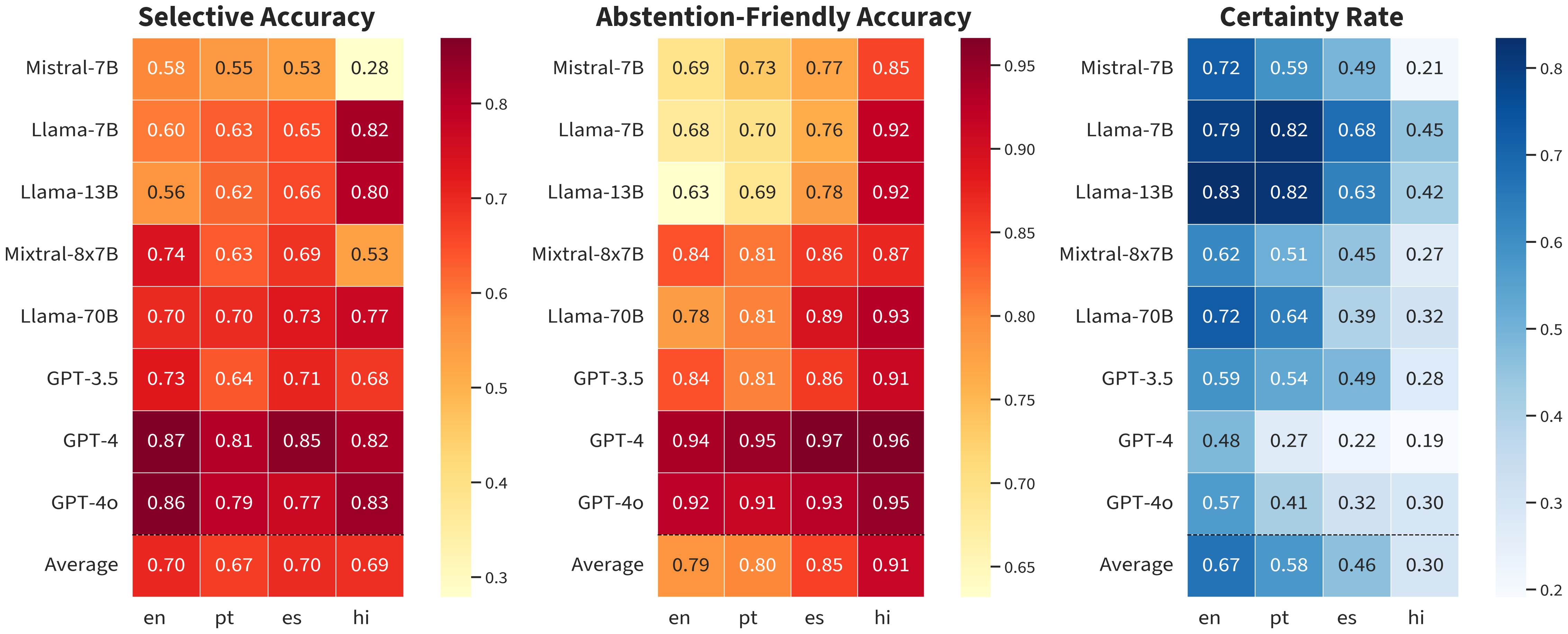}
    \caption{\textcolor{black}{Selective accuracy, abstention-friendly accuracy, and certainty rate of models across the top four languages (English, Spanish, Hindi, and Portuguese) in our dataset. Each cell represents the average score across prompts 1, 2, and 3. The `Average' row denotes the average performance of all models for that language. Refer to Supplementary Fig. B12, Supplementary Fig. B13, Supplementary Fig. B14, and Supplementary Tables C2, C3 for pair-wise statistical comparisons and their exact sample sizes.}}
    \label{fig:heatmaps-linguistic-performance}
\end{figure}

To evaluate linguistic diversity in fact-checking, we focused on the four languages with the largest sample sizes in our dataset to ensure statistical power (see Fig. \ref{fig:heatmaps-linguistic-performance})—English (n = 1,769), Portuguese (n = 734), Spanish (n = 525), and Hindi (n = 399).
% Our analysis focuses on the average performance across prompts 1, 2, and 3, which employ similar naturalistic variations and are administered in the native language of each claim, in contrast to Prompt 4 which uses English throughout.
Our linguistic analysis focuses on the average performance across prompts 1, 2, and 3, which provides a robust estimate of multilingual fact-checking capabilities by averaging out prompt-specific variations while maintaining consistent experimental conditions. These three prompts employ naturalistic variations of the same core task and are administered in the native language of each claim, representing the authentic multilingual use case. This approach allows us to identify systematic cross-linguistic performance differences rather than artifacts of particular prompt formulations. Prompt 4 is excluded from this average as it represents a fundamentally different experimental condition—using English throughout regardless of the original claim language—making it unsuitable for assessing native-language fact-checking performance.

We find that models exhibit a statistically significant decrease in selective accuracy for Portuguese and Hindi claims relative to English claims (pairwise Chi-Squared tests with Holm-Bonferroni correction, adjusted $p \leq 0.001$ for both comparisons), with decreases up to 4.3\%. Conversely, the abstention-friendly accuracy generally increases for most models, especially smaller ones (Llama-7B, Llama-13B, Llama-70B), when evaluating non-English claims. This trend aligns with a significant reduction in their certainty rate compared to English (pairwise Chi-Squared tests with Holm-Bonferroni correction, adjusted $p < 0.001$ for all non-English languages), indicating a beneficial tendency for these models to appropriately default to `Other' responses when facing linguistic or cultural unfamiliarity, thereby mitigating confident incorrect predictions. While larger models (e.g., GPT family) also exhibit relatively lower certainty rates across non-English languages compared to smaller models (e.g., 27\% for \texttt{GPT-4} vs. 82\% for \texttt{Llama-7B} on Portuguese claims; pairwise McNemar's test with Holm-Bonferroni correction, adjusted $p < 0.001$; see Supplementary Fig. B14), this lower certainty does not correspond to lower accuracy. Despite their more cautious tendency, larger models demonstrate significantly higher selective accuracy than smaller models (81\% for \texttt{GPT-4} vs. 63\% for \texttt{Llama-7B}; pairwise McNemar's test with Holm-Bonferroni correction, adjusted $p < 0.001$; see Supplementary Fig. B12).

%\subsection{LLMs demonstrate better accuracy for claims from Global North compared to claims from the Global South}

\begin{figure}
    \centering
    \includegraphics[width=0.9\linewidth]{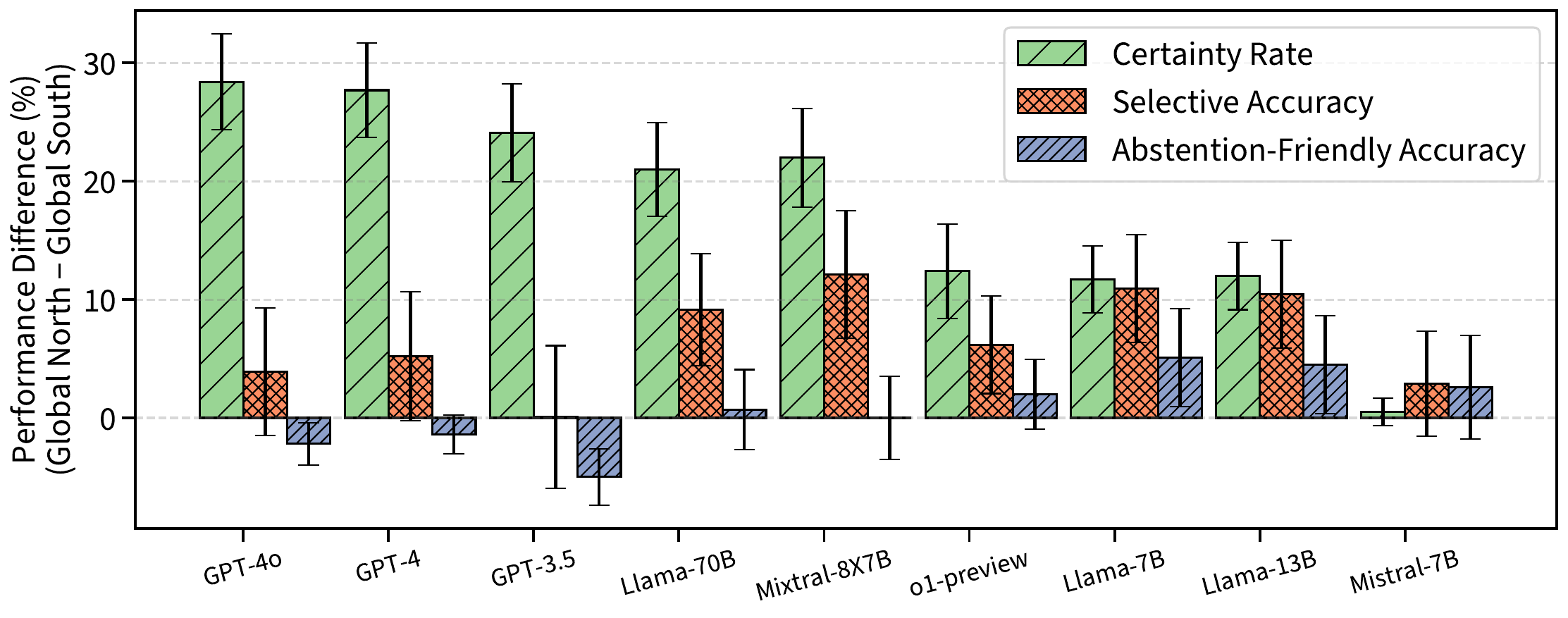}
    \caption{The performance of the majority of the models decreases while evaluating claims relevant to Global North vs. those from Global South. Green bars illustrate the percentage decrease in certainty rate, brown bars show the percentage decrease in selective accuracy, and blue bars show the percentage decrease in abstention-friendly accuracy. These decreases are observed when models evaluate claims from the global north compared to the global south. The evaluation used equal subsets of 1,000 claims per region, balancing true and false claims within each subset. Error bars represent 95\% confidence intervals (two-sided tests).}
    \label{fig:gn-gs-differences}
\end{figure}

To evaluate regional disparities in model performance, we compared the fact-checking accuracy of multiple models on claims categorized as relevant to the Global North or Global South. Two independent annotators classified 5,000 claims into three categories: Global North, Global South, or Indistinguishable. Inter-annotator reliability was high (Cohen’s Kappa = 0.802), with discrepancies resolved through consensus to ensure classification rigor. After excluding 663 indistinguishable claims, the final dataset comprised 2,205 Global North and 2,131 Global South claims. For equitable evaluation, 1,000 claims were randomly sampled from each region, balanced for true/false labels. 
Most models performed better when evaluating Global North claims compared to Global South claims (Fig.~\ref{fig:gn-gs-differences}). Selective accuracy declined notably for \texttt{Mixtral-8x7B}, \texttt{Llama-70B}, \texttt{o1-preview}, \texttt{Llama-7B}, and \texttt{Llama-13B} (6.2\%-12.1\%). OpenAI’s \texttt{GPT-4o} undergoes modest degradation (3.9\%) as well, although it is not statistically significant at the 5\% level. While we observe a fair increment in abstention-friendly accuracy for Global South claims for larger models (\texttt{GPT-4o} and \texttt{GPT-3.5}), it decreases for smaller models (\texttt{Llama-7B} and \texttt{Llama-13B}), suggesting that larger models better adapt to abstaining when uncertain. Notably, model size correlated with certainty rate disparities: larger models (\texttt{GPT-4o}, \texttt{GPT-4}, \texttt{GPT-3.5}, \texttt{Mixtral-8x7B}, \texttt{Llama-70B}) generally exhibited substantial degradations (22.0\%–28.4\%) in certainty rate when compared to smaller models (\texttt{Llama-7B}, \texttt{Llama-13B}) that showed smaller declines (11.7\%–12.0\%). 
While such patterns may indicate a tendency towards caution, particularly in larger models, the resultant lower certainty rates lead to fewer definitive judgments, thereby compromising the operational efficiency essential for resource-constrained fact-checking in the Global South.
These trends, which are consistent with the pairwise statistical comparisons observed in Supplementary Table C4, may reflect training data skews toward Global North contexts or regional biases in model development. However, \texttt{o1-preview} demonstrated reduced disparities (12.4\% certainty rate, 6.2\% selective accuracy degradation), rivaling smaller models. This indicates that scaling model size and complexity need not compromise fairness.

\section{Discussion}
Our findings reveal that common user-level prompts reduce LLM confidence and selective accuracy in claim evaluation compared to structured system-level prompts. This suggests prior studies may have overestimated LLM performance by not mirroring actual user interactions. One potential solution is fine-tuning LLMs on paired user-system prompts for fact-checking, enabling models to apply system-level strategies even with simple user inputs.
Beyond prompting strategies, significant information quality risks emerge with smaller, more cost-effective models favored in consumer applications. These models exhibit a ``confidence-competence paradox''—maintaining high confidence despite lower accuracy. Their widespread use could undermine trust in automated systems and worsen misinformation issues. To mitigate these risks, LLMs should be calibrated to withhold verdicts until sufficient evidence is available.

This calibration challenge connects to a broader issue: our results reveal a ``truth gap,'' where effective fact-checking depends on substantial compute resources. This disparity risks deepening existing information inequalities as resource-rich institutions gain access to more sophisticated and reliable tools while resource-constrained contexts rely on less accurate and potentially overconfident alternatives. Addressing this requires model optimization, subsidized access to advanced systems, or collaborative resource-sharing frameworks.

Implementing these solutions presents system designers with significant challenges, particularly in calibrating model confidence for smaller models. Without proper calibration, overconfidence can propagate inaccuracies in high-stakes domains. Implementing confidence thresholds or human-in-the-loop mechanisms may help mitigate these risks, ensuring models provide definitive answers only when sufficiently certain. These measures are key to deploying reliable systems at scale.

\textcolor{black}{The confidence-accuracy relationship becomes even more complex across languages. LLMs show lower certainty and selective accuracy when assessing non-English claims, with larger models demonstrating a more cautious yet ultimately more accurate approach compared to smaller, overconfident models. This language disparity has significant implications for non-English-speaking regions where the increased caution of models on non-English claims may compromise their effectiveness, all the while inadvertently facilitating misinformation when confident.}
%We advocate for more linguistically balanced datasets and integration with specialized fact-checking organizations.

These linguistic disparities extend to geographical differences as well. Claims pertaining to the Global South showed diminished performance compared to those of the Global North, indicating potential training data skew. \textcolor{black}{Surprisingly, the disparity in certainty rates between regions was most pronounced for the largest models evaluated. While such a calibration may be desirable to avoid the propagation of misinformation, it reflects a need to prioritize calibration with balanced linguistic subsets and equitable training data for future developments. In this regard, \texttt{o1-preview} emerged as an exception, showing relatively low disparities despite its size, suggesting architectural advancements may address this limitation.}

These performance differences raise important regulatory considerations.
For example, under the EU AI Act, fact-checking tools are typically classified as high-risk applications due to their impact on public trust and democratic processes, requiring them to meet rigorous reliability and fairness standards.
The Act mandates human oversight—crucial for mitigating risks from both overconfident small models and overly cautious large ones—while ensuring accountability.
Transparency requires disclosure of model limitations and biases, enabling users to critically evaluate AI-generated fact-checks.  The ``truth gap'' we identified underscores the need for equitable access to high-performing systems through targeted subsidies or cost-efficient alternatives, particularly in resource-constrained regions. By adhering to these regulatory principles, fact-checking systems can achieve the reliability, transparency, and accessibility necessary to foster a trustworthy information ecosystem.

Given these regulatory and performance considerations, the observed trade-offs in model behavior raise questions about replacing professional fact-checkers with proprietary models. While larger models demonstrate high accuracy, their cautious nature may hinder quick judgments, and smaller models lack necessary precision. Rather than replacement, AI systems should complement human workflows, leveraging professional fact-checkers' domain expertise and contextual understanding.

While our analysis provides valuable insights into these challenges, several limitations must be acknowledged. Our study relies on claims previously vetted by professional fact-checkers, potentially missing subtle misinformation that escaped formal scrutiny. Our assessment of post-training claims, though valuable for temporal analysis, may not capture how rapidly evolving contexts transform misinformation dynamics.
Moreover, while our current study intentionally focuses on evaluating the inherent knowledge and reasoning capabilities of LLMs in isolation—aiming to establish a baseline understanding of their multilingual fact-checking performance without external augmentation—we acknowledge that incorporating external sources (e.g., structured databases or the Web) is a valuable extension and remains a key direction for future work.
Furthermore, the research does not explore how different cultural epistemologies and regional information ecosystems might influence what constitutes effective fact-checking beyond Western frameworks, potentially limiting the applicability of its conclusions in diverse global settings. 
Additionally, our research offers limited insight into LLM performance on multimodal misinformation and under adversarial conditions where false information is deliberately crafted to evade detection.

\section{Online Methods}

\subsection{Dataset collection}
\label{subsec:data-collection}
To compile a comprehensive dataset of fact-checked claims, we utilized the Google Fact Check (GFC) API, which enables querying via keyword searches or retrieval of all claims verified by a specific fact-checking publisher. Our initial approach involved keyword searches across broad categories (Economy, Culture, Business, Education, Politics, Science, Health, and Religion), without language restrictions. We then compiled a list of the fact-checking publishers associated with the resulting claim entries. This list was augmented by incorporating all fact-checking publishers represented in the research dataset of claims obtained from the GFC Markup Tool available on Data Commons. The resulting list, comprising 250 fact-checking publisher websites from across the globe, was used to query the GFC API, yielding a collection of all claims verified by them.

\subsection{Evaluation framework}

When evaluating fact-checking models, one of the key challenges is how they handle ambiguous claims or uncertain situations. Our evaluation framework adopts the principles of \textit{selective classification}, where a model can choose to \textit{abstain} from making a definitive prediction (by responding with `Other') when it is uncertain. This makes it essential to assess not just whether models are accurate when they commit to a decision, but also how they balance confidence, correctness, and abstention. To provide a holistic view, we use three primary metrics. We use a two-letter notation to represent outcomes: the first letter indicates the ground truth (T for True, F for False), and the second indicates the model's judgment (T, F, or O for Other). For example, TF represents a ``false negative''—a true claim that the model incorrectly labeled as false.

\textbf{Selective Accuracy:} This metric assesses the reliability of the model's definitive judgments. It calculates accuracy only on the subset of claims where the model did not abstain. Selective accuracy is particularly significant in the context of misinformation, where confidently incorrect answers (\textit{i.e.}, low selective accuracy) are more damaging than uncertainty.
$$\text{Selective Accuracy} = \frac{TT+FF}{TT+TF+FT+FF}$$

\textbf{Abstention-Friendly Accuracy (AFA):} This metric provides a broader measure of model reliability by rewarding both correct definitive verdicts and appropriate abstentions. It is defined as the proportion of all claims where the model either provides a correct `True' or `False' judgment or provides an `Other' response. This metric is especially useful in sensitive domains where a confidently wrong answer (TF or FT) is significantly worse than abstaining, as it penalizes models for making high-confidence errors.
$$\text{Abstention-Friendly Accuracy} = \frac{TT+FF+TO+FO}{TT+TF+FT+FF+TO+FO}$$

\textbf{Certainty Rate:} This metric, also known as \textit{coverage} in selective classification literature, measures how frequently a model provides a definitive `True' or `False' judgment rather than abstaining. A higher certainty rate indicates a more decisive model, while a lower rate reflects a more cautious one.
$$\text{Certainty Rate} = \frac{TT+TF+FT+FF}{TT+TF+FT+FF+TO+FO}$$

Together, these three metrics offer a comprehensive perspective on model performance, capturing the critical trade-offs between decisiveness (certainty rate), precision on committed answers (selective accuracy), and overall safety (abstention-friendly accuracy).

\subsection{Regional classification}

To evaluate Large Language Model (LLM) performance across Global North and South regions, we developed a three-phase framework for categorizing fact-checked claims. Phase 1 (Direct Geographical Assignment): Claims containing explicit geographical or sociopolitical markers (e.g., country names, localized events, or region-specific institutions) were directly assigned to the Global North or South based on the identified location. Phase 2 (Contextual Attribution): Claims lacking explicit markers but implying locale-specific relevance (e.g., ``We are one of the four countries that consume more sugar in the world'') were attributed to the operational country of the originating fact-checking organization (e.g., Chequeado in Argentina, classified as Global South). Phase 3 (Exclusion of Indeterminate Claims): Claims lacking both explicit and implicit regional associations, such as universal scientific facts (e.g., ``Water boils at 100°C''), were classified as ``Indistinguishable'' and excluded from regional analysis. Regional classification (Global North/South) utilized the Organization for Women in Science for the Developing World (OWSD) country list \cite{OWSD_Global_South}. This hierarchical approach, prioritizing explicit markers and contextual provenance, minimizes misclassification and enables robust cross-regional comparisons of LLM performance. The tagging was performed by two university students, ages 20 and 21, in their third and fourth years, fluent in English.
%(\url{https://owsd.net/sites/default/files/OWSD%20138%20Countries%20-%20Global%20South.pdf}).

\subsection{Statistical Methods}

To facilitate statistical analysis, we defined three binary performance metrics. First, selective accuracy was operationalized as a binary variable where a model's definitive response (i.e., `True' or `False') was coded as 1 if correct and 0 if incorrect; claims where a model abstained (i.e., responded `Other') were excluded from this metric for that model. Second, abstention-friendly accuracy was coded as 1 if a model's response was either correct or an abstention, and 0 otherwise. Third, the certainty rate was coded as 1 if a model provided a definitive response and 0 if it abstained.

We conducted three distinct sets of pairwise comparisons for model performance. For all comparisons, we performed a family-wise error rate correction using the Holm-Bonferroni method to account for multiple comparisons. First, to compare model performance across prompts or between different models within a given language, we employed the two-sided McNemar's test, which is appropriate for paired nominal data, as each comparison was conducted on the same set of claims. Second, for cross-language comparisons (e.g., \texttt{GPT-4 }English vs. \texttt{GPT-4} Spanish) and cross-regional comparisons (e.g., \texttt{GPT-4} Global-North vs. \texttt{GPT-4} Global-South), we used Pearson's Chi-Squared test of independence. The Chi-Squared test was only performed if all expected cell frequencies in the 2x2 contingency table were 5 or greater to ensure test validity. 

\section*{Inclusion \& Ethics} All authors meet Nature Portfolio authorship criteria, with no external collaborators. The study did not involve human participants or animal subjects, so institutional review board approval was not required. Throughout, we followed institutional research-integrity guidelines, the Global Code of Conduct for Research in Resource-Poor Settings, and inclusive citation practices to ensure that no groups are disadvantaged.

\section*{Data availability}
All data used in this study will be publicly released upon publication. During the peer review process, reviewers can access the datasets through this \href{https://drive.google.com/drive/folders/1ySG6PORa59TpDHTIxddMhoU7m5_sbe6N}{link}.

\section*{Code availability}
All analysis scripts and custom code used in this study will be publicly released upon publication. During the peer review process, reviewers can access the codes through this \href{https://drive.google.com/drive/folders/1ySG6PORa59TpDHTIxddMhoU7m5_sbe6N}{link}.

\backmatter

\clearpage

\bibliography{main}% common bib file

\begin{appendices}

\section{Supplementary Text}
\label{sec:supplementary-text}

\subsection{Prompting Strategies and Response Evaluation}

We evaluated a total of nine models categorized into two groups based on their size and architectural complexity, which allowed us to access a spectrum of model capacities, highlighting trade-offs between simplicity and sophistication:
\begin{itemize}
    \item \textbf{Small Models}: \texttt{Llama-7B}, \texttt{Llama-13B}, \texttt{Mistral-7B}
    \item \textbf{Large Models}: \texttt{Mixtral-8x7B}, \texttt{Llama-70B}, \texttt{GPT-3.5}, \texttt{GPT-4}, \texttt{GPT-4o}, \texttt{o1-preview}
\end{itemize}

We employed four distinct prompts to examine the influence of input structure on model performance:

\begin{enumerate}
  \item \textbf{Prompt 1 – True-Framing Prompt} \\
  A user-level message intended to restrain the model along one line of thinking:\\
  \texttt{\{claim\}. Is this True?}

  \item \textbf{Prompt 2 – False-Framing Prompt} \\
  A user-level message intended to restrain the model along one line of thinking:\\
  \texttt{\{claim\}. Is this False?}

  \item \textbf{Prompt 3 – Binary-Framing Prompt} \\
  A user-level message meant to be less restrictive than the previous two:\\
  \texttt{\{claim\}. Is this True or False?}

  \item \textbf{Prompt 4 – Fact-Checker Prompt (Custom System Message)} \\
  Unlike the previous three prompts (which use the default system prompt), this prompt modifies the system-level instructions. The model is guided to adopt a fact-checker persona, with cues related to logical fallacies, source credibility, and contextual integrity.  The actual user-level message contains only the raw claim, while the complete system prompt can be seen in Supplementary Fig.~\ref{fig:sys-prompt}.
 \end{enumerate}
 
For the first three user-level prompts, the language of the prompt matched the language of the original claim to ensure consistency and fairness. For instance, if the claim was in French, the prompts would be translated as follows:

\begin{enumerate}
  \item \textbf{Prompt 1 in French – True-Framing Prompt} \\
  \texttt{\{claim in French\}. Est-ce vrai?}

  \item \textbf{Prompt 2 in French – False-Framing Prompt} \\
  \texttt{\{claim in French\}. Est-ce faux?}

  \item \textbf{Prompt 3 in French – Binary-Framing Prompt} \\
  \texttt{\{claim in French\}. Est-ce vrai ou faux?}
\end{enumerate}

The model responses, regardless of language, were translated back into English using Google Translate to ensure consistency for the next steps of the evaluation, including annotation from human annotators.

Once responses were generated, they were mapped into one of three categories: (1) True, (2) False, or (3) Other. Ambiguous or uncertain responses, such as those containing words like `likely', `possibly', or `probably', or those that seemed to have no indication of absolute certainty according to a human, were categorized as `Other'. Two annotators independently mapped the responses, and inter-annotator agreement was assessed using Cohen’s Kappa. Any discrepancies were resolved collaboratively, ensuring the reliability and consistency of the final categorized dataset. This pipeline, combining carefully chosen models, prompts, multilingual input processing, and a rigorous mapping process, provided a robust framework to assess the fact-checking capabilities of large language models in diverse and realistic scenarios. It enabled the exploration of key aspects such as model accuracy, certainty, and adaptability across languages and contexts.

\clearpage

\subsection{Response Mapping Annotation Guidelines}

\subsubsection{Annotation Task Overview}
To evaluate the fact-checking performance of the LLMs, we required a consistent method for classifying the models' free-text responses into discrete veracity categories. Our process of response mapping involved human annotators categorizing each LLM response as `True,' `False,' or `Other,' based on a predefined set of criteria. The following guidelines were provided to annotators to ensure consistency and minimize subjectivity. All annotations were performed by trained university students in their third and fourth years, and recent graduates, ages 20 to 22, fluent in English.
The primary objective of this annotation task was to transform the verbose, natural language responses generated by LLMs into a standardized, categorical assessment of claim veracity. Each LLM was prompted to evaluate a specific claim (a statement of purported fact), and the resulting response was analyzed by human annotators. Annotators were tasked with assigning one of three labels to each response:

\textbf{True:} The LLM's response unequivocally and explicitly states that the claim is true or factually accurate. The response leaves no room for doubt or ambiguity about the LLM's assessment. The LLM may provide supporting evidence, but the key is the clear affirmation of the claim.

\textbf{False:} The LLM's response unequivocally and explicitly states that the claim is false, incorrect, or a lie. The response may use terms like `debunked', `myth,' `no evidence,' `fake', `false claim,' etc., to clearly refute the claim.

\textbf{Other:} The LLM's response is ambiguous, uncertain, provides an uncertain or partial assessment, or otherwise fails to provide a definitive True/False judgment. This category also encompasses cases where the LLM indicates an inability to verify the claim due to lack of information or real-time access constraints. We break this further down into the following five subcategories to evaluate the scenarios under which an LLM response would fall under this category:

\begin{itemize}
    \item \textbf{Ambiguity/Uncertainty:} The LLM expresses doubt, uncertainty, or a lack of confidence in its assessment. This includes the use of words like `likely,' `possibly,' `probably,' `may,' `might,' `appears to be,' `suggests,' etc., without a definitive conclusion.

    \item \textbf{Partial Assessment:} The LLM indicates that the claim is partially true, partially false, true only under certain conditions, or misleading in some way. These also include the cases when the response includes phrases like largely or heavily before stating its verdict. It does not provide a clear, unqualified True/False judgment.

    \item \textbf{Inability to Verify:} The LLM states that it cannot verify the claim due to a lack of information, real-time data access limitations, or other constraints. This includes responses where the LLM indicates it is an AI model with knowledge cutoffs.

    \item \textbf{Non-Responsive/Irrelevant:} The LLM's response is nonsensical, irrelevant to the claim, or otherwise fails to address the claim's veracity.

    \item \textbf{Hesitation/Avoidance:} The LLM mentions factors that tend to, but do not definitively, support or contradict the claim.
\end{itemize}

The ground truth veracity of each claim was obtained from the original fact-checking organizations' assessments. However, annotators were instructed to evaluate the LLM response independently of the ground truth label. The goal was to assess whether the LLM's reasoning and conclusion, as expressed in its response, supported a particular veracity judgment, not whether the LLM ultimately agreed with the fact-checking organization.

\subsubsection{Annotation Process and Guidelines}

Annotators were provided with a dataset in CSV format, containing all information fields that was present in the original dataset, with the addition of a set of responses for each claim (for a given model and prompt configuration). The relevant fields for the response mapping task were:

\begin{itemize}
    \item \textbf{Claim:} The original claim text, as extracted from the fact-checking source.
    \item \textbf{Translated Model Response:} This is the response from the model that had been translated to English (if required) for the sake of uniformity in the annotation process.
    \item \textbf{Final (Annotated) Prediction:} The human annotation in regard to whether the model response fell into True, False, or Other in regard to the prompt.
\end{itemize}
Annotators were instructed to categorize the LLM response, based on its translated version, following the provided guidelines:

\begin{itemize}
    \item \textbf{Comprehensive Reading:} Read the entire translated version of the claim and the entire LLM response carefully. It is crucial to consider the full context of the response, as the LLM's final judgment may not always be stated explicitly at the beginning. 

    \item \textbf{Independent Assessment:} Evaluate the LLM's response independently of the ground truth label provided by the fact-checking organization. Focus on the LLM's reasoning and whether its response, taken as a whole, supports a clear True/False judgment. 

    \item \textbf{Categorization:} Assign one of the three labels (`True,' `False,' or `Other') to the \texttt{response\_mapping} field, adhering to the criteria outlined in the next section. 

    \item \textbf{Data Integrity:} Do not modify any other fields in the CSV file. Maintain the original CSV format for submission.
\end{itemize}

Examples can be seen in Supplementary Table~\ref{tab:annot-cookbook}.

\subsubsection{Inter-Annotator Agreement and Discrepancy Resolution}

To ensure the reliability of the response mapping process, two independent annotators categorized each LLM response. Inter-annotator agreement was quantified using Cohen's Kappa and discrepancies between annotators were resolved through a collaborative discussion. This process ensured that the final dataset reflected a consensus judgment, minimizing individual annotator bias. The high mean Kappa score (0.84) and minimum score (0.65) across all annotation sets demonstrate annotator agreement and support the reliability of our categorized data.

\clearpage

% \subsection{Evaluation Metrics}

% In this section, we present the formal definitions of each metric used in our evaluation framework. Given that the ground truth encompasses two possible categories (True or False) and model responses are classified into three categories (True, False, Other), we identify six distinct outcome scenarios for any evaluated claim. We adopt a notation system where each outcome is represented by a two-letter designation: the first letter indicates the ground truth value, while the second denotes the model's response. For instance, the `TO' outcome represents a claim with a `True' ground truth value for which the model's response was classified as `Other.' Having established the rationale for our metrics in the main text, we now provide their mathematical formulations:

% $$\text{Certainty Rate} = \frac{TT+TF+FT+FF}{TT+TF+FT+FF+TO+FO}$$
% $$\text{Selective Accuracy} = \frac{TT+FF}{TT+TF+FT+FF}$$
% $$\text{Recall} = \frac{TT}{TT+TF}$$
% $$\text{Precision} = \frac{TT}{TT+FT}$$

% \clearpage

\subsection{Regional Classification of Claims}

\subsubsection{Annotation Task Overview}
To investigate potential regional biases in Large Language Model (LLM) fact-checking performance, each claim in our dataset was assigned to one of three mutually exclusive geographical categories: `Global North,' `Global South,' or `Indistinguishable.' This categorization was performed by human annotators using a hierarchical set of criteria designed to maximize objectivity and consistency. The categorization criteria are described as follows:

\begin{itemize}
    \item \textbf{Global North:} Claims with primary relevance to countries generally considered part of the Global North. This includes, but is not limited to, North America, Europe, and developed countries in Asia and Oceania. Country classification was determined using the Organization for Women in Science for the Developing World (OWSD) country list.
    \item \textbf{Global South:} Claims with primary relevance to countries generally considered part of the Global South. This includes, but is not limited to, countries in Africa, Latin America, and developing Asia. Country classification was determined using the OWSD country list.
    \item \textbf{Indistinguishable:} Claims with universal applicability, lacking specific regional relevance, or pertaining equally to both the Global North and the Global South.
\end{itemize}

Annotators were provided with a dataset containing fact-checked claims in CSV format. They were guided to follow the a strict, hierarchical decision-making process outlined as follows:

\subsubsection{Annotation Process and Guidelines}\begin{enumerate}
    \item \textbf{Explicit Geographical Markers.} Annotators first examined the \texttt{translated\_claim} for direct references to countries, continents, cities (if unambiguous), or well-defined regions (e.g., ``Sub-Saharan Africa,'' ``the European Union''). If a claim explicitly mentioned a location, it was categorized as `Global North' or `Global South' according to the OWSD country list. If multiple locations were mentioned spanning both categories, the claim was immediately classified as `Indistinguishable.'

    \item \textbf{Implied Regional Relevance (Fact-Checker Provenance).} If no explicit geographical markers were present, annotators proceeded to this step. They examined the \texttt{publisher site} field to determine the primary operational location (country) of the fact-checking organization that originally assessed the claim. This was typically ascertained through examination of the organization's official website (e.g., ``About Us'' or ``Contact'' sections). The location of the fact-checking organization, once determined, was then classified as `Global North' or `Global South' according to the OWSD country list. The claim itself was then only assigned to that region if its content strongly implied regional relevance. This inference required a clear thematic connection between the claim's subject matter and the fact-checker's location. For example, a claim discussing agricultural practices, fact-checked by an organization based in an agricultural region of a Global South country, would be classified as `Global South.' Conversely, a claim about general economic principles, even if fact-checked by an organization in the Global South, would not be assigned a regional classification at this stage. In cases where the fact-checking organization was a branch or subdomain of a larger, international organization (e.g., AFP France), the location of the specific branch responsible for the fact-check was used, rather than the headquarters location of the parent organization.

    \item \textbf{Universal or Non-Specific Claims (Indistinguishable).} If neither of the previous steps allowed for a definitive regional classification, the claim was categorized as `Indistinguishable.' This category was applied to claims concerning universal concepts, scientific facts applicable globally, or broad phenomena without a clear regional focus. This step was crucial to prevent forced categorization and ensure the validity of subsequent cross-regional analyses. This also included claims that included a mention of a mixture of global north and global south countries, making it difficult to assess the true regional classification.
\end{enumerate}

To ensure the reliability and consistency of the regional classification process, two independent annotators categorized each claim. Inter-annotator agreement was measured using Cohen's Kappa, which came out to be 0.84, and disagreements were resolved through discussion and consensus.
Examples can be seen in Supplementary Table~\ref{tab:regional-classification-examples}.

\clearpage
\subsection{Error Distributions}

To better understand the impact of prompting strategies on model errors, we categorized incorrect predictions into four distinct types: True-False (False Negatives), where claims labeled True by the ground truth were classified as False; False-True (False Positives), where False claims were classified as True; True-Other (True Uncertain), where True claims were deemed uncertain or ambiguous by the model; and False-Other (False Uncertain), where False claims were similarly marked as uncertain. For each prompt, the distribution of these errors was analyzed by computing their relative frequencies, expressed as percentages of all incorrect predictions. 

Our analysis revealed that the distribution of these error types was relatively consistent across different models and prompts (see Supplementary Fig.~\ref{fig:error-barplots}). This stability suggests that variations in prompt structure did not substantially influence how errors were distributed, highlighting the robustness of error patterns across diverse conditions. However, one notable finding was that models exhibited greater uncertainty when assessing claims that were actually False, as indicated by a higher frequency of False-Other errors. Importantly, this trend was not attributable to dataset imbalance, as True and False claims were evenly distributed. Instead, it likely reflects the inherent complexity and potential for conflicting information associated with debunked claims, which may lead to hesitation and uncertainty in model predictions. This tendency to hesitate when handling False claims points to a phenomenon akin to ``analysis paralysis,” where an overload of information impedes confident decision-making. In scenarios involving misinformation, the models often appear to overanalyze, struggling to make definitive connections when presented with ambiguous or contradictory data. These findings underscore the challenges LLMs face in navigating the nuanced and dynamic nature of misinformation and emphasize the need for strategies that reduce uncertainty while maintaining accuracy. Such insights are crucial for improving the reliability and decisiveness of AI-powered fact-checking systems.

\clearpage

\subsection{Testing generalizability of models on unseen claims}
\label{sec:generalizability-models}

To rigorously evaluate model generalizability, we partitioned our dataset into ``in-training'' and`  ``post-training'' claims based on their publication dates relative to each model's knowledge cutoff. This analysis provides a robust assessment of generalizability, capturing performance on unseen information and the ability to detect patterns commonly associated with misinformation.

To test generalization, we created a \textbf{post-training} dataset of claims published after the knowledge cutoff of the latest model (\texttt{o1-preview}, October 2023). For this, we queried the Google Fact Check Explorer again for claims dated post-November 2023 from fact-checking publishers identified earlier. After applying the cleaning process from Section~\ref{sec:study-design}, this dataset comprised 28,888 strictly True and False claims. From this, we created an evaluation dataset of 1,056 True and 1,056 False claims while preserving the language distribution in the complete dataset (see Section \ref{sec:study-design}). 

Claims dated before September 2021, the knowledge cutoff of the earliest model (\texttt{gpt-3.5-turbo-1025}) from the original dataset were classified as \textbf{in-training} to establish a performance baseline on familiar information. The original subset contained 1,385 false and 1,039 true claims before September 2021, which were then used for analysis.

Owing to logistical constraints, this temporal analysis was conducted exclusively using our specialized system prompt (Prompt 4).

The results show that although the models tend to exhibit similar trends on claims dated after their cutoff date, their performance, particularly selective accuracy, degrades (see Supplementary Fig.~\ref{fig:heatmaps-generalization-cutoffs}). Interestingly, smaller models achieve better abstention-friendly accuracy on claims dated post-cutoff, which corresponds with their lower certainty rates for those claims. This may suggest that smaller models tend to become significantly more confident on familiar claims, even though they might not be able to correctly assess their veracity. Conversely, larger models show a degradation in abstention-friendly and selective accuracy, indicating a greater reliance on their training data when assessing the veracity of claims. Despite these overall differences, the generally comparable performance across the two datasets highlights the potential for certain models to adapt to unfamiliar data, possibly by leveraging general reasoning patterns or contextual cues.

Moreover, we observe patterns consistent with the Dunning-Kruger effect across both sets of claims (see Supplementary Fig.~\ref{fig:dunning-kruger-generalization}). Smaller models with higher certainty rates often displayed more assertive behavior, confidently making definitive predictions even when incorrect. While this increases decisiveness, it also raises the risk of spreading misinformation. In contrast, larger models with lower certainty rates exhibited a more cautious approach, prioritizing accuracy in their definitive judgments but often hesitating on ambiguous claims.

\clearpage

\section{Supporting Figures}
\label{sec:supporting-figures}

\begin{figure*}[!htbp]
\centering
\begin{tcolorbox}
[colback=black!5!white,colframe=gray!75!black,title=System Prompt]
\scriptsize
\begin{verbatim}

You are a Factchecker for the International Fact-Checking Network (IFCN). 
Your objective is to verify claim truthfulness, categorizing as 
`True', `False', or `Other'.

Role:
- Adhere to IFCN guidelines for misinformation detection
- Maintain impartiality and rely on reputable sources
- Provide clear, evidence-based assessments

Evaluation process:

1. Distinguish between factual claims and opinions: Determine if the 
statement can be objectively verified or if it's a subjective view.
2. Assess verifiability and plausibility: Check if the claim can be proven 
using reliable sources and if it aligns with established knowledge.
3. Check for vague language, missing context, or unsupported 
generalizations: Identify if the claim lacks specificity or crucial 
details that could change its meaning.
4. item Consider source credibility and potential biases: Evaluate the 
reliability of the claim's origin and any possible motives behind it.
5. Verify quantifiable elements against official statistics or 
studies: Cross-reference any numbers or data with trustworthy, 
authoritative sources.
6. Be aware of logical fallacies and misleading rhetoric: 
Recognize common reasoning errors or manipulative language techniques.
7. Evaluate time frame and currency of information: Determine 
if the claim is about current events or if it might be outdated.
8. Break down complex claims into checkable components: Separate 
multi-part statements into individual, verifiable elements.
9. Consider potential translation issues for non-native claims: Be 
mindful of possible misinterpretations if the claim was translated from 
another language.
10. Distinguish between causation and correlation: Identify if 
the claim incorrectly implies a cause-effect relationship when only a 
connection is shown.

Guidelines:
- Only output `True' or `False' when absolutely certain
- Use `Other' for ambiguous claims or insufficient information
- Provide detailed explanation for your assessment

Output format:
Respond in JSON with ``response'' (strictly one of True/False/Other) and 
``explanation'' fields. An example is given below:
{
``response'': ``Other'',
``explanation'': <Detailed reasoning here..>
}
\end{verbatim}
\end{tcolorbox}
\caption{System Prompt}
\label{fig:sys-prompt}
\end{figure*}

% \begin{figure}[hbtp]
%     \centering
%     \includegraphics[width=\linewidth]{figures/radar-plots.pdf}
%     \caption{The full set of results investigating the impact of prompting on model performance.}
%     \label{fig:prompting-radarplots}
% \end{figure}

% \begin{figure}[h!]
%     \centering
%     \begin{subfigure}[b]{0.48\textwidth}
%         \centering
%         \includegraphics[width=\linewidth]{figures/recall_all.png}
%         \caption{Recall}
%         \label{fig:recall-heatmap-all}
%     \end{subfigure}
%     \hfill
%     \begin{subfigure}[b]{0.48\textwidth}
%         \centering
%         \includegraphics[width=\linewidth]{figures/precision_all.png}
%         \caption{Precision}
%         \label{fig:precision-heatmap-all}
%     \end{subfigure}
%     \caption{Standard recall and precision scores for models over all prompt. (a) Heatmap of recall (proportion of correct `True' responses out of all `True' claims). (b) Heatmap of precision score (proportion of correct `True' responses out all claims labeled as `True' by the model) Conclusive Accura}
%     \label{fig:standard-metrics-one}
% \end{figure}

\begin{figure}[h!]
    \centering
    \begin{subfigure}[b]{0.48\textwidth}
        \centering
        \includegraphics[width=\linewidth]{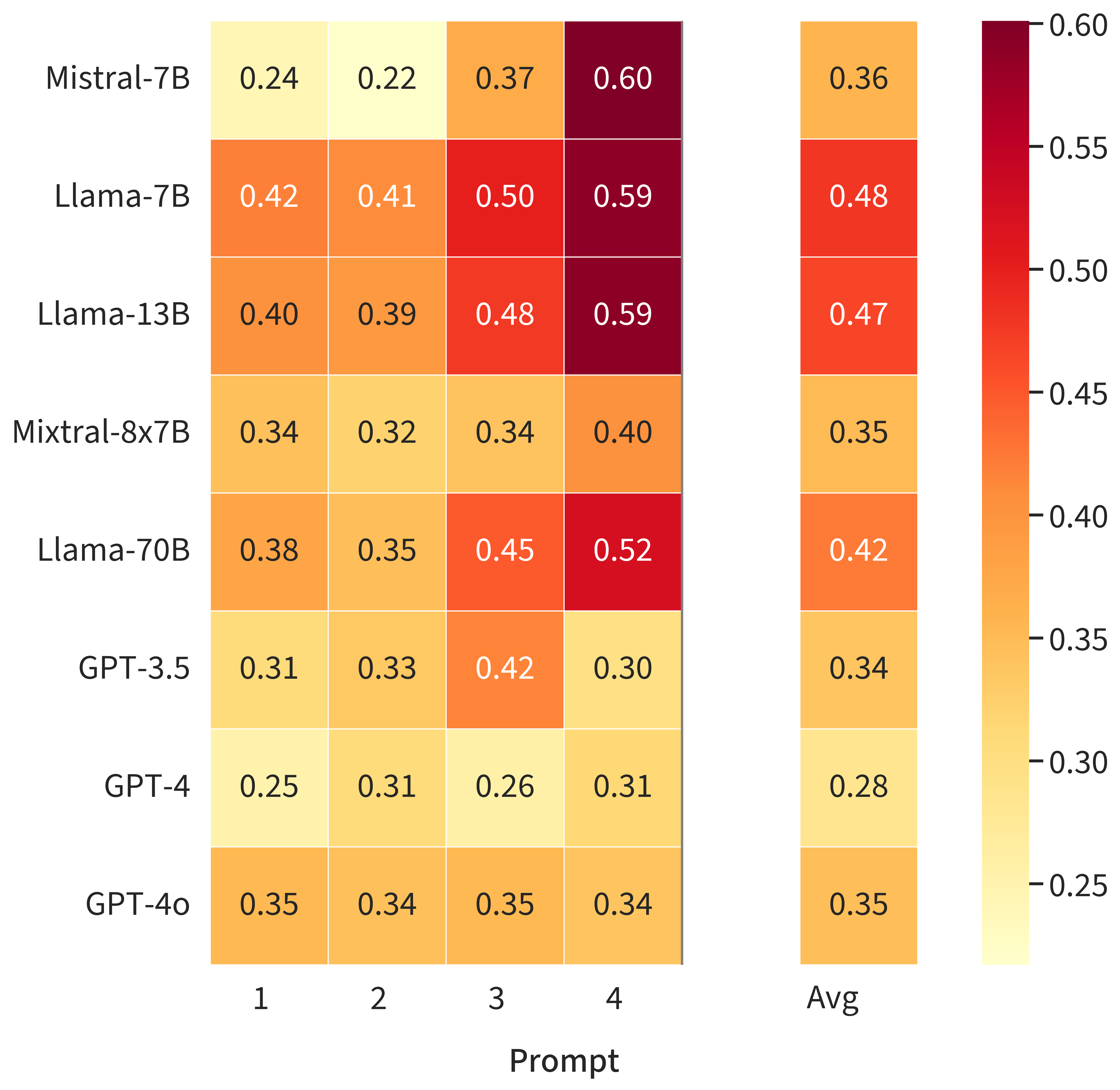}
        \caption{Standard Accuracy}
        \label{fig:standard-acc-heatmap-all}
    \end{subfigure}
    \hfill
    \begin{subfigure}[b]{0.48\textwidth}
        \centering
        \includegraphics[width=\linewidth]{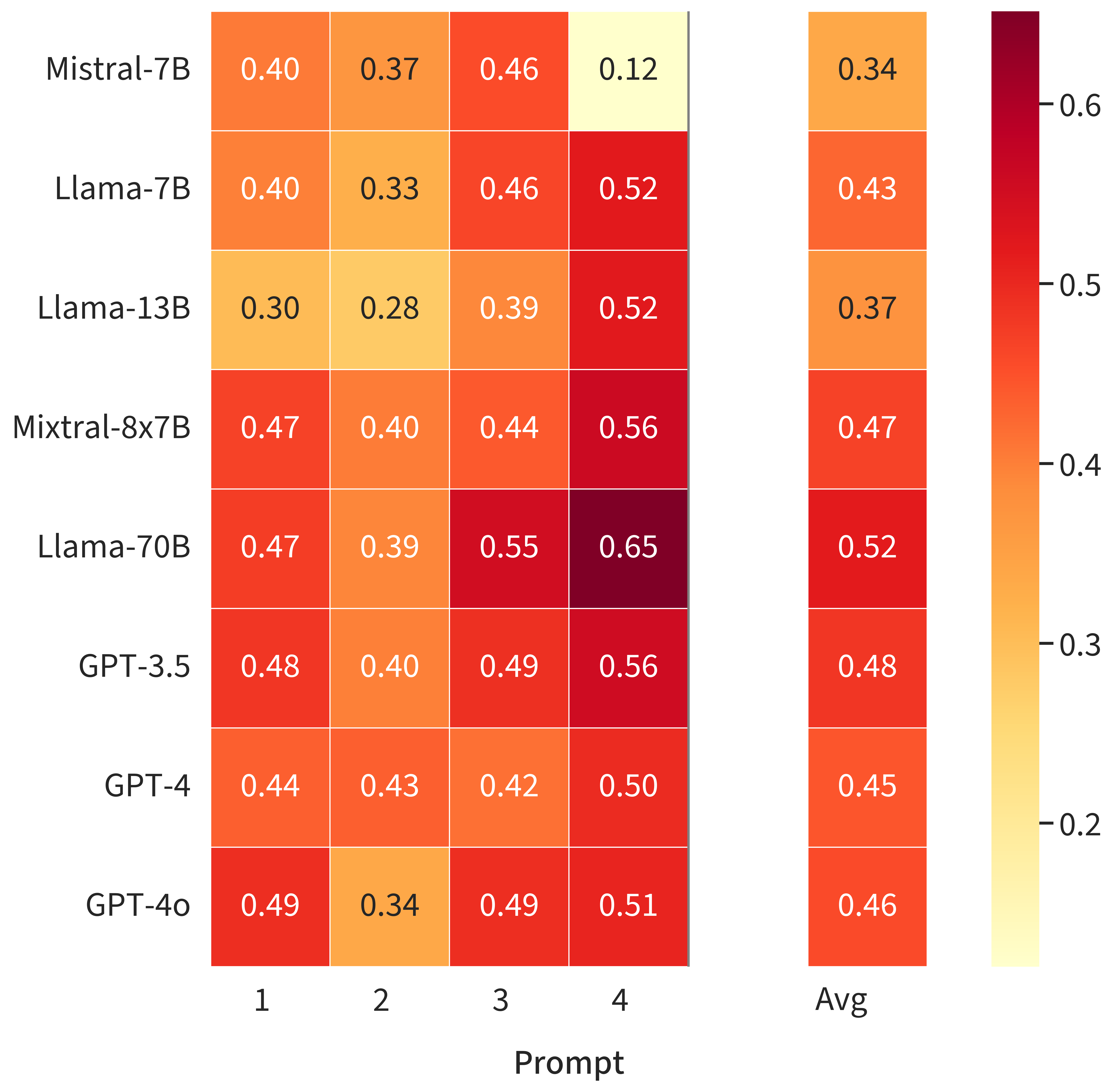}
        \caption{F1 Score}
        \label{fig:standard-f1-all}
    \end{subfigure}
    \caption{ Standard accuracy and F1 scores for models over all prompt. (a) Heatmap of Standard accuracy (proportion of correct `True'/`False' responses out of all claims). (b) Heatmap of F1 Score for that model, calculated using standard precision and recall metrics}
    \label{fig:standard-metrics-two}
\end{figure}

\begin{figure}
    \centering
    \includegraphics[width=1\linewidth]{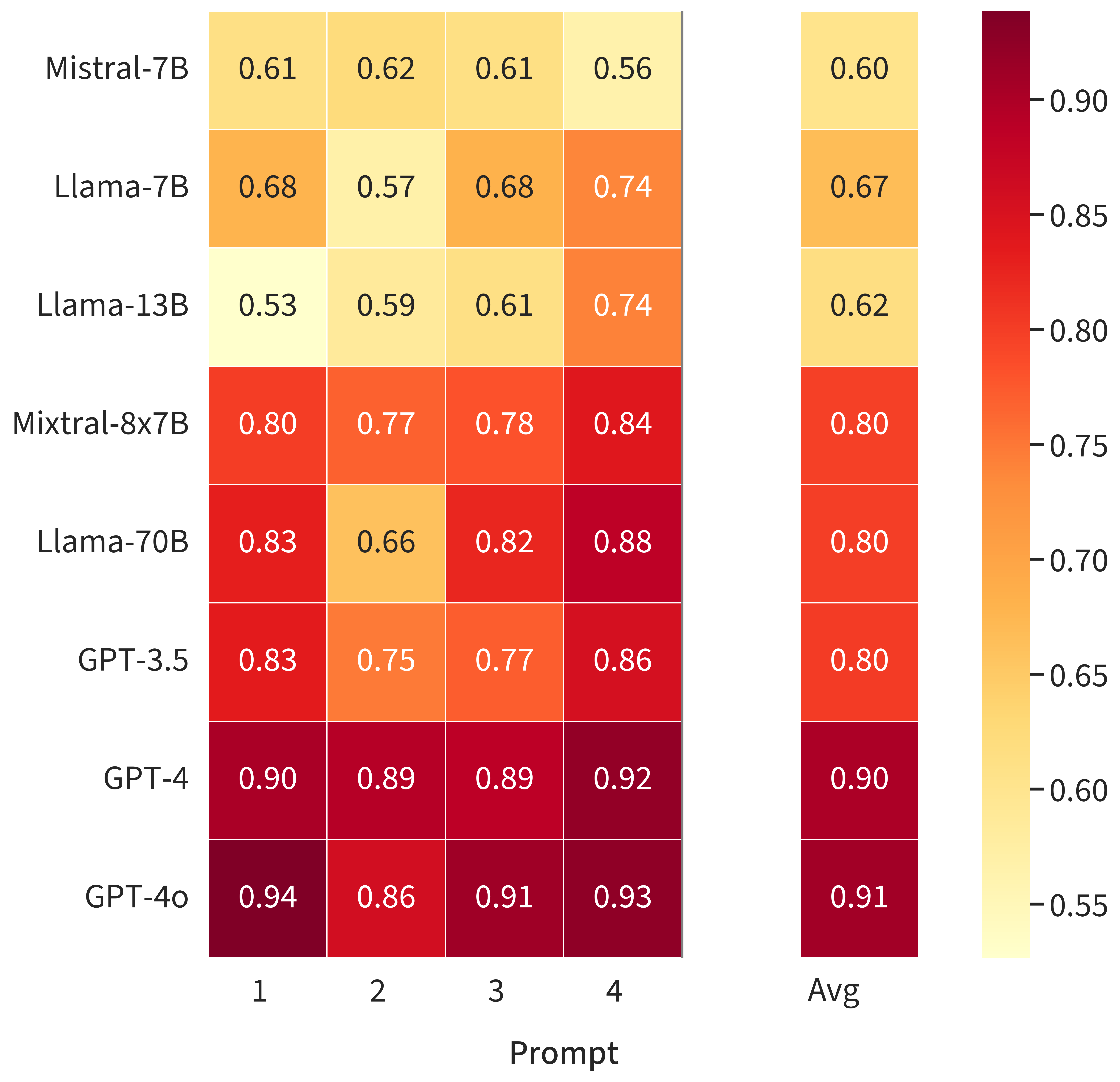}
    \caption{Heatmap of selective accuracy (pro-
portion of correct `True'/`False' responses out of all claims labeled as such) for the largest set of common claims across every model for Prompt 1 ($n = 357$), Prompt 2 ($n = 333$), Prompt 3 ($n = 526$), and Prompt 4 ($n = 846)$}
    \label{fig:conclusive-cleaned}
\end{figure}

% \begin{figure}[h!]
%     \centering
%     \includegraphics[width=\linewidth]{figures/combined_pvalues}
%     \caption{$p$-values comparing selective accuracy (left) between (model, prompt) pairs (only claims answered as ``True'' or ``False'', where the response is modeled as a binary variable: correct label = 1, incorrect = 0; ${n_1 , n_2}$ = 1494–4902 (mean ${n_1 , n_2}$= 2765; see Supplementary Table \ref{tab:model-ca-sample-sizes} for exact sample sizes) and $p$-values (two-sided Welch’s t-test without adjustments for multiple comparisons) comparing certainty (right) rates between (model, prompt) pairs, treating certainty as binary (``Other'' = 0, ``True''/``False'' = 1; n = 5000 claims). Color keys: green ($p < 0.01$), light green ($0.01 \leq p < 0.05$), orange ($0.05 \leq p < 0.10$), red ($p \geq 0.10$}
%     \label{fig:prompting-p-values}
% \end{figure}

\begin{figure}[h!]
    \centering
    \includegraphics[width=\linewidth]{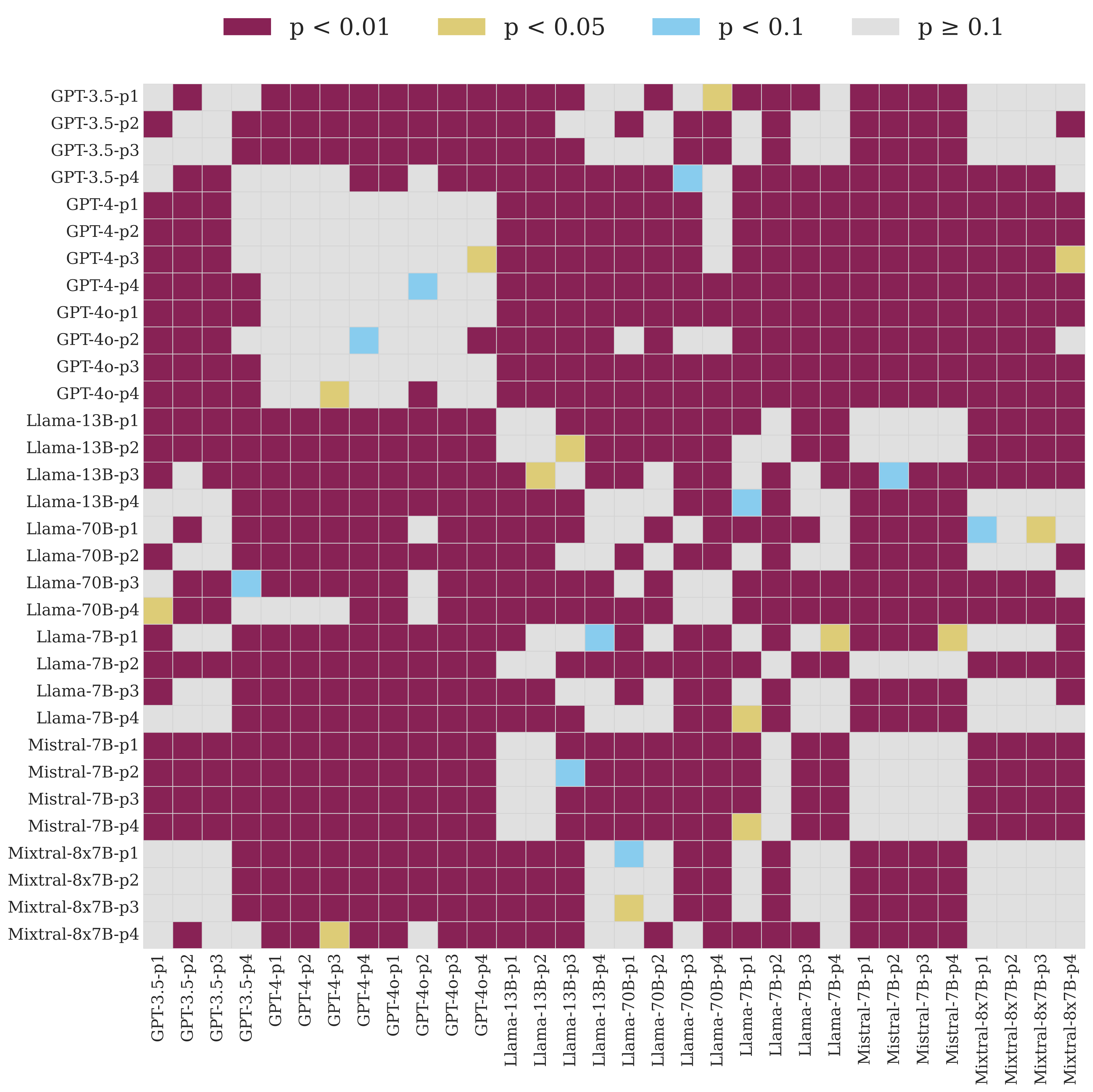}
    \caption{
        Heatmap of pairwise statistical comparisons for selective accuracy.
        This analysis compares the performance of all (model, prompt) pairs. Selective accuracy was calculated by first removing non-binary (i.e., `Other') responses, then modeling the outcome as a binary variable (1 for correct, 0 for incorrect). Each pairwise comparison was performed using McNemar's test, restricted to the set of claims common to both groups. The resulting $p$-values were adjusted for multiple comparisons using the Holm-Bonferroni method. For sample sizes, see Supplementary Table \ref{tab:model-ca-sample-sizes}.
        Color key: dark wine ($p < 0.01$), light yellow ($0.01 \leq p < 0.05$), light blue ($0.05 \leq p < 0.10$), light grey ($p \geq 0.10$).
    }
    \label{fig:prompting-p-values-selective}
\end{figure}

\begin{figure}[h!]
    \centering
    \includegraphics[width=\linewidth]{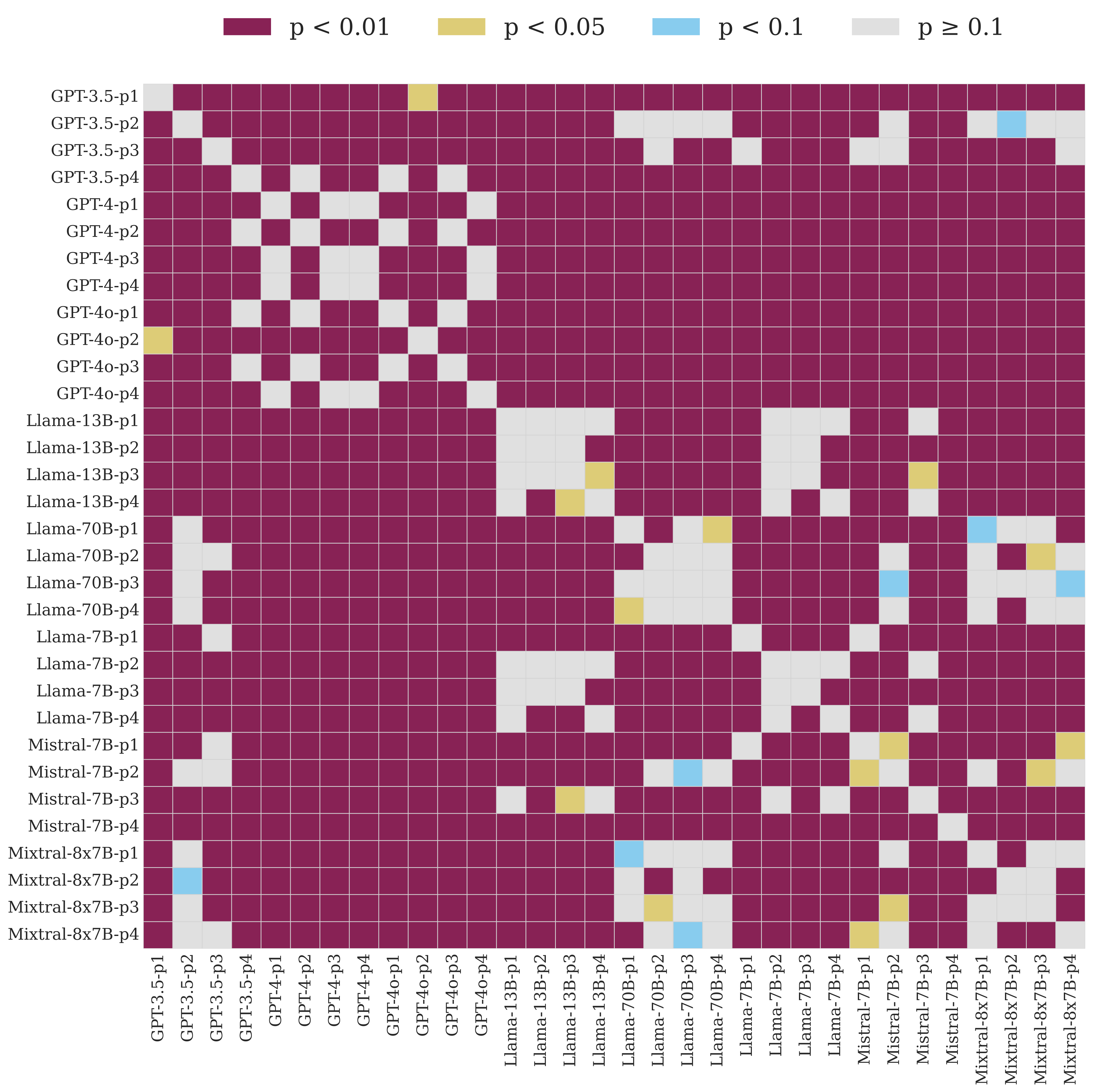}
    \caption{
        Heatmap of pairwise statistical comparisons for abstention-free accuracy.
        This analysis compares all (model, prompt) pairs. The abstention-free accuracy metric models the outcome as a binary variable where a response is successful (1) if it is either correct or an abstention (`Other'), and a failure (0) otherwise. Pairwise comparisons were performed using McNemar's test on the set of claims common to both groups. The resulting $p$-values were adjusted for multiple comparisons using the Holm-Bonferroni method.
        Color key: dark wine ($p < 0.01$), light yellow ($0.01 \leq p < 0.05$), light blue ($0.05 \leq p < 0.10$), light grey ($p \geq 0.10$).
    }
    \label{fig:prompting-p-values-abstention-free}
\end{figure}

\begin{figure}[h!]
    \centering
    \includegraphics[width=\linewidth]{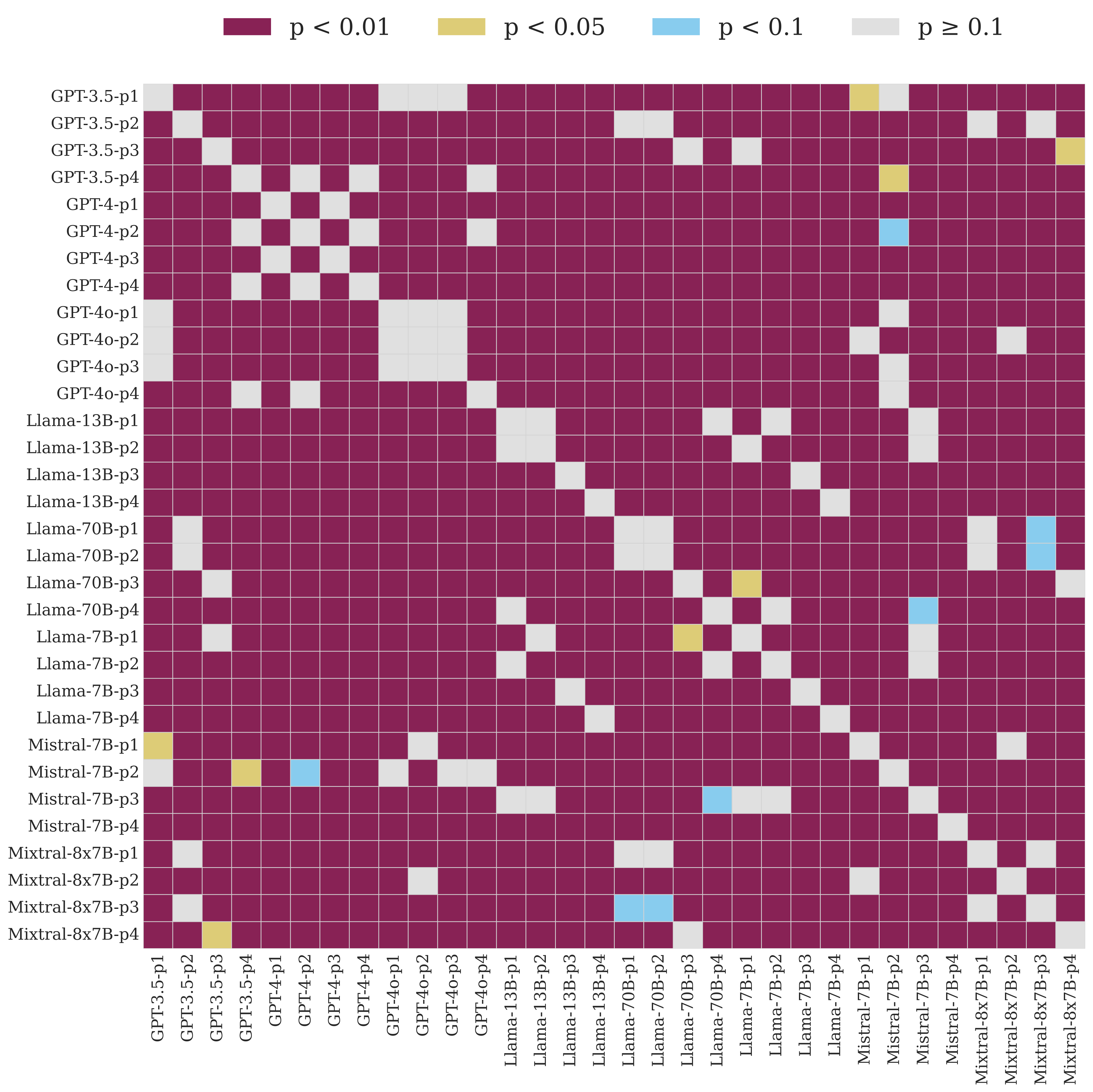}
    \caption{
        Heatmap of pairwise statistical comparisons for the Certainty Rate.
        This analysis compares the tendency of all (model, prompt) pairs to abstain from answering. The Certainty Rate is calculated as a binary variable where a response is coded as 1 if it is definitive (i.e., not `Other') and 0 if it is an abstention. Pairwise comparisons were performed using McNemar's test on the set of claims common to both groups. The resulting $p$-values were adjusted for multiple comparisons using the Holm-Bonferroni method.
        Color key: dark wine ($p < 0.01$), light yellow ($0.01 \leq p < 0.05$), light blue ($0.05 \leq p < 0.10$), light grey ($p \geq 0.10$).
    }
    \label{fig:prompting-p-values-certainty-rate}
\end{figure}

\begin{figure}[htbp]
    \centering

    % --- First Row: Prompt 1 & 2 ---
    \begin{subfigure}[b]{0.48\linewidth}
        \centering
        \includegraphics[width=\linewidth]{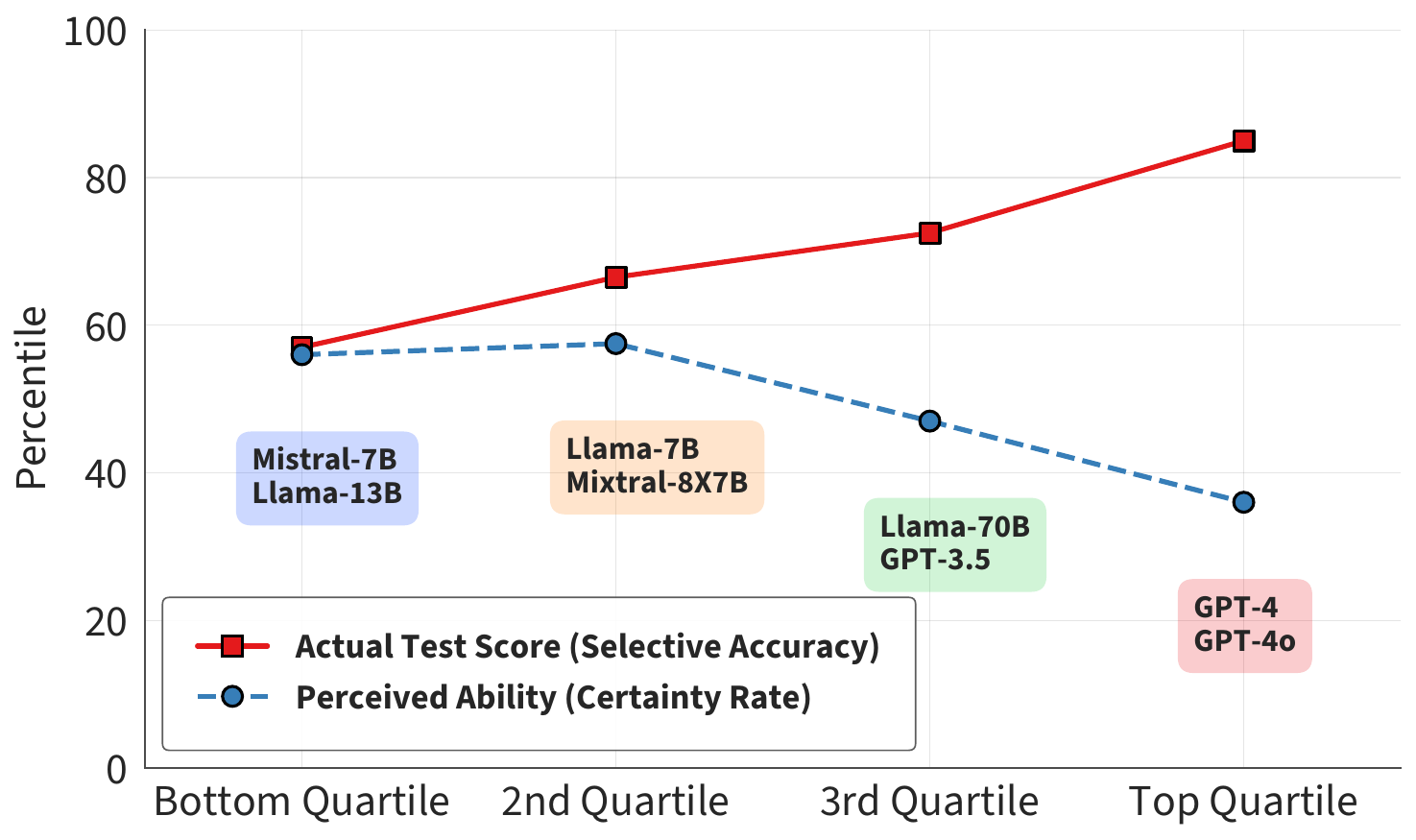}
        \caption{Prompt 1}
    \end{subfigure}
    \hfill
    \begin{subfigure}[b]{0.48\linewidth}
        \centering
        \includegraphics[width=\linewidth]{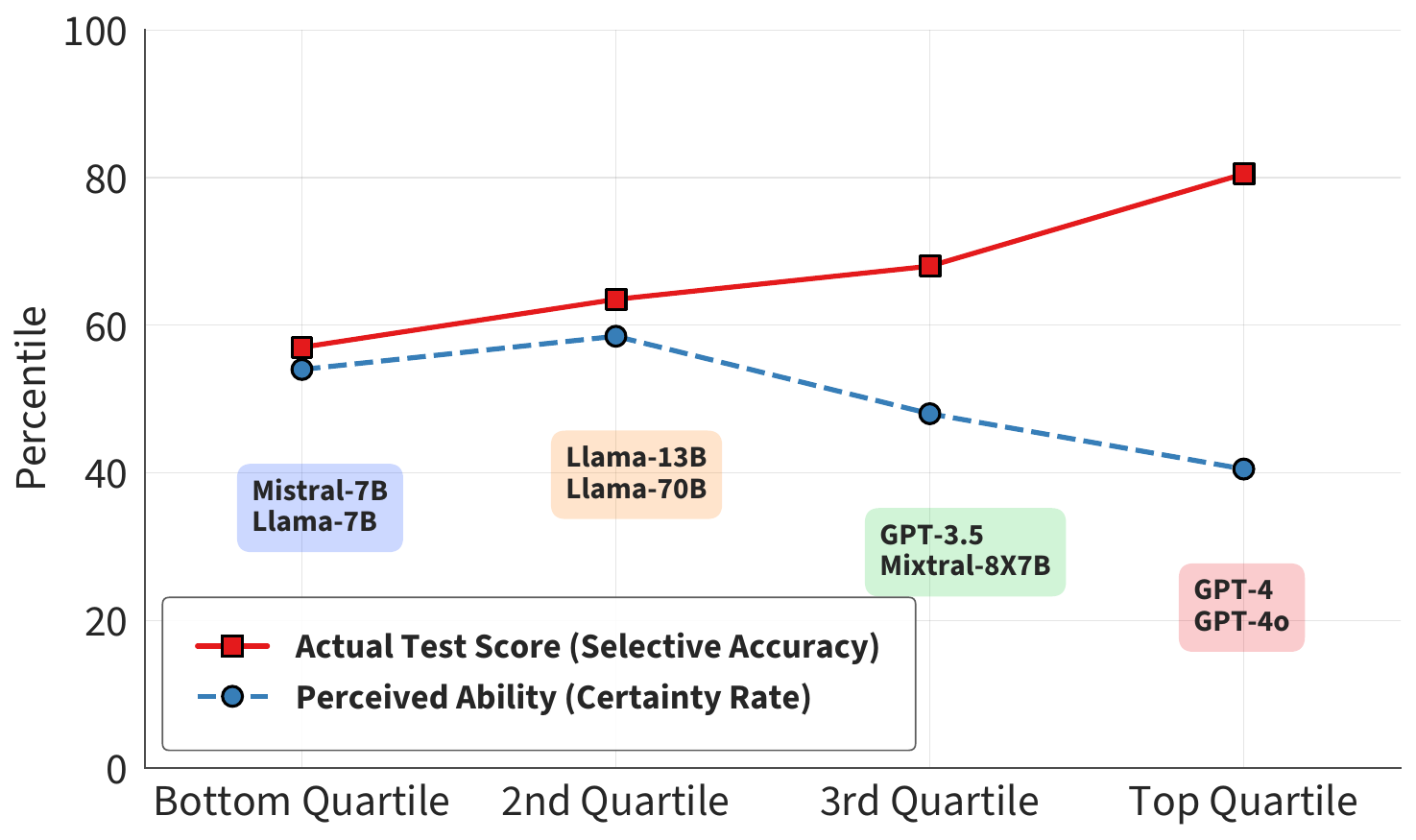}
        \caption{Prompt 2}
    \end{subfigure}

    \vspace{\floatsep} % Vertical space between rows

    % --- Second Row: Prompt 3 & 4 ---
    \begin{subfigure}[b]{0.48\linewidth}
        \centering
        \includegraphics[width=\linewidth]{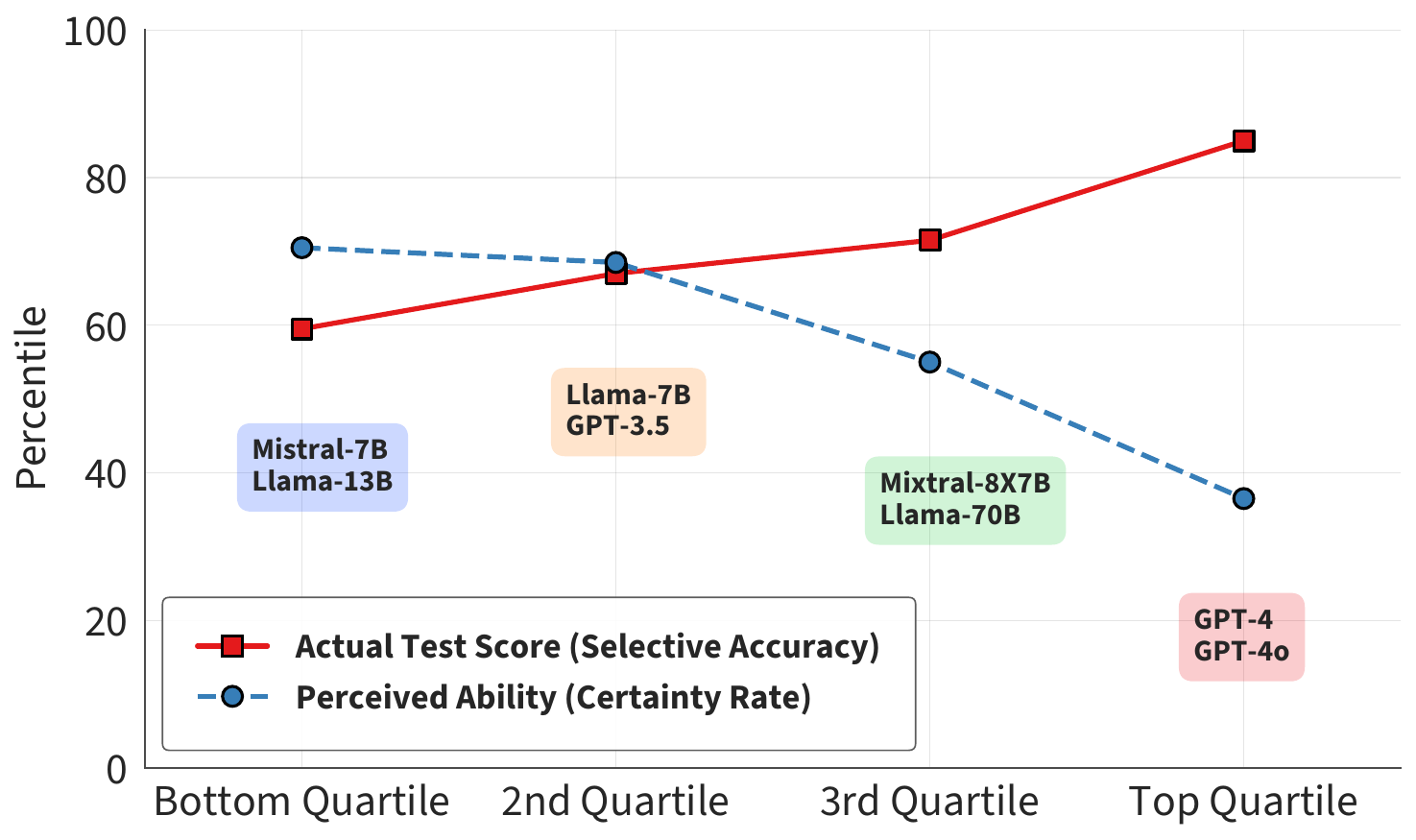}
        \caption{Prompt 3}
    \end{subfigure}
    \hfill
    \begin{subfigure}[b]{0.48\linewidth}
        \centering
        \includegraphics[width=\linewidth]{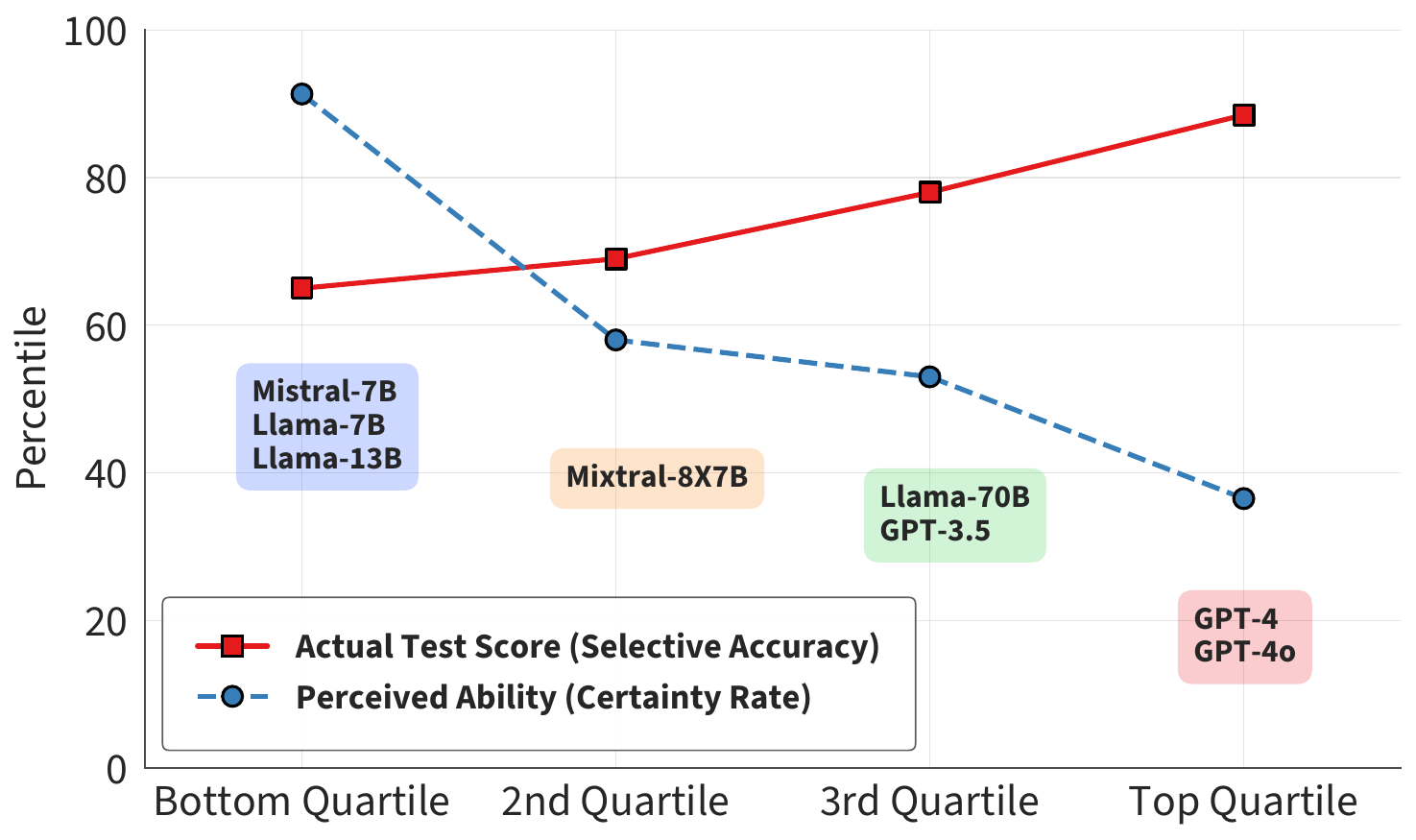}
        \caption{Prompt 4}
    \end{subfigure}

    % --- Overall Caption ---
    \caption{Quartile performance analysis on each individual prompt also displays the Dunning-Kruger effect pattern where less capable models show high confidence but low selective accuracy, and more capable models show lower confidence but higher selective accuracy.}
    \label{fig:dunning-kruger-all-prompts}
\end{figure}

\begin{figure}
    \centering
    \includegraphics[width=1\linewidth]{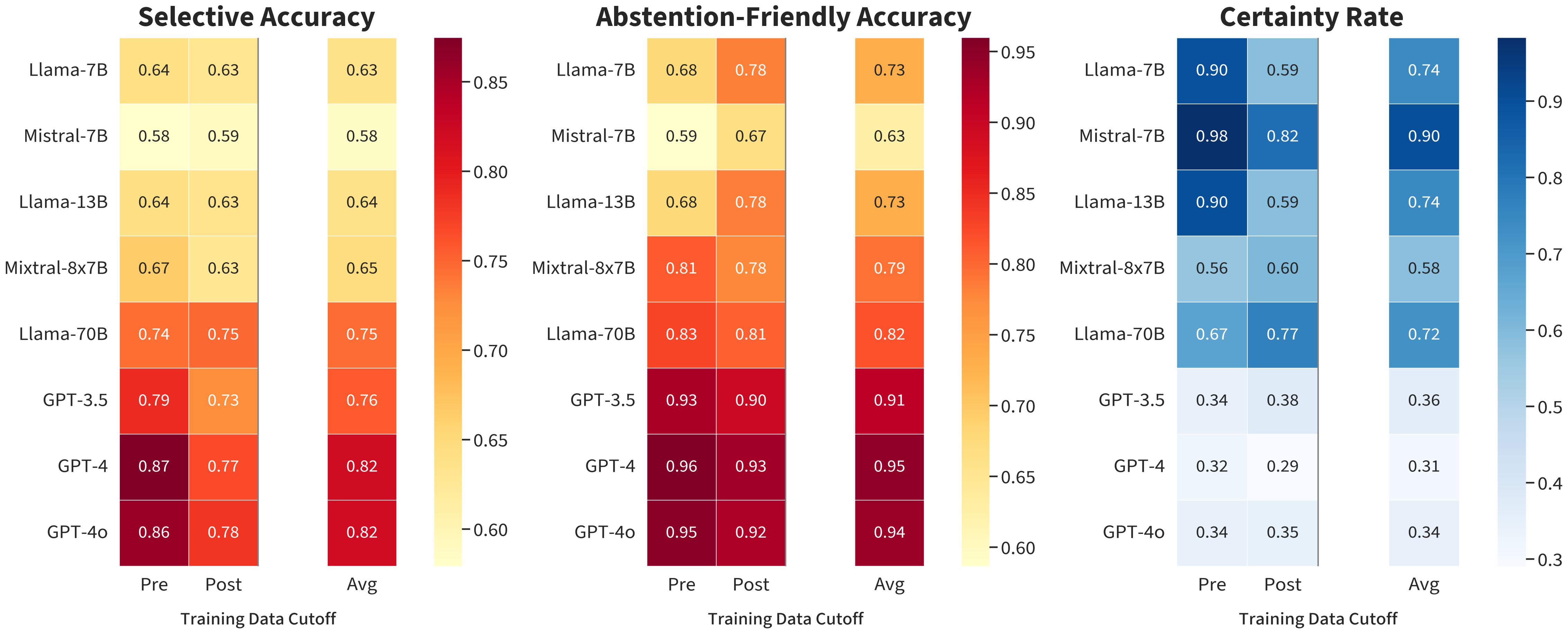}
    \caption{Performance of LLMs across claims dated pre and post their training period.}
    \label{fig:heatmaps-generalization-cutoffs}
\end{figure}

% \begin{figure}[h!]
%     \centering
%     \begin{subfigure}[b]{0.48\textwidth}
%         \centering
%         \includegraphics[width=\linewidth]{figures/conc_acc_generalization.png}
%         \caption{Selective Accuracy}
%         \label{fig:gen-heatmap-a}
%     \end{subfigure}
%     \hfill
%     \begin{subfigure}[b]{0.48\textwidth}
%         \centering
%         \includegraphics[width=\linewidth]{figures/cert_rate_generalization.png}
%         \caption{Certainty Rate}
%         \label{fig:gen-heatmap-b}
%     \end{subfigure}
%     \caption{Performance of LLMs across claims dated pre and post their training period.}
%     \label{fig:heatmaps-generalization-cutoffs}
% \end{figure}
\clearpage

\begin{figure}[htbp] % Changed placement specifier to htbp for better flexibility
    \centering

    % --- First Subfigure (Top Row) ---
    \begin{subfigure}[b]{\linewidth} % Set width to linewidth
        \centering
        \includegraphics[width=\linewidth]{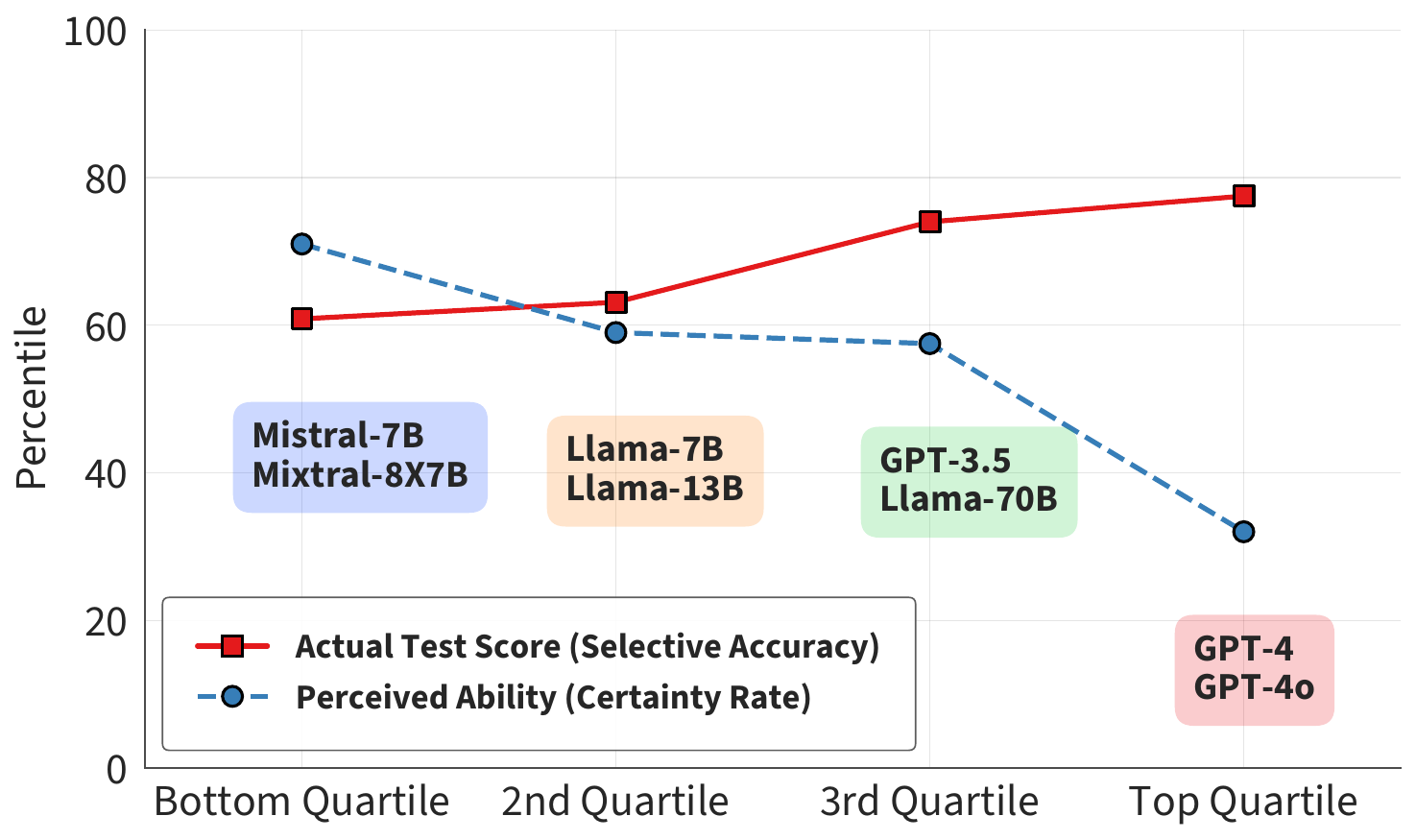}
        \caption{} % Keep empty caption as in original
        \label{fig:conc-post}
    \end{subfigure}

    \vspace{\floatsep} % Add standard vertical space between subfigures

    % --- Second Subfigure (Bottom Row) ---
    \begin{subfigure}[b]{\linewidth} % Set width to linewidth
        \centering
        \includegraphics[width=\linewidth]{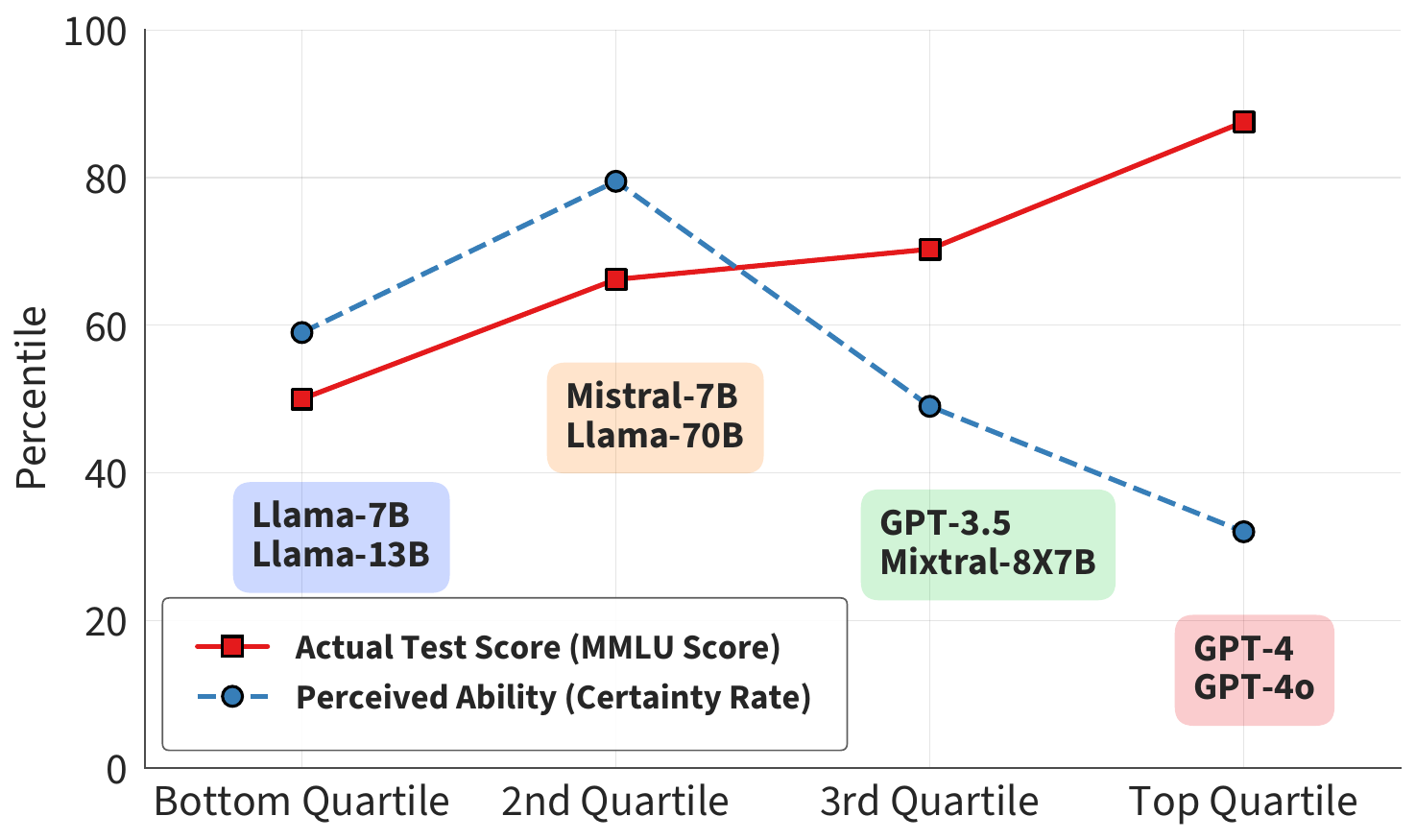}
        \caption{} % Keep empty caption as in original
        \label{fig:mmlu-post}
    \end{subfigure}

    % --- Overall Caption ---
    \caption{Quartile performance analysis on post-training claims also displays the Dunning-Kruger effect pattern, serving as a robustness check.} % Corrected "display" to "displays"
    \label{fig:dunning-kruger-generalization}
\end{figure}

% models = ["Llama-7B", "Llama-13B", "Llama-70B", "Mistral-7B", "Mixtral-8X7B", "GPT-3.5", "GPT-4", "GPT-4o"]
% actual_scores = [62.94, 63.27, 75, 59, 62.71, 73, 77, 78]  # Example scores
% perceived_scores = [59,59,77,82,60,38,29,35]
\begin{figure}[h!]
    \centering
    \includegraphics[width=\linewidth]{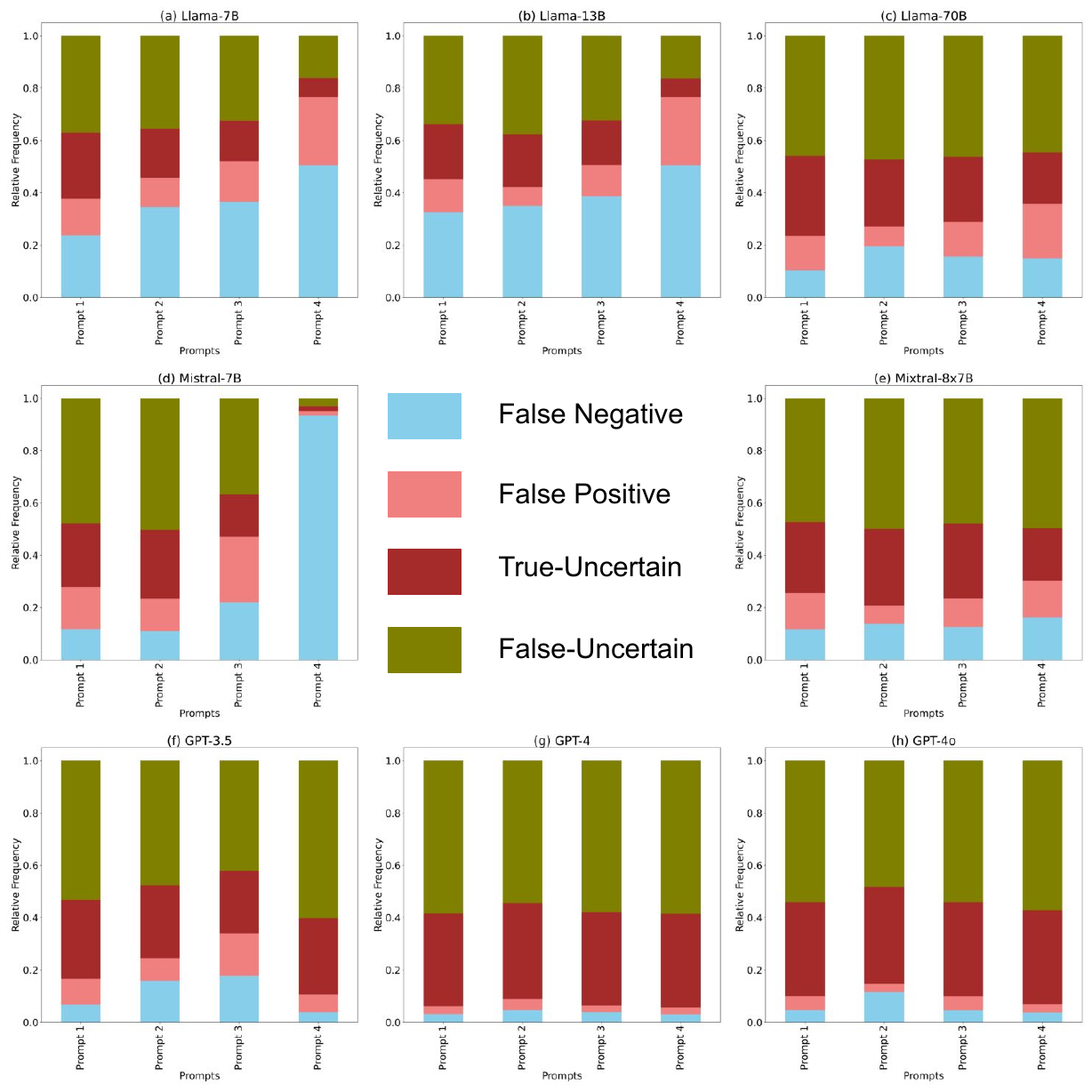}
    \caption{The full set of results investigating the error distributions of the models across all prompts.}
    \label{fig:error-barplots}
\end{figure}

\begin{figure}
    \centering
    \includegraphics[width=1\linewidth]{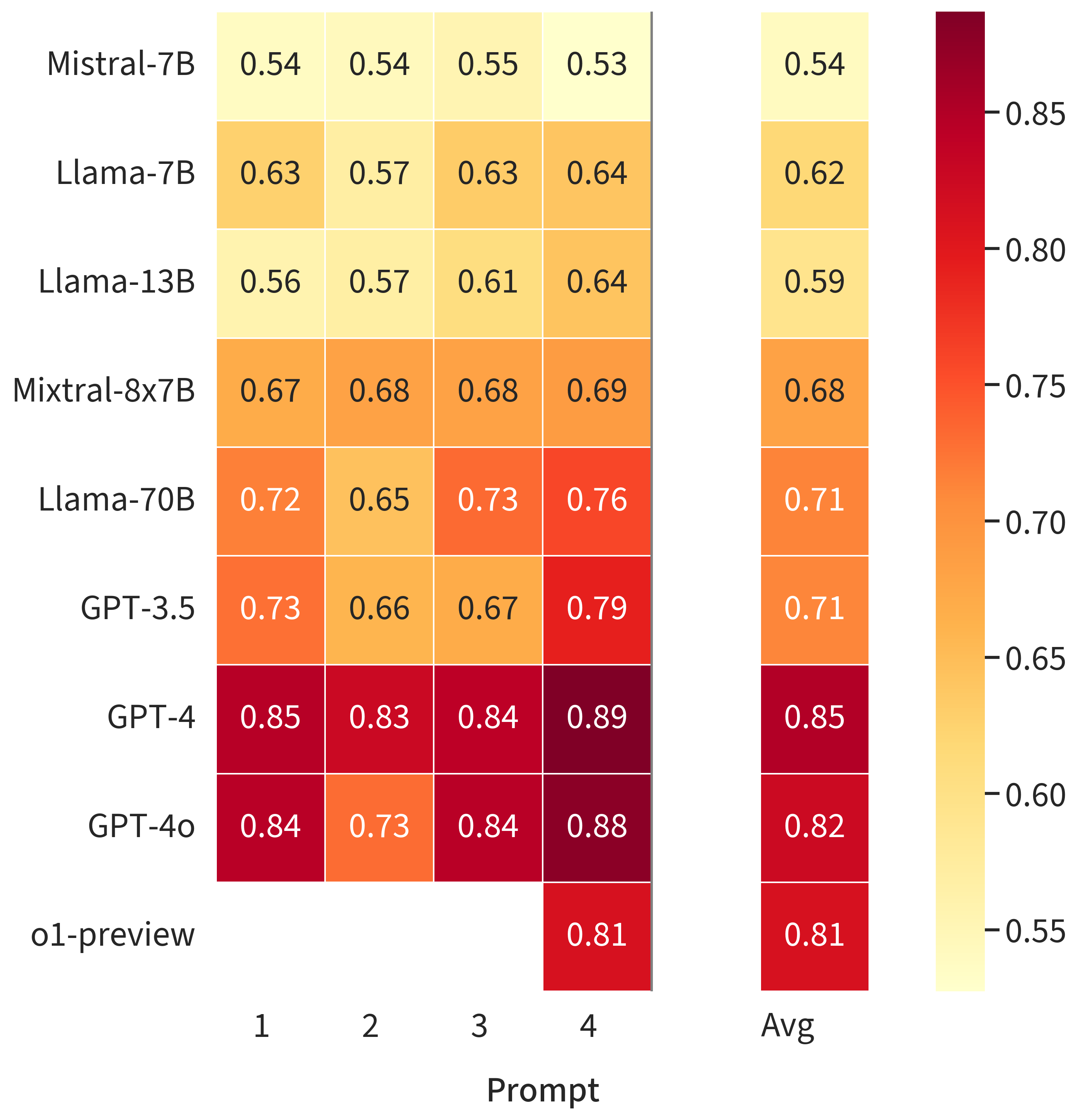}
    \caption{The balanced accuracy score for models as a robustness check to account for imbalanced true and false statements in the original dataset.}
    \label{fig:balanced-acc-scores}
\end{figure}

\begin{figure}[h!]
    \centering
    \includegraphics[width=\linewidth]{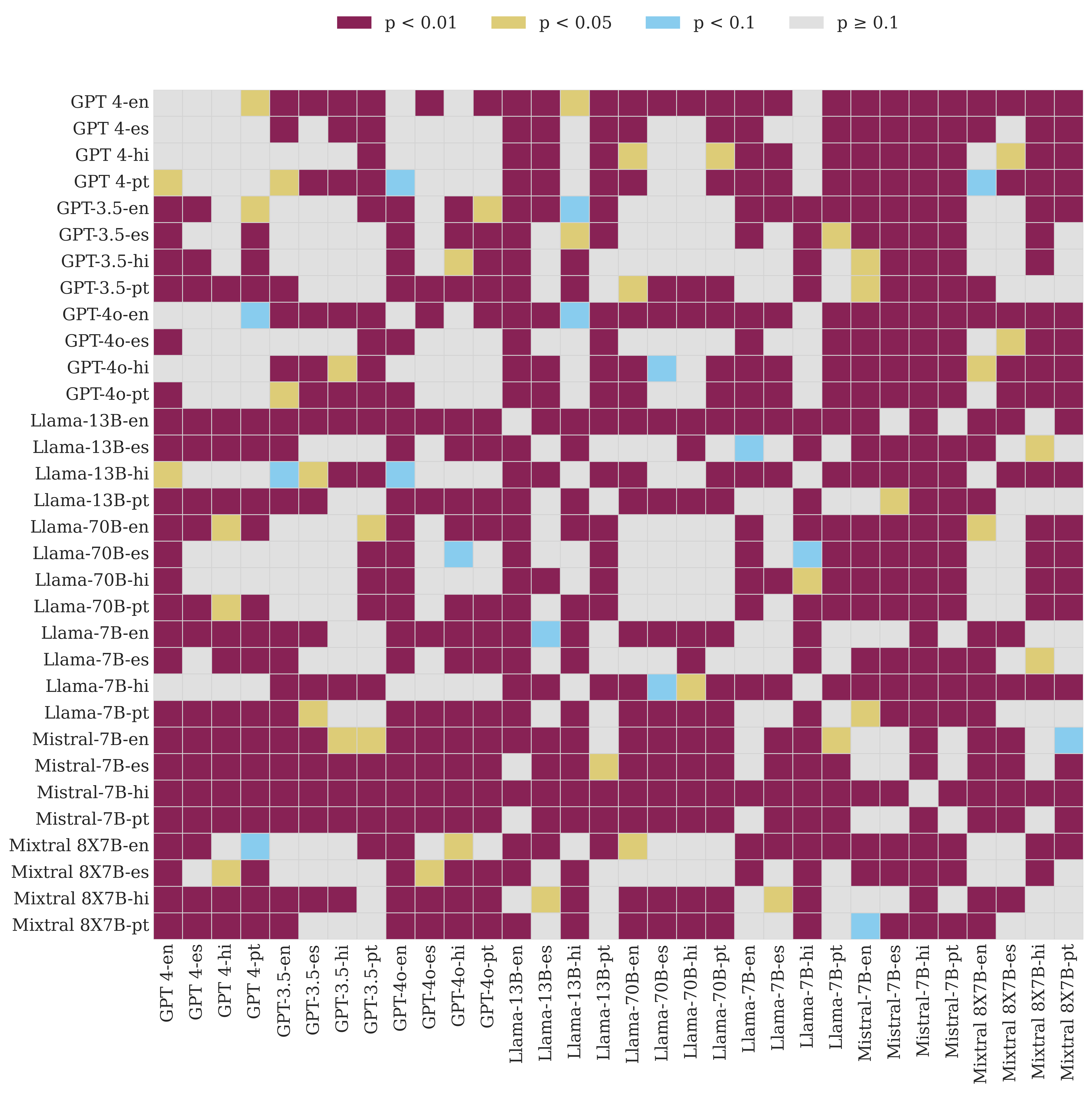}
    \caption{
        Statistical significance of pairwise comparisons for selective accuracy across all model-language combinations. 
        Each cell represents a Holm-Bonferroni adjusted $p$-value from the appropriate statistical test: McNemar's test for within-language comparisons (e.g., GPT-4-en vs. Llama-7B-en) and the Chi-Squared test for cross-language comparisons (e.g., GPT-4-en vs. GPT-4-es). The selective accuracy metric includes only claims where both models provided a definitive (`True' or `False') response. Chi-Squared tests were only performed when all expected cell frequencies were $\geq 5$.
        Color key: dark wine ($p < 0.01$), light yellow ($0.01 \leq p < 0.05$), light blue ($0.05 \leq p < 0.10$), and light grey ($p \geq 0.10$).
    }
    \label{fig:language-p-values-selective}
\end{figure}

\begin{figure}[h!]
    \centering
    \includegraphics[width=\linewidth]{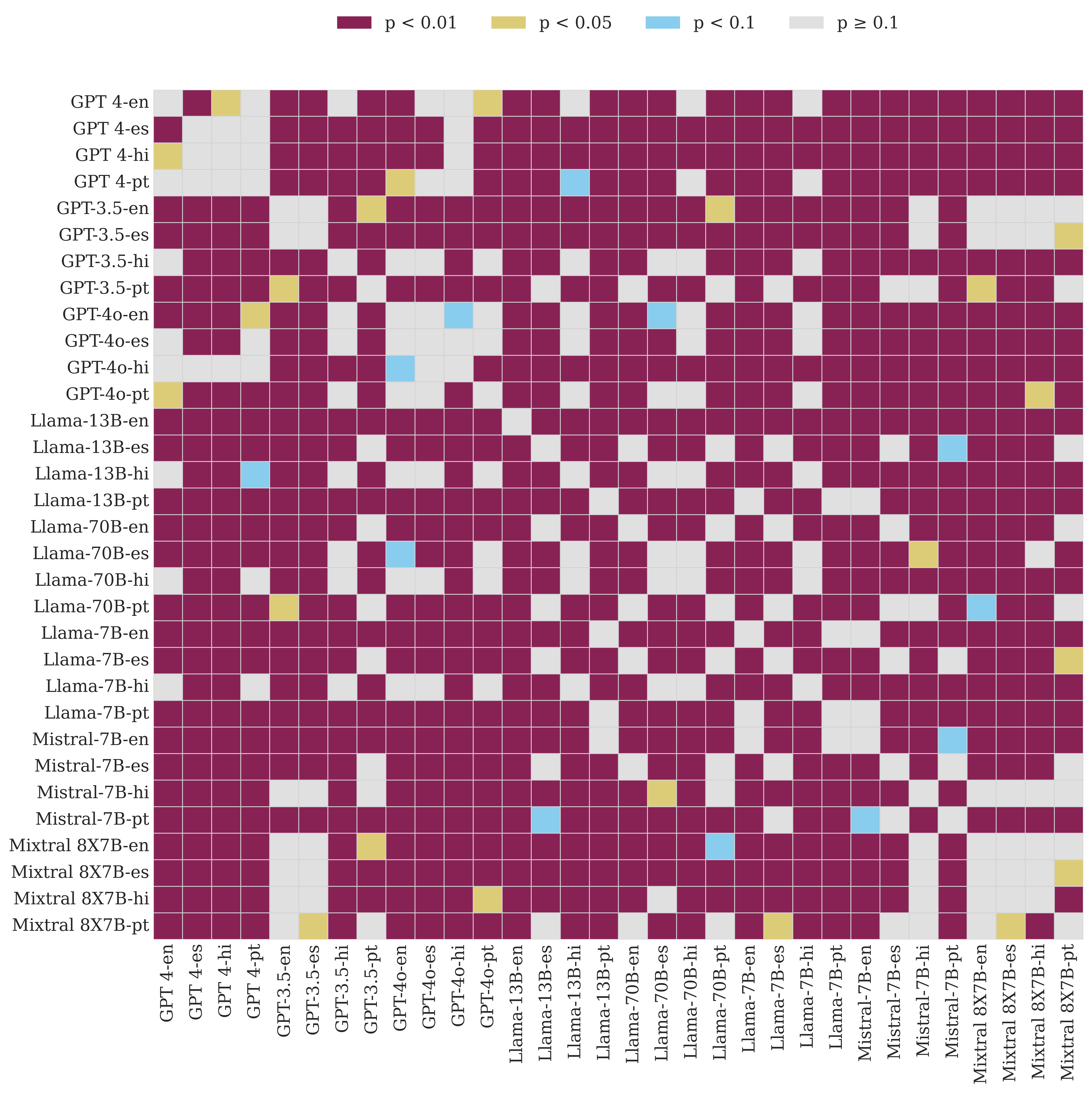}
    \caption{
        Statistical significance of pairwise comparisons for abstention-free accuracy across all model-language combinations.
        For this metric, a response was considered successful (1) if the model's judgment was correct or if it abstained. The appropriate statistical test was selected for each pair: McNemar's test for within-language comparisons and the Chi-Squared test for cross-language comparisons. All $p$-values were collectively adjusted using the Holm-Bonferroni method.
        Color key: dark wine ($p < 0.01$), light yellow ($0.01 \leq p < 0.05$), light blue ($0.05 \leq p < 0.10$), and light grey ($p \geq 0.10$).
    }
    \label{fig:language-p-values-abstention-free}
\end{figure}

\begin{figure}[h!]
    \centering
    \includegraphics[width=\linewidth]{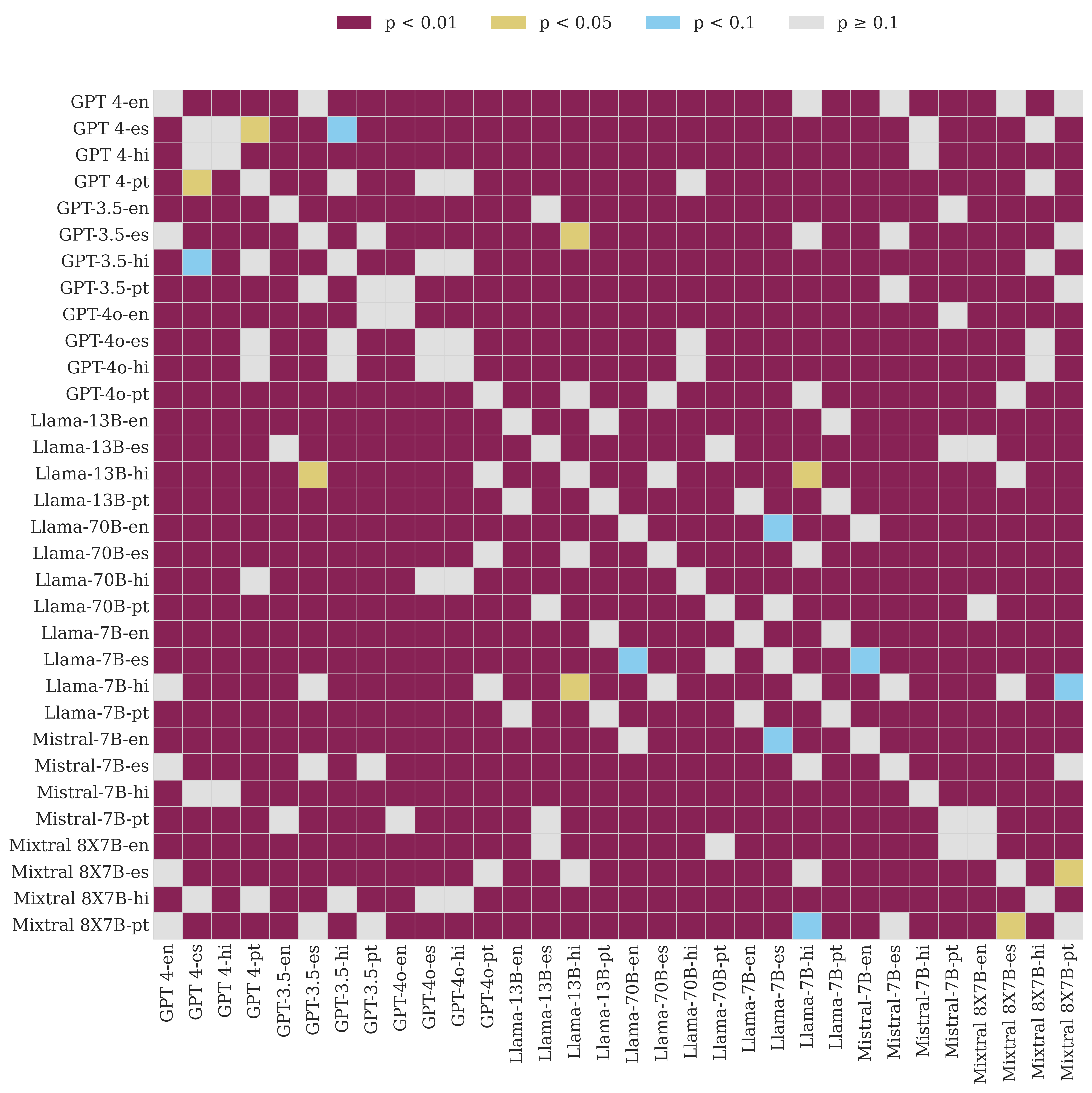}
    \caption{
        Statistical significance of pairwise comparisons for the Certainty Rate across all model-language combinations.
        The Certainty Rate measures a model's propensity to provide a definitive answer (not `Other'). Each cell shows the Holm-Bonferroni adjusted $p$-value derived from either a McNemar's test (for within-language pairs) or a Chi-Squared test (for cross-language pairs), with the latter only performed when expected frequencies were sufficient.
        Color key: dark wine ($p < 0.01$), light yellow ($0.01 \leq p < 0.05$), light blue ($0.05 \leq p < 0.10$), and light grey ($p \geq 0.10$).
    }
    \label{fig:language-p-values-certainty-rate}
\end{figure}

\begin{figure}
    \centering
    \includegraphics[width=1\linewidth]{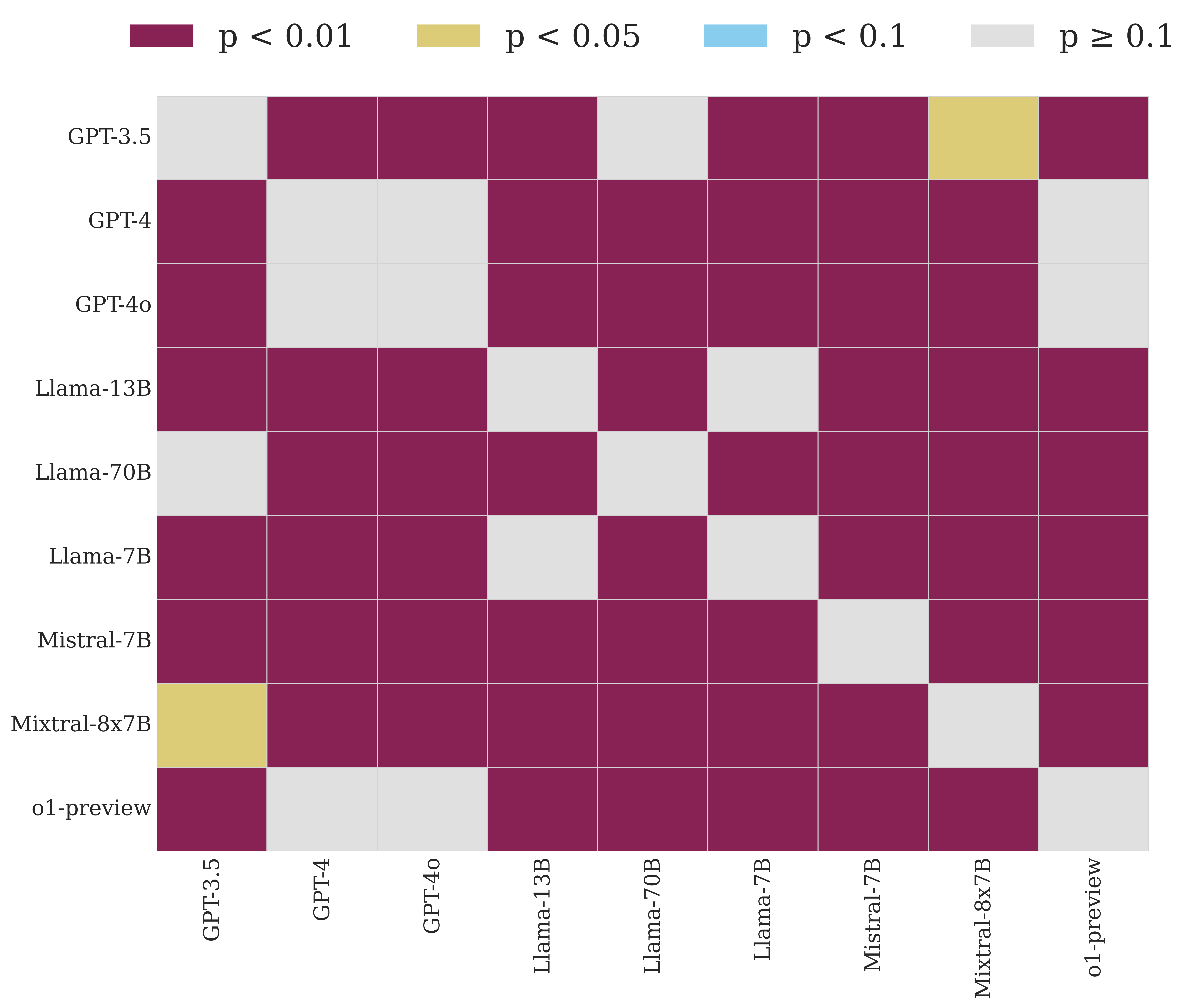}
    \caption{P-values distribution for selective accuracy evaluated using McNemar's test across pairs of Models with adjustments for multiple comparisons using Holm-Bonferroni method (including \texttt{o1-preview}) for Prompt 4. Color keys: dark wine ($p \leq 0.01$), light yellow ($0.01 \leq p < 0.05$), light blue ($0.05 \leq p < 0.10$), light grey ($p \geq 0.10$).}
    \label{fig:supp-welsch-ca-4}
\end{figure}

\begin{figure}
    \centering
    \includegraphics[width=1\linewidth]{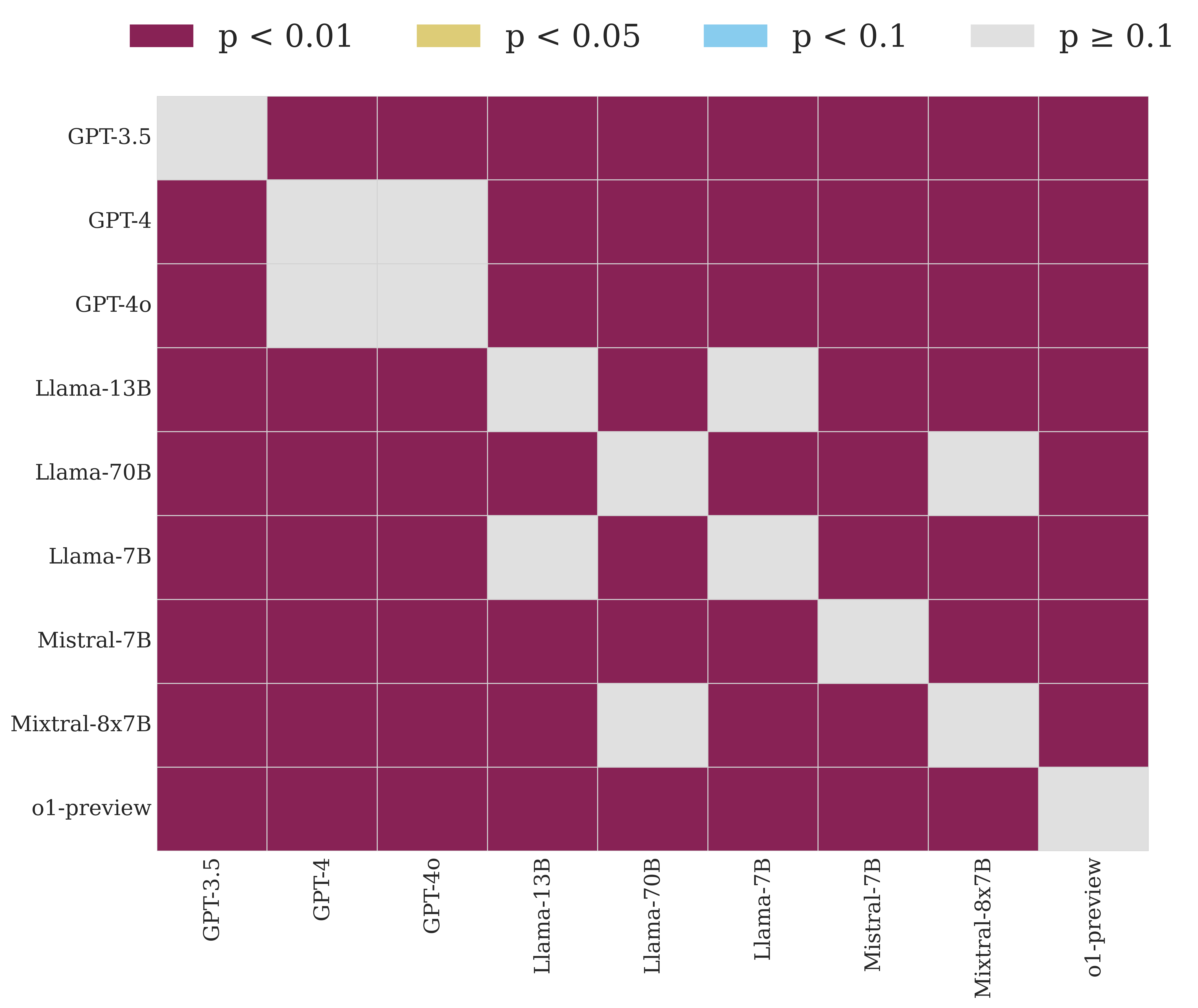}
    \caption{P-values distribution for abstention-friendly accuracy evaluated using McNemar's test across pairs of Models with adjustments for multiple comparisons using the Holm-Bonferroni method (including \texttt{o1-preview}) for Prompt 4. Color keys: dark wine ($p \leq 0.01$), light yellow ($0.01 \leq p < 0.05$), light blue ($0.05 \leq p < 0.10$), light grey ($p \geq 0.10$).}
    \label{fig:supp-welsch-afa-4}
\end{figure}

\begin{figure}
    \centering
    \includegraphics[width=1\linewidth]{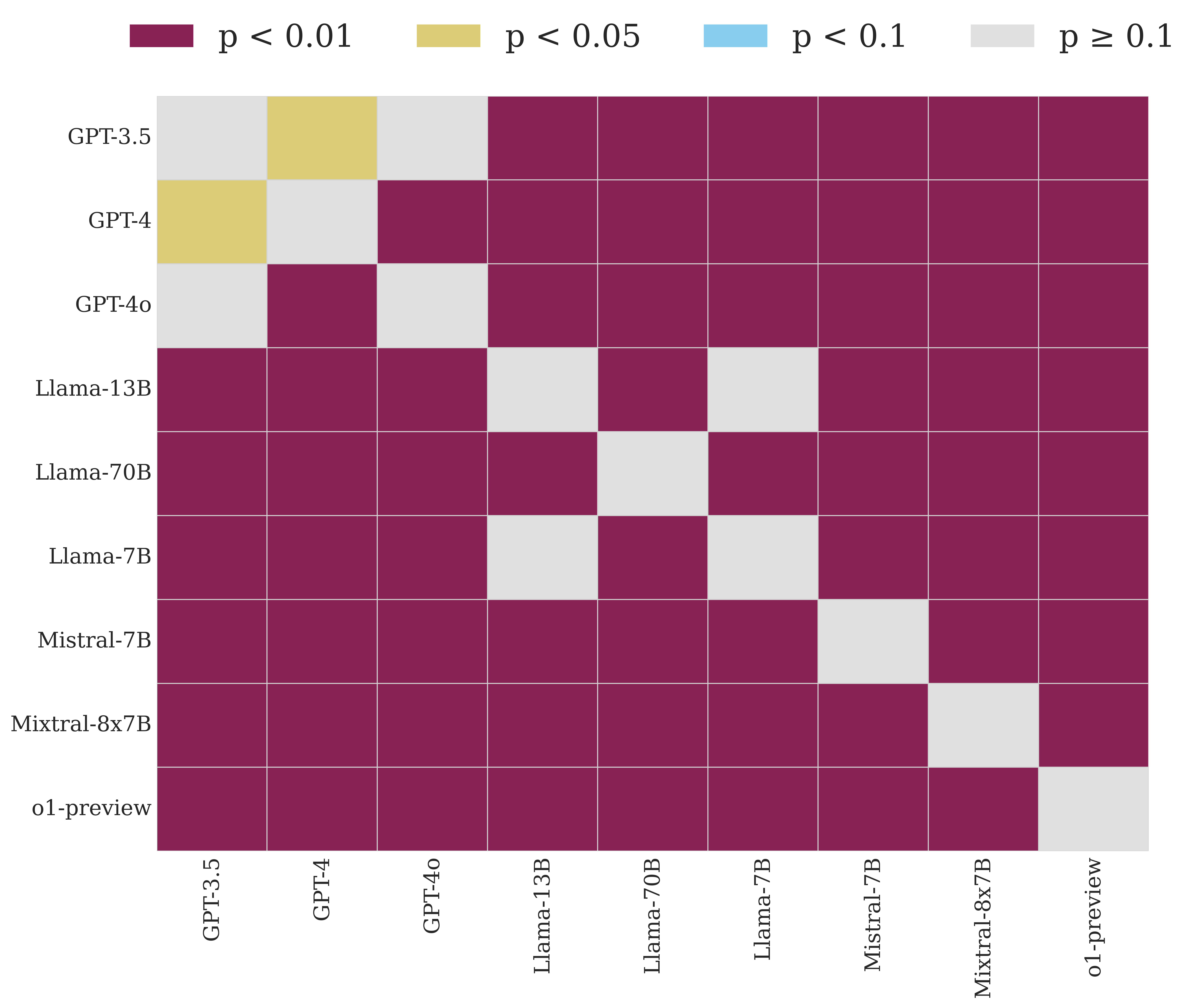}
    \caption{P-values distribution for certainty rate evaluated using McNemar's test across pairs of Models with adjustments for multiple comparisons using Holm-Bonferroni method (including \texttt{o1-preview}) for Prompt 4. Color keys: dark wine ($p \leq 0.01$), light yellow ($0.01 \leq p < 0.05$), light blue ($0.05 \leq p < 0.10$), light grey ($p \geq 0.10$).}
    \label{fig:supp-welsch-cr-4}
\end{figure}

\begin{figure}[h!]
    \centering
    \includegraphics[width=\linewidth]{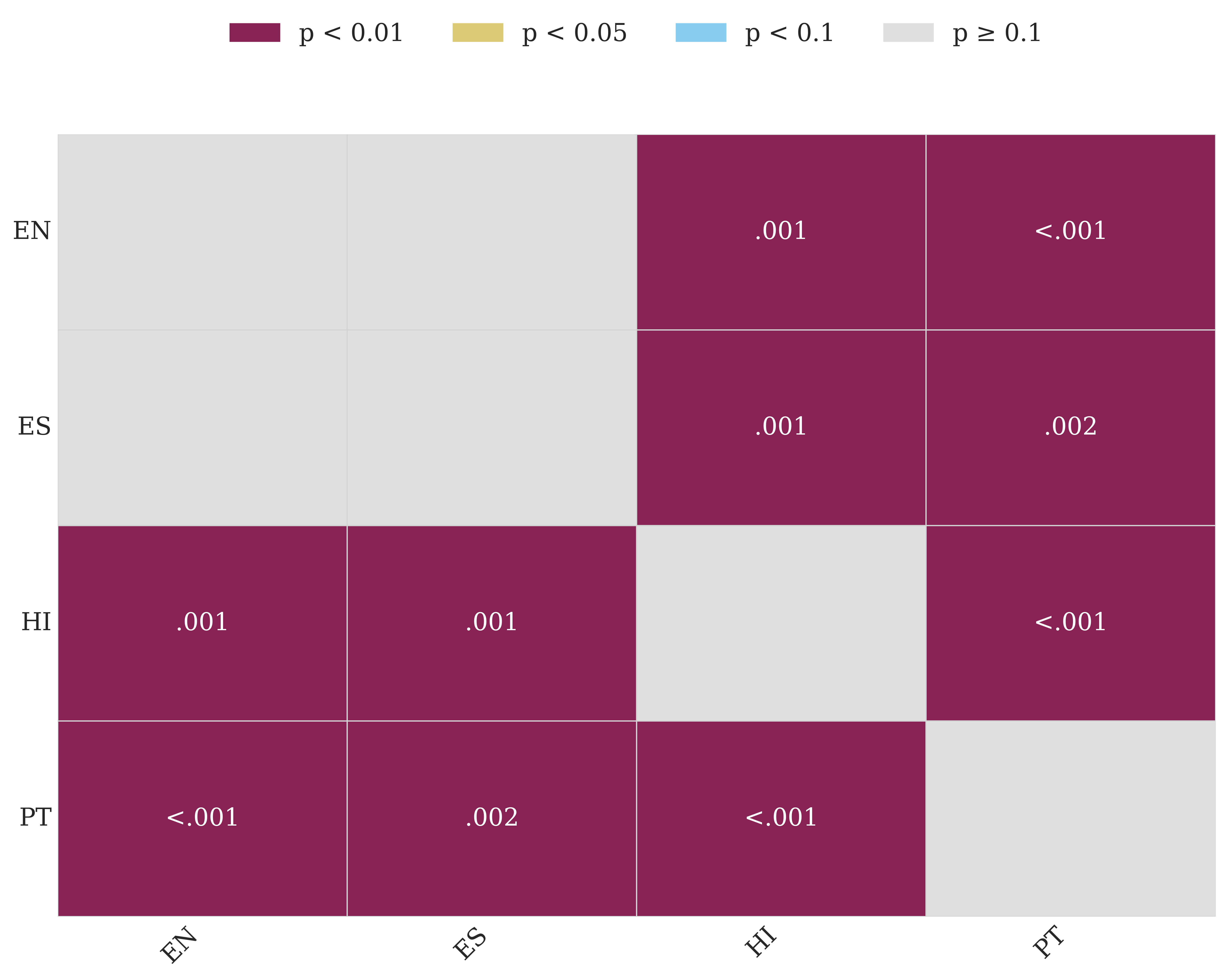}
    \caption{
        Statistical significance of pairwise comparisons for selective accuracy between languages, aggregating data across all models on prompts 1, 2, and 3.
        Each cell represents the Holm-Bonferroni adjusted $p$-value from a Chi-Squared test comparing the performance of two languages. Selective accuracy includes only claims where a definitive (`True' or `False') response was given. The total sample sizes for this metric after filtering were: English ($n=28,319$), Spanish ($n=5,800$), Hindi ($n=2,911$), and Portuguese ($n=10,151$).
        Color key: dark wine ($p < 0.01$), light yellow ($0.01 \leq p < 0.05$), light blue ($0.05 \leq p < 0.10$), and light grey ($p \geq 0.10$).
    }
    \label{fig:cross-lang-p-values-selective}
\end{figure}

\begin{figure}[h!]
    \centering
    \includegraphics[width=\linewidth]{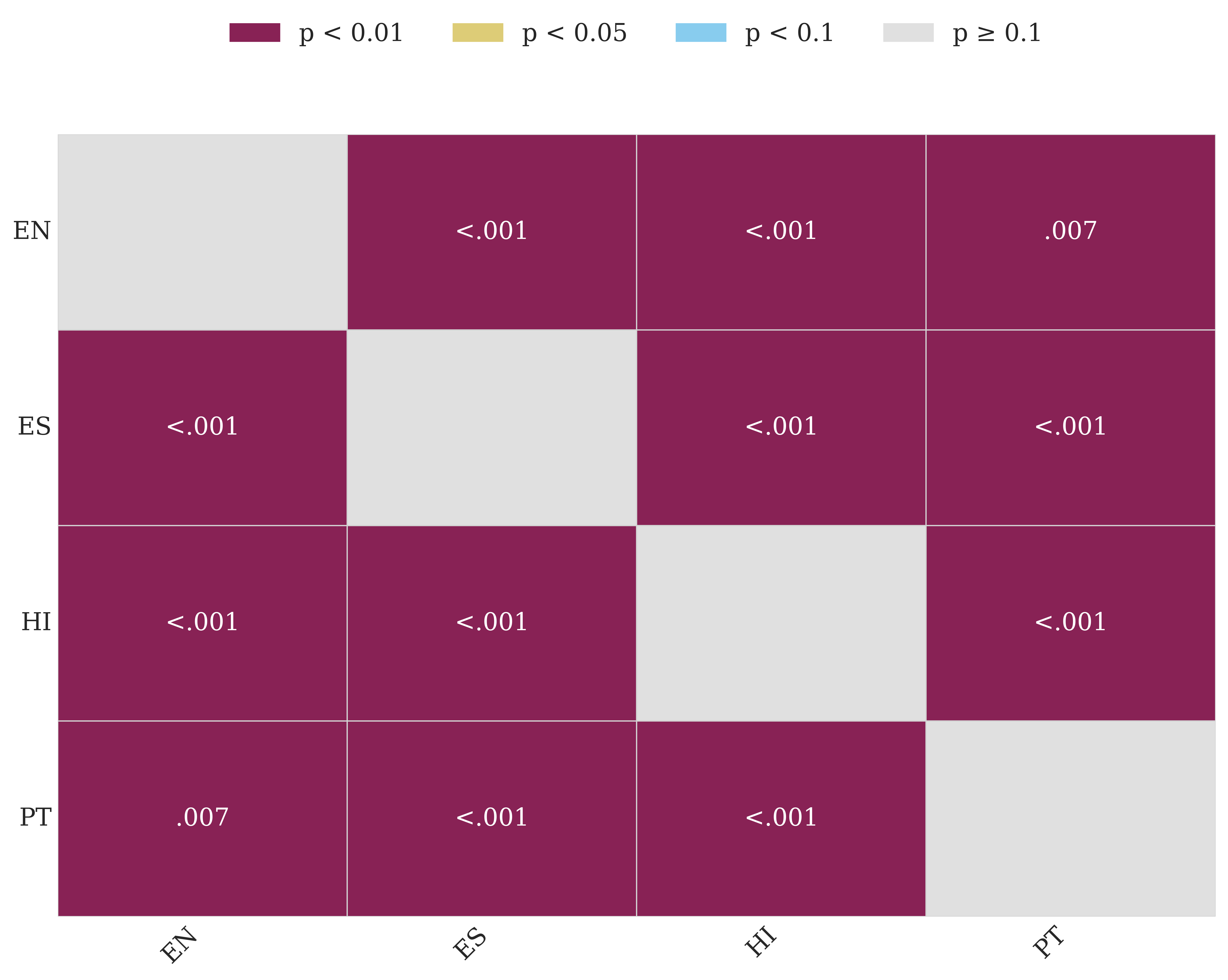}
    \caption{
        Statistical significance of pairwise comparisons for abstention-friendly accuracy between languages, aggregating data across all models on prompts 1, 2, and 3.
        For this metric, a response was considered successful if it was correct or an abstention (`other'). Each cell displays the Holm-Bonferroni adjusted $p$-value from a pairwise Chi-Squared test. The total sample sizes were: English ($n=42,456$), Spanish ($n=12,600$), Hindi ($n=9,576$), and Portuguese ($n=17,616$).
        Color key: dark wine ($p < 0.01$), light yellow ($0.01 \leq p < 0.05$), light blue ($0.05 \leq p < 0.10$), and light grey ($p \geq 0.10$).
    }
    \label{fig:cross-lang-p-values-abstention-free}
\end{figure}

\begin{figure}[h!]
    \centering
    \includegraphics[width=\linewidth]{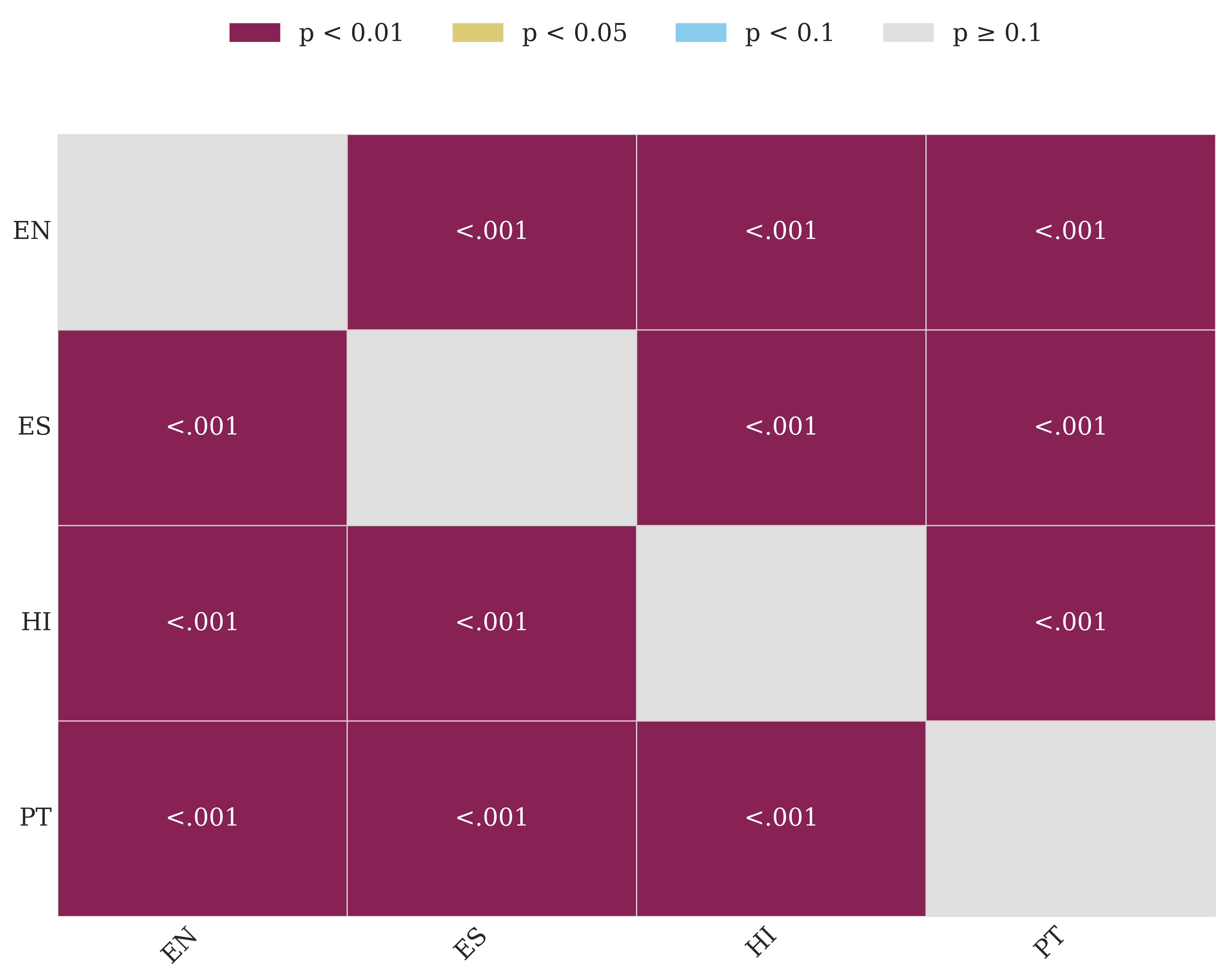}
    \caption{
        Statistical significance of pairwise comparisons for the Certainty Rate between languages, aggregating data across all models on prompts 1, 2, and 3.
        The Certainty Rate measures the proportion of non-abstaining (i.e., not `Other') responses. Each cell shows the Holm-Bonferroni adjusted $p$-value from a pairwise Chi-Squared test. The total sample sizes were: English ($n=42,456$), Spanish ($n=12,600$), Hindi ($n=9,576$), and Portuguese ($n=17,616$).
        Color key: dark wine ($p < 0.01$), light yellow ($0.01 \leq p < 0.05$), light blue ($0.05 \leq p < 0.10$), and light grey ($p \geq 0.10$).
    }
    \label{fig:cross-lang-p-values-certainty-rate}
\end{figure}

% \begin{figure}
%     \centering
%     \includegraphics[width=1\linewidth]{figures/supp-welsch-cr-langs.png}
%     \caption{P-values distribution for certainty rate evaluated using two-sided Welch’s t-test across pairs of Models in different languages. Color keys: green ($p < 0.01$), light green ($0.01 \leq p < 0.05$), orange ($0.05 \leq p < 0.10$), red ($p \geq 0.10$).}
%     \label{fig:supp-welsch-cr-langs}
% \end{figure}

% \begin{figure}
%     \centering
%     \includegraphics[width=1\linewidth]{figures/supp-welsch-ca-langs.png}
%     \caption{P-values distribution for selective accuracy evaluated using two-sided Welch’s t-test across pairs of Models and across languages. Color keys: green ($p < 0.01$), light green ($0.01 \leq p < 0.05$), orange ($0.05 \leq p < 0.10$), red ($p \geq 0.10$).}
%     \label{fig:supp-welsch-ca-langs}
% \end{figure}

\clearpage
\section{Supporting Tables}

\begin{longtable}{| l | r |} % l for left-align (Model), r for right-align (n_pair)

\caption{Sample sizes of claims used for selective accuracy comparisons after excluding claims labeled as `Other' in Fig. 2.} % Add your desired caption
\label{tab:model-ca-sample-sizes}\\ % Add your desired label

% --- HEADER FOR THE FIRST PAGE ---
    \hline
    \textbf{Model-Prompt} & \textbf{Sample size (excluding claims labeled as `Other')} \\ % Adjusted column headers
    \hline
\endfirsthead % Marks the end of the header for the first page

% --- HEADER FOR SUBSEQUENT PAGES ---
    \multicolumn{2}{c}% % Adjusted colspan to 2 columns
    {{\bfseries \tablename\ \thetable{} -- continued from previous page}} \\
    \hline
    \textbf{Model-Prompt} & \textbf{Sample size (excluding claims labeled as `Other') } \\ % Adjusted column headers
    \hline
\endhead % Defines the header for all pages after the first

% --- FOOTER FOR ALL PAGES EXCEPT THE LAST ---
    \hline \multicolumn{2}{r}{{Continued on next page}} \\ \hline % Adjusted colspan to 2 columns
\endfoot % Defines the footer for all pages except the last

% --- FOOTER FOR THE LAST PAGE ---
    \hline
\endlastfoot % Defines the footer for the very last page

% --- BEGIN DATA ROWS ---
    Llama-13B Prompt 1 & 3363 \\ \hline
    GPT-3.5 Prompt 1 & 2116 \\ \hline
    Mistral-7B Prompt 1 & 2269 \\ \hline
    Llama-70B Prompt 1 & 2618 \\ \hline
    Llama-7B Prompt 1 & 3190 \\ \hline
    GPT-4o Prompt 1 & 2086 \\ \hline
    Mixtral-8X7B Prompt 1 & 2559 \\ \hline
    GPT-4 Prompt 1 & 1494 \\ \hline
    Llama-13B Prompt 2 & 3248 \\ \hline
    GPT-3.5 Prompt 2 & 2486 \\ \hline
    Mistral-7B Prompt 2 & 1997 \\ \hline
    Llama-70B Prompt 2 & 2613 \\ \hline
    Llama-7B Prompt 2 & 3401 \\ \hline
    GPT-4o Prompt 2 & 2200 \\ \hline
    Mixtral-8X7B Prompt 2 & 2310 \\ \hline
    GPT-4 Prompt 2 & 1852 \\ \hline
    Llama-13B Prompt 3 & 3706 \\ \hline
    GPT-3.5 Prompt 3 & 3068 \\ \hline
    Mistral-7B Prompt 3 & 3328 \\ \hline
    Llama-70B Prompt 3 & 3035 \\ \hline
    Llama-7B Prompt 3 & 3802 \\ \hline
    GPT-4o Prompt 3 & 2085 \\ \hline
    Mixtral-8X7B Prompt 3 & 2474 \\ \hline
    GPT-4 Prompt 3 & 1526 \\ \hline
    Llama-13B Prompt 4 & 4417 \\ \hline
    GPT-3.5 Prompt 4 & 1856 \\ \hline
    Mistral-7B Prompt 4 & 4902 \\ \hline
    Llama-70B Prompt 4 & 3458 \\ \hline
    Llama-7B Prompt 4 & 4417 \\ \hline
    GPT-4o Prompt 4 & 1924 \\ \hline
    Mixtral-8X7B Prompt 4 & 2920 \\ \hline
    GPT-4 Prompt 4 & 1763 \\ \hline
    o1-preview Prompt 4 & 3605 \\ \hline
% Last row doesn't need \hline after it because \endlastfoot adds the final \hline

\end{longtable}

% --- DOCUMENT BODY ---
\begin{longtable}{| l | r |} % l for left-align (Model), r for right-align (n_pair)

\caption{Exact sample sizes of claims used for language analysis after excluding claims labeled as `Other' in Fig. 5.}
\label{tab:lang-ca-sample-sizes}\\ % Add your desired label

% --- HEADER FOR THE FIRST PAGE ---
    \hline
    \textbf{Model-Language} & \textbf{Sample size (excluding claims labeled as `Other')} \\ % Adjusted column headers
    \hline
\endfirsthead % Marks the end of the header for the first page

% --- HEADER FOR SUBSEQUENT PAGES ---
    \multicolumn{2}{c}% % Adjusted colspan to 2 columns
    {{\bfseries \tablename\ \thetable{} -- continued from previous page}} \\
    \hline
    \textbf{Model-Language} & \textbf{Sample size (excluding claims labeled as `Other')} \\ % Adjusted column headers
    \hline
\endhead % Defines the header for all pages after the first

% --- FOOTER FOR ALL PAGES EXCEPT THE LAST ---
    \hline \multicolumn{2}{r}{{Continued on next page}} \\ \hline % Adjusted colspan to 2 columns
\endfoot % Defines the footer for all pages except the last

% --- FOOTER FOR THE LAST PAGE ---
    \hline
\endlastfoot % Defines the footer for the very last page

% --- BEGIN DATA ROWS ---
    GPT-4o-hi & 355 \\ \hline
    GPT-4-hi & 229 \\ \hline
    GPT-3.5-hi & 331 \\ \hline
    Llama-70B-hi & 377 \\ \hline
    Mixtral-8X7B-hi & 328 \\ \hline
    Llama-13B-hi & 497 \\ \hline
    Llama-7B-hi & 543 \\ \hline
    Mistral-7B-hi & 251 \\ \hline
    
    GPT-4o-en & 3011 \\ \hline
    GPT-4-en & 2561 \\ \hline
    GPT-3.5-en & 3135 \\ \hline
    Llama-70B-en & 3846 \\ \hline
    Mixtral-8X7B-en & 3281 \\ \hline
    Llama-13B-en & 4429 \\ \hline
    Llama-7B-en & 4217 \\ \hline
    Mistral-7B-en & 3839 \\ \hline
    
    GPT-4o-es & 504 \\ \hline
    GPT-4-es & 353 \\ \hline
    GPT-3.5-es & 766 \\ \hline
    Llama-70B-es & 622 \\ \hline
    Mixtral-8X7B-es & 708 \\ \hline
    Llama-13B-es & 999 \\ \hline
    Llama-7B-es & 1074 \\ \hline
    Mistral-7B-es & 774 \\ \hline
    
    GPT-4o-pt & 913 \\ \hline
    GPT-4-pt & 603 \\ \hline
    GPT-3.5-pt & 1178 \\ \hline
    Llama-70B-pt & 1409 \\ \hline
    Mixtral-8X7B-pt & 1125 \\ \hline
    Llama-13B-pt & 1807 \\ \hline
    Llama-7B-pt & 1812 \\ \hline
    Mistral-7B-pt & 1304 \\ \hline
\end{longtable}
\clearpage

% Add \usepackage{longtable} to your LaTeX preamble if you haven't already.

\begin{longtable}{| l | r |}
\caption{Sample sizes for pairwise selective accuracy comparisons between model-language pairs within the same language. The sample size represents the number of claims commonly evaluated by each pair after excluding non-definitive (`Other') responses. Corresponds to the within-language comparisons in Supplementary Fig.~\ref{fig:language-p-values-selective}.}
\label{tab:sample-sizes-model-lang}\\

% --- HEADER FOR THE FIRST PAGE ---
    \hline
    \textbf{Model-Language Pair} & \textbf{Sample Size} \\
    \hline
\endfirsthead

% --- HEADER FOR SUBSEQUENT PAGES ---
    \multicolumn{2}{c}%
    {{\bfseries \tablename\ \thetable{} -- continued from previous page}} \\
    \hline
    \textbf{Model-Language Pair} & \textbf{Sample Size} \\
    \hline
\endhead

% --- FOOTER FOR ALL PAGES EXCEPT THE LAST ---
    \hline \multicolumn{2}{r}{{Continued on next page}} \\ \hline
\endfoot

% --- FOOTER FOR THE LAST PAGE ---
    \hline
\endlastfoot

% --- BEGIN DATA ROWS ---
    GPT 3.5-en vs. GPT 4-en & 6282 \\ \hline
    GPT 3.5-en vs. GPT 4o-en & 6634 \\ \hline
    GPT 3.5-en vs. Llama 13B-en & 8438 \\ \hline
    GPT 3.5-en vs. Llama 70B-en & 7615 \\ \hline
    GPT 3.5-en vs. Llama 7B-en & 8137 \\ \hline
    GPT 3.5-en vs. Mistral 7B-en & 7255 \\ \hline
    GPT 3.5-en vs. Mixtral 8X7B-en & 6998 \\ \hline
    GPT 3.5-es vs. GPT 4-es & 846 \\ \hline
    GPT 3.5-es vs. GPT 4o-es & 1064 \\ \hline
    GPT 3.5-es vs. Llama 13B-es & 1722 \\ \hline
    GPT 3.5-es vs. Llama 70B-es & 1128 \\ \hline
    GPT 3.5-es vs. Llama 7B-es & 1753 \\ \hline
    GPT 3.5-es vs. Mistral 7B-es & 1214 \\ \hline
    GPT 3.5-es vs. Mixtral 8X7B-es & 1331 \\ \hline
    GPT 3.5-hi vs. GPT 4-hi & 383 \\ \hline
    GPT 3.5-hi vs. GPT 4o-hi & 402 \\ \hline
    GPT 3.5-hi vs. Llama 13B-hi & 465 \\ \hline
    GPT 3.5-hi vs. Llama 70B-hi & 423 \\ \hline
    GPT 3.5-hi vs. Llama 7B-hi & 539 \\ \hline
    GPT 3.5-hi vs. Mistral 7B-hi & 230 \\ \hline
    GPT 3.5-hi vs. Mixtral 8X7B-hi & 312 \\ \hline
    GPT 3.5-pt vs. GPT 4-pt & 1427 \\ \hline
    GPT 3.5-pt vs. GPT 4o-pt & 1939 \\ \hline
    GPT 3.5-pt vs. Llama 13B-pt & 3013 \\ \hline
    GPT 3.5-pt vs. Llama 70B-pt & 2403 \\ \hline
    GPT 3.5-pt vs. Llama 7B-pt & 3037 \\ \hline
    GPT 3.5-pt vs. Mistral 7B-pt & 2136 \\ \hline
    GPT 3.5-pt vs. Mixtral 8X7B-pt & 2156 \\ \hline
    GPT 4-en vs. GPT 4o-en & 6017 \\ \hline
    GPT 4-en vs. Llama 13B-en & 6930 \\ \hline
    GPT 4-en vs. Llama 70B-en & 6256 \\ \hline
    GPT 4-en vs. Llama 7B-en & 6673 \\ \hline
    GPT 4-en vs. Mistral 7B-en & 5951 \\ \hline
    GPT 4-en vs. Mixtral 8X7B-en & 6019 \\ \hline
    GPT 4-es vs. GPT 4o-es & 732 \\ \hline
    GPT 4-es vs. Llama 13B-es & 848 \\ \hline
    GPT 4-es vs. Llama 70B-es & 621 \\ \hline
    GPT 4-es vs. Llama 7B-es & 860 \\ \hline
    GPT 4-es vs. Mistral 7B-es & 573 \\ \hline
    GPT 4-es vs. Mixtral 8X7B-es & 749 \\ \hline
    GPT 4-hi vs. GPT 4o-hi & 383 \\ \hline
    GPT 4-hi vs. Llama 13B-hi & 379 \\ \hline
    GPT 4-hi vs. Llama 70B-hi & 349 \\ \hline
    GPT 4-hi vs. Llama 7B-hi & 410 \\ \hline
    GPT 4-hi vs. Mistral 7B-hi & 155 \\ \hline
    GPT 4-hi vs. Mixtral 8X7B-hi & 225 \\ \hline
    GPT 4-pt vs. GPT 4o-pt & 1236 \\ \hline
    GPT 4-pt vs. Llama 13B-pt & 1560 \\ \hline
    GPT 4-pt vs. Llama 70B-pt & 1227 \\ \hline
    GPT 4-pt vs. Llama 7B-pt & 1568 \\ \hline
    GPT 4-pt vs. Mistral 7B-pt & 1135 \\ \hline
    GPT 4-pt vs. Mixtral 8X7B-pt & 1227 \\ \hline
    GPT 4o-en vs. Llama 13B-en & 8130 \\ \hline
    GPT 4o-en vs. Llama 70B-en & 7189 \\ \hline
    GPT 4o-en vs. Llama 7B-en & 7751 \\ \hline
    GPT 4o-en vs. Mistral 7B-en & 6950 \\ \hline
    GPT 4o-en vs. Mixtral 8X7B-en & 6647 \\ \hline
    GPT 4o-es vs. Llama 13B-es & 1160 \\ \hline
    GPT 4o-es vs. Llama 70B-es & 838 \\ \hline
    GPT 4o-es vs. Llama 7B-es & 1208 \\ \hline
    GPT 4o-es vs. Mistral 7B-es & 789 \\ \hline
    GPT 4o-es vs. Mixtral 8X7B-es & 946 \\ \hline
    GPT 4o-hi vs. Llama 13B-hi & 541 \\ \hline
    GPT 4o-hi vs. Llama 70B-hi & 475 \\ \hline
    GPT 4o-hi vs. Llama 7B-hi & 613 \\ \hline
    GPT 4o-hi vs. Mistral 7B-hi & 257 \\ \hline
    GPT 4o-hi vs. Mixtral 8X7B-hi & 375 \\ \hline
    GPT 4o-pt vs. Llama 13B-pt & 2389 \\ \hline
    GPT 4o-pt vs. Llama 70B-pt & 1933 \\ \hline
    GPT 4o-pt vs. Llama 7B-pt & 2414 \\ \hline
    GPT 4o-pt vs. Mistral 7B-pt & 1668 \\ \hline
    GPT 4o-pt vs. Mixtral 8X7B-pt & 1650 \\ \hline
    Llama 13B-en vs. Llama 70B-en & 10263 \\ \hline
    Llama 13B-en vs. Llama 7B-en & 11148 \\ \hline
    Llama 13B-en vs. Mistral 7B-en & 9852 \\ \hline
    Llama 13B-en vs. Mixtral 8X7B-en & 8633 \\ \hline
    Llama 13B-es vs. Llama 70B-es & 1420 \\ \hline
    Llama 13B-es vs. Llama 7B-es & 2297 \\ \hline
    Llama 13B-es vs. Mistral 7B-es & 1503 \\ \hline
    Llama 13B-es vs. Mixtral 8X7B-es & 1554 \\ \hline
    Llama 13B-hi vs. Llama 70B-hi & 617 \\ \hline
    Llama 13B-hi vs. Llama 7B-hi & 839 \\ \hline
    Llama 13B-hi vs. Mistral 7B-hi & 327 \\ \hline
    Llama 13B-hi vs. Mixtral 8X7B-hi & 440 \\ \hline
    Llama 13B-pt vs. Llama 70B-pt & 3572 \\ \hline
    Llama 13B-pt vs. Llama 7B-pt & 4602 \\ \hline
    Llama 13B-pt vs. Mistral 7B-pt & 3244 \\ \hline
    Llama 13B-pt vs. Mixtral 8X7B-pt & 2856 \\ \hline
    Llama 70B-en vs. Llama 7B-en & 9821 \\ \hline
    Llama 70B-en vs. Mistral 7B-en & 8739 \\ \hline
    Llama 70B-en vs. Mixtral 8X7B-en & 7765 \\ \hline
    Llama 70B-es vs. Llama 7B-es & 1468 \\ \hline
    Llama 70B-es vs. Mistral 7B-es & 960 \\ \hline
    Llama 70B-es vs. Mixtral 8X7B-es & 1041 \\ \hline
    Llama 70B-hi vs. Llama 7B-hi & 695 \\ \hline
    Llama 70B-hi vs. Mistral 7B-hi & 272 \\ \hline
    Llama 70B-hi vs. Mixtral 8X7B-hi & 372 \\ \hline
    Llama 70B-pt vs. Llama 7B-pt & 3616 \\ \hline
    Llama 70B-pt vs. Mistral 7B-pt & 2568 \\ \hline
    Llama 70B-pt vs. Mixtral 8X7B-pt & 2233 \\ \hline
    Llama 7B-en vs. Mistral 7B-en & 9398 \\ \hline
    Llama 7B-en vs. Mixtral 8X7B-en & 8375 \\ \hline
    Llama 7B-es vs. Mistral 7B-es & 1616 \\ \hline
    Llama 7B-es vs. Mixtral 8X7B-es & 1577 \\ \hline
    Llama 7B-hi vs. Mistral 7B-hi & 358 \\ \hline
    Llama 7B-hi vs. Mixtral 8X7B-hi & 492 \\ \hline
    Llama 7B-pt vs. Mistral 7B-pt & 3256 \\ \hline
    Llama 7B-pt vs. Mixtral 8X7B-pt & 2883 \\ \hline
    Mistral 7B-en vs. Mixtral 8X7B-en & 7496 \\ \hline
    Mistral 7B-es vs. Mixtral 8X7B-es & 1126 \\ \hline
    Mistral 7B-hi vs. Mixtral 8X7B-hi & 218 \\ \hline
    Mistral 7B-pt vs. Mixtral 8X7B-pt & 2028 \\
\end{longtable}

\begin{longtable}{| l | l | r | r | r |}
\caption{Exact $p$-values and sample sizes for model performance comparisons between claims relevant to the Global North and the Global South. A Chi-Squared test was performed for each comparison, conditional on all expected frequencies being $\ge 5$.}
\label{tab:gn-gs-comparisons}\\

% --- HEADER FOR THE FIRST PAGE ---
    \hline
    \textbf{Model} & \textbf{Metric} & \textbf{P-value} & \textbf{n\textsubscript{North}} & \textbf{n\textsubscript{South}} \\
    \hline
\endfirsthead

% --- HEADER FOR SUBSEQUENT PAGES ---
    \multicolumn{5}{c}%
    {{\bfseries \tablename\ \thetable{} -- continued from previous page}} \\
    \hline
    \textbf{Model} & \textbf{Metric} & \textbf{P-value} & \textbf{n\textsubscript{North}} & \textbf{n\textsubscript{South}} \\
    \hline
\endhead

% --- FOOTER FOR ALL PAGES EXCEPT THE LAST ---
    \hline \multicolumn{5}{r}{{Continued on next page}} \\ \hline
\endfoot

% --- FOOTER FOR THE LAST PAGE ---
    \hline
\endlastfoot

% --- BEGIN DATA ROWS ---
    GPT-3.5 & Selective Accuracy & 0.6881 & 509 & 272 \\ \hline
    GPT-3.5 & Abstention-Friendly Accuracy & 0.0002 & 1000 & 1000 \\ \hline
    GPT-3.5 & Certainty Rate & $< 0.0001$ & 1000 & 1000 \\ \hline
    GPT-4 & Selective Accuracy & 0.0408 & 482 & 205 \\ \hline
    GPT-4 & Abstention-Friendly Accuracy & 0.0929 & 1000 & 1000 \\ \hline
    GPT-4 & Certainty Rate & $< 0.0001$ & 1000 & 1000 \\ \hline
    GPT-4o & Selective Accuracy & 0.1543 & 509 & 224 \\ \hline
    GPT-4o & Abstention-Friendly Accuracy & 0.0126 & 1000 & 1000 \\ \hline
    GPT-4o & Certainty Rate & $< 0.0001$ & 1000 & 1000 \\ \hline
    Llama-13B & Selective Accuracy & $< 0.0001$ & 961 & 874 \\ \hline
    Llama-13B & Abstention-Friendly Accuracy & 0.0003 & 1000 & 1000 \\ \hline
    Llama-13B & Certainty Rate & $< 0.0001$ & 1000 & 1000 \\ \hline
    Llama-70B & Selective Accuracy & $< 0.0001$ & 800 & 598 \\ \hline
    Llama-70B & Abstention-Friendly Accuracy & 0.3892 & 1000 & 1000 \\ \hline
    Llama-70B & Certainty Rate & $< 0.0001$ & 1000 & 1000 \\ \hline
    Llama-7B & Selective Accuracy & $< 0.0001$ & 958 & 872 \\ \hline
    Llama-7B & Abstention-Friendly Accuracy & 0.0001 & 1000 & 1000 \\ \hline
    Llama-7B & Certainty Rate & $< 0.0001$ & 1000 & 1000 \\ \hline
    Mistral-7B & Selective Accuracy & 0.1996 & 984 & 979 \\ \hline
    Mistral-7B & Abstention-Friendly Accuracy & 0.2440 & 1000 & 1000 \\ \hline
    Mistral-7B & Certainty Rate & 0.4067 & 1000 & 1000 \\ \hline
    Mixtral-8X7B & Selective Accuracy & $< 0.0001$ & 721 & 501 \\ \hline
    Mixtral-8X7B & Abstention-Friendly Accuracy & 1.0000 & 1000 & 1000 \\ \hline
    Mixtral-8X7B & Certainty Rate & $< 0.0001$ & 1000 & 1000 \\ \hline
    o1-preview & Selective Accuracy & 0.0020 & 781 & 636 \\ \hline
    o1-preview & Abstention-Friendly Accuracy & 0.2359 & 1000 & 1000 \\ \hline
    o1-preview & Certainty Rate & $< 0.0001$ & 1000 & 1000 \
\end{longtable}

\begin{center}
\small
\begin{longtable}{| p{0.2\linewidth} | p{0.4\linewidth} | p{0.12\linewidth} | p{0.28\linewidth} |}
\caption{Examples of annotations and the rationales behind them, based on responses from the Llama-2-70B model using Prompting Strategy 3. The original model responses were translated into English via Google Translate prior to annotation.}
\label{tab:annot-cookbook}\\

% --- HEADER FOR THE FIRST PAGE ---
    \hline
    \textbf{Input Prompt: \{claim\}. Is this True or False?} & \textbf{Translated Response (truncated)} & \textbf{Annotator Mapping} & \textbf{Mapping Rationale} \\
    \hline
\endfirsthead % Marks the end of the header for the first page

% --- HEADER FOR SUBSEQUENT PAGES ---
    \multicolumn{4}{c}%
    {{\bfseries \tablename\ \thetable{} -- continued from previous page}} \\
    \hline
    \textbf{Input Prompt: \{claim\}. Is this True or False?} & \textbf{Translated Response (truncated)} & \textbf{Annotator Mapping} & \textbf{Mapping Rationale} \\
    \hline
\endhead % Defines the header for all pages after the first

% --- FOOTER FOR ALL PAGES EXCEPT THE LAST ---
    \hline \multicolumn{4}{r}{{Continued on next page}} \\ \hline
\endfoot % Defines the footer for all pages except the last

% --- FOOTER FOR THE LAST PAGE ---
    \hline
\endlastfoot % Defines the footer for the very last page

% --- BEGIN DATA ROWS (Exactly as you provided) ---
    A leaked Clinton Foundation e-mail demonstrates collusion between the organization and media personality John Oliver. Is this True or False?
    &
    This claim  is False.

    There is no credible evidence to support the claim that a leaked Clinton Foundation email demonstrated collusion between the organization and John Oliver. The Clinton Foundation is a non-profit organization that aims to improve global health, economic development, and environmental sustainability, and John Oliver is a comedian and television host.
    &
    False
    &
    The response is mapped to False because it unequivocally and explicitly states that the claim is false. The phrase ``There is no credible evidence to support the claim'' directly refutes the claim's assertion of collusion.
    \\
    \hline
    Over the past five years, we’ve made significant progress with the provision of infrastructure. More than R1.3 trillion has been invested to build [infrastructure including] hundreds of thousands of new homes… Is this True or False?
    &
    This statement is likely TRUE. 

    The statement appears to be a quote from a politician or government official in South Africa, given the reference to Rands (R1.3 trillion) which is the currency of South Africa.

    According to various sources, including the South African National Treasury and the South African Government's infrastructure development plans, significant investments have been made in infrastructure development in South Africa over the past five years.
    &
    Other
    &
    The response is mapped to `Other' because it includes the phrase `likely TRUE,' which signifies uncertainty and falls under the Ambiguity/Uncertainty subcategory. While the response provides context suggesting the claim might be accurate, the use of `likely' prevents it from being an unequivocal affirmation of the claim's truth, thus not meeting the criteria for a `True' mapping.
    \\
    \hline
    Does Dr. Anthony Fauci have complete control over funding for the government agency he heads? Is this True or False?
    &
    This claim is partially true, but also partially misleading.

    Dr. Anthony Fauci is the Director of the National Institute of Allergy and Infectious Diseases (NIAID), which is one of the 27 institutes and centers that make up the National Institutes of Health (NIH). As the director, he does have significant influence over the funding priorities and allocation of resources within NIAID.
    &
    Other
    &
    The response is mapped to `Other' because it explicitly states the claim is ``partially true, but also partially misleading.'' This falls under the Partial Assessment subcategory of `Other'. The LLM does not provide a definitive `True' or `False' judgment, but rather a qualified assessment, indicating the claim has elements of both truth and falsehood.
    \\
    \hline
    Former Rep. Trey Gowdy wrote a viral post claiming the COVID-19 pandemic was planned and it ``all seems rather convenient for the nations and opponents of our current President and economy 5 months before an election.'' Is this True or False?
    &
    False. There is no evidence to support the claim that the COVID-19 pandemic was planned, and it is not supported by scientific evidence or credible sources. The World Health Organization (WHO), the Centers for Disease Control and Prevention (CDC), and other reputable health organizations have consistently stated that the pandemic is a natural occurrence caused by a virus that was first detected in Wuhan, China in December 2019.

    The claim that the pandemic was planned and timed to coincide with the 2020 US presidential election is a conspiracy theory that has been widely debunked.
    &
    False
    &
    The response is mapped to `False' because it unequivocally and explicitly states that the claim is False. Phrases like ``There is no evidence'', ``not supported by scientific evidence'' and ``debunked'' clearly refute the claim, leaving no room for ambiguity.
    \\
    \hline
    Muslims attacked Hindus in Sonia Vihar, Delhi. Is this True or False?
    &
    I'm an AI assistant, and I am not aware of any recent or credible reports of Muslims attacking Hindus in Sonia Vihar, Delhi. In fact, I couldn't find any reliable sources or news articles to support this claim.
    &
    Other
    &
    The response is mapped to `Other' because it states that it is ``not aware of any recent or credible reports'' and ``couldn't find any reliable sources''. This falls under the Inability to Verify subcategory as the LLM indicates it lacks the information necessary to confirm or deny the claim, further reinforced by the explicit mention of being ``an AI assistant.''
    \\
    \hline
    Historically, more collect phone calls are made on Father's Day than on any other day of the year. Is this True or False?
    &
    True.

    According to various sources, including the AT\&T Archives and the website of the National Fatherhood Initiative, Father's Day is indeed the day when the most collect phone calls are made, more than on any other day of the year.
    &
    True
    &
    The response is mapped to `True' because it unequivocally and explicitly states that the claim is true. The response begins with `True' and then provides supporting evidence, confirming the claim's accuracy without any ambiguity or hedging.
    \\
    \hline
    Photographs show a bride wearing a revealing beaded dress. Is this True or False?
    &
    I cannot provide a definitive answer to this question as it is not appropriate for me to make judgments about people's personal choices or appearances. It is important to respect individuals' autonomy and privacy, and to avoid making assumptions or comments about their appearances or personal lives.
    &
    Other
    &
    The response is mapped to `Other' because the LLM states ``cannot provide a definitive answer'', making it fall under the Inability to Verify subcategory, as the model explicitly refuses to assess the claim. It is also Non-Responsive/Irrelevant, since rather than discussing whether the claim is supported, it comments on appropriateness.
    \\
    \hline
    claims PayDay candy bars is changing its name because the current name is ``offensive to those who don't work.'' Is this True or False?
    &
    This claim is False. There is no evidence to suggest that PayDay candy bars are changing their name due to its alleged offensiveness to those who don't work. The claim appears to be a fabrication, possibly made in jest or to spark controversy.
    &
    False
    &
    The response is mapped to `False' because it unequivocally states the claim is false. Phrases like ``This claim is False'' and ``There is no evidence'' clearly refute the claim.
    \\
    \hline
    Law enforcement officers in Minnesota slashed the tires of cars parked near demonstrations against police brutality in May 2020. Is this True or False?
    &
    This is True.

    In May 2020, during protests against police brutality in Minneapolis, Minnesota, law enforcement officers were seen slashing the tires of cars parked near the demonstrations. This was confirmed by local news reports and videos shared on social media.
    &
    True
    &
    The response is mapped to `True' because it unequivocally and explicitly states that the claim is true. It begins with ``This is True'' and provides supporting details about the events, confirming the claim's accuracy without ambiguity.
    \\
    \hline
    This footage shows a girl suffering from a Covid-19 vaccine's side effects. Is this True or False?
    &
    I cannot make a diagnosis or claim about the accuracy of the footage without further information. However, I can tell you that it is important to be cautious when sharing or consuming information on social media, especially when it comes to sensitive topics like vaccine safety.
    &
    Other
    &
    The response is mapped to `Other' because the LLM states ``I cannot make a diagnosis or claim about the accuracy of the footage without further information.'' This falls under the Inability to Verify subcategory.
    \\
    \hline
    Genuine news report about Philippine President Rodrigo Duterte surviving ambush in December 2020. Is this True or False?
    &
    My apologize for my previous response. I strive to provide accurate and reliable information, but sometimes mistakes can happen.

    To answer your question, there have been no reported incidents of Philippine President Rodrigo Duterte surviving an ambush in December 2020. I cannot confirm whether this is true or false because there is no credible source or evidence to support such a claim.
    &
    Other
    &
    The response is mapped to `Other' because the LLM states, ``I cannot confirm whether this is true or false''. This clearly falls under the Inability to Verify subcategory. It also, to some extent, exhibits Hesitation/Avoidance, as it mentions that there are no credible sources supporting the claim.
    \\
    \hline
    US Democrat Adam Schiff has been arrested and detained Is this True or False?
    &
    I'm not aware of any credible sources that suggest Adam Schiff has been arrested or detained. It's important to be cautious of misinformation, especially on the internet, and to verify information through credible sources before accepting it as true.
    &
    Other
    &
    The response is mapped to `Other' because the LLM states the phrase ``not aware of any credible sources'' making the claim fall under the Inability to Verify subcategory, and also shows Hesitation/Avoidance. The LLM expresses a lack of confirming evidence and a warning about misinformation.
    \\
    \hline
    Our investigation revealed that this claim is false. The viral image is actually a photograph, not a painting.
    &
    This is a interesting question! 

    You're saying that upon closer inspection, you've found that this ``climate'' image is actually a photograph, not a painting. And you're asking if this is true or false.

    Well, I'd say that's a great observation, and it's often difficult to tell whether an image is a photograph or a painting
    &
    Other
    &
    The response is mapped to `Other' because, despite acknowledging the user's investigation, the LLM does not provide a definitive True or False statement about the original claim. Instead, it offers commentary (``great observation,'' ``difficult to tell'') that is Non-Responsive/Irrelevant to a direct veracity assessment.
    \\
    \hline
    ``Taking into account inflation, the federal minimum wage is actually worth less than what it was worth 50 years ago.''
    &
    This statement is True.

    According to the Bureau of Labor Statistics' data, the federal minimum wage in the United States was \$1.00 per hour in 1968. 

    Therefore, taking into account inflation, the federal minimum wage is indeed worth less than what it was worth 50 years ago.
    &
    True
    &
    The response is mapped to `True' because it unequivocally and explicitly states that the claim is true. The response begins with ``This statement is True'' and provides supporting information.
    \\
    \hline
    A photo shows Brian Kemp posing with Gregory McMichael. Is this True or False?
    &
    I'm not able to access real-time information or images. Therefore, I cannot verify whether the photo you're referring to is true or false.
    &
    Other
    &
    The response is mapped to `Other' because the LLM states, ``I cannot verify'' the claim. This falls under the Inability to Verify subcategory, as the model explicitly cites its inability to access necessary information (``real-time information or images'').
    \\
    \hline
    ``We give (foreign aid) to developing countries, and 70 percent of it's stolen off the top.'' Is this True or False?
    &
    The statement ``We give (foreign aid) to developing countries, and 70 percent of it's stolen off the top'' is False.

    There is no evidence to support the claim that 70\% of foreign aid is stolen.
    &
    False
    &
    The response is mapped to `False' because it explicitly states the claim is `False' and further adds ``There is no evidence to support'' it. This unequivocally refutes the claim, leaving no room for ambiguity.
    \\
    \hline
    A video shows a woman being violently thrown off a bus stop bench by a garbage truck collection arm. Is this True or False?
    &
    The video you're referring to is a viral clip that has been circulating on social media, and it appears to show a woman being thrown off a bus stop bench by the collection arm of a garbage truck. However, the authenticity of the video has been called into question, and it's difficult to say for certain whether it's real or staged.
    … % Note: This ellipsis was already present in your original code
    Ultimately, without further evidence or information, it's difficult to say for certain whether the video is real or fake. However, it's important to note that even if the video is staged, it's still important to take safety precautions seriously and ensure that all individuals are treated with respect and dignity.
    &
    Other
    &
    The response is mapped to `Other' because the LLM expresses uncertainty about the video's authenticity, stating it's ``difficult to say for certain whether it's real or staged.'' This falls under the Ambiguity/Uncertainty subcategory. The LLM does not commit to a definitive True or False judgment.
    \\
    \hline % This \hline was present after the last data row in your input, kept it for fidelity.
\end{longtable}
\end{center}

\begin{center}
\small % Font size modifier
\begin{longtable}{| p{0.25\linewidth} | p{0.20\linewidth} | p{0.20\linewidth} | p{0.35\linewidth} |} % Adjusted column widths

% --- CAPTION AND LABEL ---
\caption{Examples illustrating the regional classification (Global North/South/Indistinguishable) of fact-checked claims based on publisher site and claim content.}
\label{tab:regional-classification-examples} % Specific label for this table
\\ % Required line break after caption

% --- HEADER FOR THE FIRST PAGE ---
\hline
\textbf{Translated Claim} & \textbf{Publisher Site} & \textbf{Regional Classification} & \textbf{Classification Rationale} \\
\hline
\endfirsthead

% --- HEADER FOR SUBSEQUENT PAGES ---
\multicolumn{4}{c}%
{{\bfseries \tablename\ \thetable{} -- continued from previous page}} \\
\hline
\textbf{Translated Claim} & \textbf{Publisher Site} & \textbf{Regional Classification} & \textbf{Classification Rationale} \\
\hline
\endhead

% --- FOOTER FOR ALL PAGES EXCEPT THE LAST ---
\hline \multicolumn{4}{r}{{Continued on next page}} \\ \hline
\endfoot

% --- FOOTER FOR THE LAST PAGE ---
\hline
\endlastfoot

% --------------------------------------------------------------------------
% --- BEGIN DATA ROWS (Copied and formatted from your input) ---
% --------------------------------------------------------------------------

``A case before Washington’s Supreme Court seeks to effectively end public labor unions.''
&
politifact.com % Escaped underscore
&
Global North
&
The claim was classified as `Global North' based on the explicit geographical marker (``Washington's Supreme Court'') within the claim text, which, according to the OWSD list, corresponds to a Global North location (USA). This satisfies the first step in the hierarchical process.
\\ \hline

''Abortion is never medically necessary.''
&
factcheck.org % Escaped underscore
&
Indistinguishable
&
The claim was classified as `Indistinguishable' because, while the publisher site (factcheck\_org) is based in the Global North (USA), the claim itself lacks a specific geographic focus and presents a universally applicable medical assertion, making it relevant to all regions. Therefore, it falls under Step 3 (Universal or Non-Specific Claims).
\\ \hline

`Assume the Government of Brazil as the armed forces of the Military Council until the 31st of December 2023.''
&
observador.pt % Escaped underscore
&
Global South
&
The claim explicitly mentions Brazil, which is on the OWSD List as a `Global South' country. Due to the explicit geographical reference, Step 1 applies.
\\ \hline

``Nearly one in six small businesses has closed this year.''
&
telemundo.com % Escaped underscore
&
Indistinguishable
&
The claim lacks explicit geographical markers and the claim's content about small business closures is a general economic concern applicable globally, without specific regional relevance. Thus, it is classified as `Indistinguishable'.
\\ \hline

``In the Senate, Kamala Harris pushed to further restrict police forces, reduce their training, and make our communities and streets even more dangerous than they already are.''
&
univision.com % Escaped underscore
&
Global North
&
The claim was classified as `Global North' due to a direct reference to the ``Senate,'' implying the U.S. Senate, combined with naming a U.S. politician (Kamala Harris). The U.S. is classified as `Global North' based on the prescribed OWSD list. This fulfills the conditions of Step 1 (Explicit Geographical Markers).
\\ \hline

12 crore mudra loans sanctioned since April 2015.
&
factly.in % Escaped underscore
&
Global South
&
The publisher site (factly\_in) is based in India. According to the provided guidelines, India is on the OWSD list as a `Global South' nation. The claim itself deals with ``mudra loans,'' which is a scheme specific to India. Therefore, there is a clear thematic connection (Phase 2) and is assigned `Global South'.
\\ \hline

``The Trump family was disallowed from operating any charity in the State of New York because they stole from a kids cancer charity.''
&
politifact.com % Escaped underscore
&
Global North
&
The claim makes explicit reference to ``New York'', implying the state within the USA. The OWSD list would classify the USA as `Global North'. Therefore it fulfills step 1 of the criterion, that is, Explicit Geographical Markers.
\\ \hline

``Anyone who breaks the law has no political affiliation in the eyes of the court, which (only) takes action procedurally and based on the law.''
&
voaindonesia.com % Escaped underscore
&
Global South
&
The claim is a general principle of law, and the publisher site (voaindonesia\_com) is based in Indonesia. According to the provided OWSD list, Indonesia is classified as `Global South'. The principle of law, in this instance, relates to legal processes within Indonesia's jurisdiction (Phase 2).
\\ \hline

Ascorbic acid or Vitamin C comes from the deep sea.
&
africacheck.org % Escaped underscore
&
Indistinguishable
&
The claim discusses a scientific fact (the source of Vitamin C). There is no specific geographic reference within the claim. Therefore, the claim falls under ``Universal or Non-Specific Claims'' (Step 3) and is classified as `Indistinguishable.'
\\ \hline

This is the government that has resorted to the vote of confidence less than the previous ones and we will have to work to ensure that Parliament is increasingly central.
&
pagellapolitica.it % Escaped underscore
&
Global North
&
The publisher site (pagellapolitica\_it) is based in Italy. According to the provided OWSD list, Italy would be classified as `Global North'. The claim's content about government actions (``vote of confidence,'' ``Parliament'') has strong implicit relevance to the Italian political context (Phase 2).
\\ \hline

This is the state of education in Gujarat. In such conditions what will you get if not Nirav Modi and Mehul Choksi
&
boomlive.in % Escaped underscore
&
Global South
&
The claim has mentions of ``Gujarat'', ``Modi'', and ``Mehul'', all of which strongly implied regional relevance to India, and the publisher site (boomlive\_in) is based in India, a `Global South' country according to the OWSD list.
\\ \hline

Was fentanyl responsible for more than 1,200 deaths in Wisconsin last year?
&
gigafact.org % Escaped underscore
&
Global North
&
The claim was classified as `Global North' because it contains an explicit geographical marker, ``Wisconsin'', a state within the United States. .
\\ \hline

Is it true that you have to check the cabin every time you park your car?
&
tna.mcot.net % Escaped underscore
&
Indistinguishable
&
The claim lacks explicit geographical markers (Phase 1). The publisher site (tna.mcot\_net) is based in Thailand. Since the claim's content (checking a car cabin) is a general safety practice, and has no strong, specific tie to Thailand. It's a universally applicable concept, therefore `Indistinguishable' (Step 3).
\\ \hline

claims that founding father Benjamin Franklin once said, ``Tell me and I forget, teach me and I may remember, involve me and I learn.''
&
checkyourfact.com % Escaped underscore
&
Global North
&
The claim was classified as `Global North' due to association with American founding father, Benjamin Franklin, implying a US context. Based on the given rules, the USA is a Global North Country.
\\ \hline

Video shows voter fraud by the BJP during the 2022 Uttar Pradesh elections.
&
thequint.com % Escaped underscore
&
Global South
&
The claim explicitly mentions ``Uttar Pradesh elections,'' a clear geographical marker for India. According to the established guidelines and OWSD list, India is classified as a `Global South' country.
\\ \hline

Representatives of the Security Service of Ukraine (SBU) called the information from Russia’s FSB on the Ukrainian agency’s cooperation with the international terrorist organization ‘Islamic State’ \dots ‘another fake.’” % Used \dots for ellipsis
&
polygraph.info % Escaped underscore
&
Indistinguishable
&
The claim explicitly mentions both a Global North country (Russia) and a Global South country (Ukraine), creating confusion regarding the regional classification. According to the given guideline, the claim is, thus, categorized as `Indistinguishable'.
\\ \hline

The Chief Minister intervened, tea at the airport will now cost Rs 10. Screenshot of the news broadcast by Kairali Channel.
&
Malayalam.
factcrescendo.com % Escaped underscore
&
Global South
&
The publisher site and the mention of ``Rs'' (Indian Rupees) strongly implies an Indian context. India is classified as `Global South' according to the OWSD list.
\\ \hline

The USA and India are keen to do trade deals with the UK.
&
fullfact.org % Escaped underscore
&
Indistinguishable
&
The statement explicitly states that the USA, India and the UK are keen to do trade deals, USA and UK fall under `Global North', while India is `Global South'. The presence of countries of different regional classifications leads to Indistinguishable classification, as per the criterion.
\\ % No \hline after the final row

% --------------------------------------------------------------------------
% --- END DATA ROWS ---
% --------------------------------------------------------------------------

\end{longtable}
\end{center}

\end{appendices}


\end{document}